\documentclass[a4paper,11pt]{article}
\pdfoutput=1 

\usepackage[english]{babel}
\usepackage{jheppub}
\usepackage[T1]{fontenc}
\usepackage{graphicx,calc}
\usepackage{epsfig}
\usepackage{amsmath}
\usepackage{amssymb}
\usepackage{float}
\usepackage{placeins}
\usepackage{braket}
\usepackage{slashed}
\usepackage{mathdots}
\usepackage{lipsum}
\usepackage{feynmf}
\usepackage{bbm}
\usepackage{color}
\usepackage{multicol}
\usepackage{caption}
\usepackage{array}
\usepackage[export]{adjustbox}
\usepackage{subcaption}
\usepackage{multicol}
\usepackage{comment}
\usepackage{colortbl}
\usepackage[most]{tcolorbox}
\definecolor{myred}{RGB}{168,4,4}
\definecolor{myblue}{RGB}{48,53,149}

\makeatletter
\newcommand\xleftrightarrow[2][]{%
  \ext@arrow 9999{\longleftrightarrowfill@}{#1}{#2}}
\newcommand\longleftrightarrowfill@{%
  \arrowfill@\leftarrow\relbar\rightarrow}
\makeatother

\allowdisplaybreaks

\title{A first computation of three-loop master integrals for the production of two off-shell vector bosons with different masses}

\author[]{Dhimiter Canko}
\author[]{and Mattia Pozzoli}
\affiliation[]{Dipartimento di Fisica e Astronomia, Università di Bologna \\
INFN, Sezione di Bologna, \\
via Irnerio 46, I-40126 Bologna, Italy}
\emailAdd{dhimiter.canko2@unibo.it}
\emailAdd{mattia.pozzoli@unibo.it}

\abstract{We present analytic results on physical kinematics for four integral families that are relevant to the production of two off-shell vector bosons with different masses. Our study consists of a ladder-box, a tennis-court, and two reducible ladder-box-like families. The results for the master integrals of these families are expressed up to order six in the dimensional regulator in terms of real-valued multiple polylogarithms. Furthermore, a semi-numeric solution is provided, employing generalized power series expansions using the package \texttt{DiffExp}.}


\begin{document} 
\maketitle
\flushbottom

\section{Introduction}
\label{Introduction}

As a fundamental component in the calculation of multi-loop scattering amplitudes in perturbative quantum field theory, Feynman integrals play a pivotal role in the process of obtaining high-precision theoretical predictions for measurable observables at particle colliders such as the Large Hadron Collider. In addition to their practical importance in predicting particle interactions, Feynman integrals are also of great mathematical interest due to  the fact that their solutions involve several complex multivalued functions, such as multiple polylogarithms~\cite{Goncharov:1998kja, Goncharov:2010jf}, elliptic integrals~\cite{Broedel:2018qkq}, and other special functions associated with surfaces of higher genus~\cite{Bourjaily:2022bwx}. It is also possible that the theory may contain other, more complicated special functions that have not yet appeared in current computations.

Feynman integrals are usually calculated analytically or numerically within the framework of dimensional regularisation~\cite{Weinzierl:2022eaz, Badger:2023eqz}, where the space-time dimension is shifted from four to $d=4-2\varepsilon$ in order to quantify and handle infrared and ultraviolet divergences. The method of differential equations~\cite{Kotikov:1990kg, Bern:1993kr, Remiddi:1997ny, Gehrmann:1999as} is frequently employed for analytical calculations of Feynman integrals. In this approach, Feynman integrals are grouped into integral families based on their propagator structure and kinematic characteristics (number and type of external particles), and first-order differential equations are constructed with respect to kinematic invariants for a basis of Feynman integrals, known as master integrals, using integration-by-parts identities~\cite{Tkachov:1981wb,Chetyrkin:1981qh}. These identities establish a linear relation between any Feynman integral within a given family and the master integrals, with the coefficients of these relations being algebraic functions of the kinematic invariants and the dimensional regulator $\varepsilon$. Integration-by-parts identities can be solved using Laporta's algorithm~\cite{Laporta:2000dsw}.


Over the last decade, two novel methods have been developed and have enhanced the applicability and resolution of the method of differential equations. The first method consists in selecting a basis of master integrals such that the system of differential equations is brought in the canonical form~\cite{Henn:2013pwa}. In general, this form is characterized by the factorization of $\varepsilon$ out of the matrices of the differential equations, usually called connection matrices, and the presence of singularities of logarithmic type only. In the best-understood cases, this form is characterized by differential equations that can be rewritten in terms of logarithmic one-forms. 
The second novel method is the utilization of finite field modular arithmetic for the numerical solution of integration-by-parts identities. This method allows to work numerically in intermediate steps of the computations, and later reconstruct analytic expressions in the final results, which are usually much simpler~\cite{vonManteuffel:2014ixa, Peraro:2016wsq, Peraro:2019svx,  Klappert:2019emp}, i.e.~as it happens for the matrices of the canonical DEs. These new developments have prompted the calculation of numerous multi-loop Feynman integral families, establishing the current frontier at families related to five-particle kinematics for two loops\footnote{First steps for the computation of six-particle families have also been made recently at two loops~\cite{Henn:2024ngj}.}~\cite{Papadopoulos:2015jft, Gehrmann:2018yef, Chicherin:2018mue, Chicherin:2018old, Abreu:2020jxa, Chicherin:2020oor, Canko:2020ylt, Abreu:2021smk, Chicherin:2021dyp, Kardos:2022tpo, Abreu:2023rco, Badger:2022hno, FebresCordero:2023pww, Badger:2024fgb, Abreu:2024flk} and to four-particle kinematics at three loops~\cite{Smirnov:2003vi, Henn:2013tua, Henn:2013nsa, DiVita:2014pza, Henn:2020lye, Canko:2020gqp, Canko:2021xmn, Henn:2023vbd, Syrrakos:2023mor, Long:2024bmi, Gehrmann:2024tds, Henn:2024pki}. Families with up to two massive (same mass) propagators and up to three massive (two same masses and one distinct) external particles have been considered in the former case. In contrast, only families with massless propagators and up to two massive external particles, that share the same mass, have been studied in the latter case. More recently also problems with five-particle kinematics at three loops~\cite{Liu:2024ont} have been tackled.

In this work, we extend the current three-loop frontier by computing, in terms of multiple polylogarithms, four three-loop Feynman integral families describing four-particle processes where two of the external particles have unequal masses. In particular, we calculate in physical kinematics one ladder-box, one tennis-court, and two reducible ladder-box-like families. The internal propagators are massless and for the two irreducible families the massive external particles are attached to the same loop line. These families are required for the study of three-loop scattering amplitudes for processes such as $q \bar{q}' \to V_1 V_2 \to (l_1 \bar{l}_1')(l_2 \bar{l}_2')$ and $g g \to V_1 V_2 \to (l_1 \bar{l}_1')(l_2 \bar{l}_2')$ (for massless quarks propagating within the loops), where the intermediate vector bosons can be arbitrary electroweak gauge bosons, e.g. $\gamma^*\gamma^*,  W^+W^-, ZZ, W^{\pm} Z, W^{\pm} \gamma^*, Z \gamma^*$. Ultimately, these families are key for making N3LO QCD predictions for the process $pp \to V_1 V_2$\footnote{The gluon channel $gg \to V_1 V_2$ is formally NNLO}, respectively, which will be important for future comparisons with experimental data. This is true since future high-luminosity LHC runs at a center-of-mass energy of 14 TeV, combined with the clear signatures left by the production of vector bosons due to their leptonic decays, will allow observables to be measured with an accuracy of well below one percent~\cite{Gehrmann:2021qex}. This level of precision can be achieved theoretically only if three-loop corrections are included in the computation. For the aforementioned processes, two-loop calculations of the integral families~\cite{Henn:2014lfa, Caola:2014lpa} and the scattering amplitudes~\cite{Caola:2014iua, Caola:2015ila, Gehrmann:2015ora, vonManteuffel:2015msa} were accomplished approximately a decade ago, while, numerous results for NNLO and NLO cross-sections have emerged in recent years~\cite{Grazzini:2015wpa, Grazzini:2015hta, Caola:2015psa, Grazzini:2016swo, Campbell:2016ivq, Grazzini:2016ctr, Grazzini:2017ckn, Heinrich:2017bvg, Re:2018vac, Kallweit:2018nyv, Grazzini:2018owa, Grazzini:2019jkl, Kallweit:2020gva, Poncelet:2021jmj, Lombardi:2021rvg, Degrassi:2024fye}.

The structure of this paper is organized in the following way. In section~\ref{Kinematics_and_Families}, we introduce our notation for the kinematics and the four families under consideration. In section~\ref{Pure_Bases_and_DEs}, we describe the methods employed for constructing pure bases for these families, comment on the form and the features of their canonical differential equations, and explain the approach we used for analytically obtaining boundary conditions for their pure bases elements. In section~\ref{Results}, we validate and present our analytic and semi-numeric (obtained using \texttt{DiffExp}) results for the bases elements, commenting on the evaluation efficiency of the two results. Finally, in section~\ref{Conclusions}, we conclude and discuss future extensions of our work and possible developments in the same direction.
\section{Kinematics and families}
\label{Kinematics_and_Families}
In this section, we describe the setup of our calculation. We define the scattering kinematics and the physical scattering region. We then introduce our conventions for the Feynman integrals that we compute. In particular, we define two superfamilies of integrals, to which the four families we study in this work belong. We define the four families based on their top sector and we justify our choice to study those four.

\subsection{Notation}
\label{Kinematics}
We consider the scattering of four particles with momenta $(p_1,p_2,p_3,p_4)$. The first two of these particles are massless, i.e.~$p_1^2=p_2^2=0$, and the remaining two massive, with masses $p_3^2 = m_3^2$ and $p_4^2=m_4^2$, respectively. By selecting the momenta to be incoming, the momentum conservation takes the form $p_4=-p_1-p_2-p_3$. All the kinematic invariants can be expressed in terms of four independent Mandelstam variables, chosen to be
\begin{equation}
s_{12}=(p_1+p_2)^2, \qquad s_{23}=(p_2+p_3)^2, \qquad m_3^2=p_3^2,  \qquad m_4^2=p_4^2.
\end{equation}
In this scattering process, the physical region described in terms of the variables selected above is defined by the following inequalities
\begin{equation}
\label{physMand}
\begin{split}
&m_3^2 > 0, \qquad m_4^2 > 0, \qquad s_{12} >(m_3+m_4)^2,\\
&\frac{m_3^2 + m_4^2 - s_{12} - R}{2} \leq s_{23} \leq \frac{m_3^2 + m_4^2 - s_{12} + R}{2},
\end{split}
\end{equation}
where 
\begin{equation}
\label{R_invs}
R=\sqrt{s_{12}^2 + (m_4^2 - m_3^2)^2 - 2 s_{12} (m_4^2 + m_3^2)}.
\end{equation}
As in~\cite{Henn:2014lfa}, the square root $R$ can be rationalized by transforming into the variables $(x,y,z)$ defined by
\begin{equation}
\frac{s_{12}}{m_3^2}= (1+x)(1+x y), \qquad \frac{s_{23}}{m_3^2}=-x z, \qquad \frac{m_4^2}{m_3^2}=x^2 y,
\end{equation}
with $R=m_3^2 x(1-y)$. We will refer to these variables as rationalized ones. Using these variables to describe our problem, we can then set $m_3^2=1$ given that the dependence of the Feynman integrals on this variable can be subsequently recovered through the application of dimensional arguments. In this parametrization, the physical region is convex and assumes the following form\footnote{More specifically, Eq.~\eqref{physMand} splits the kinematic space into two regions, the one of Eq.~\eqref{physxyz} and the one where $x > 0$, $y > 1$ and $1 < z < y$, but by selecting the branch where $R$ is positive, the second region is omitted.}
\begin{equation}
\label{physxyz}
x > 0, \qquad y > 0, \qquad y < z < 1.
\end{equation}

\subsection{Integral families}
\label{Families}

\begin{figure}[t]
    \centering
    \includegraphics[scale=0.27]{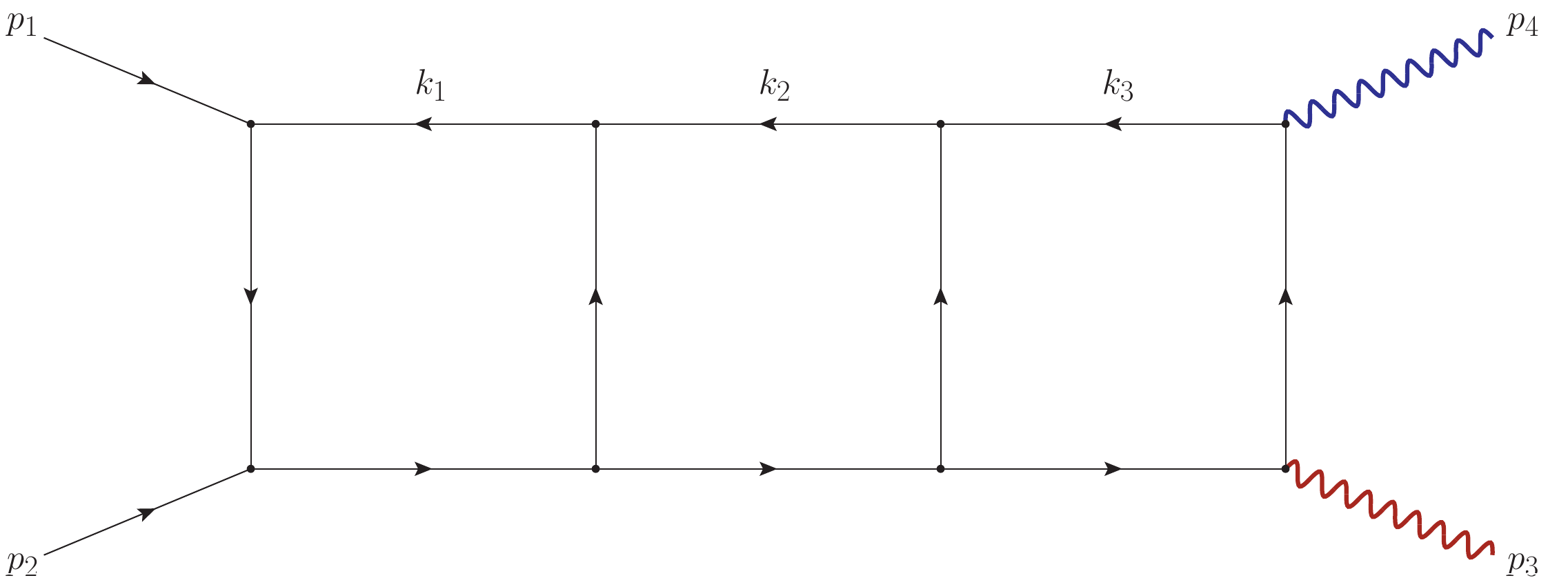}
    \includegraphics[scale=0.22]{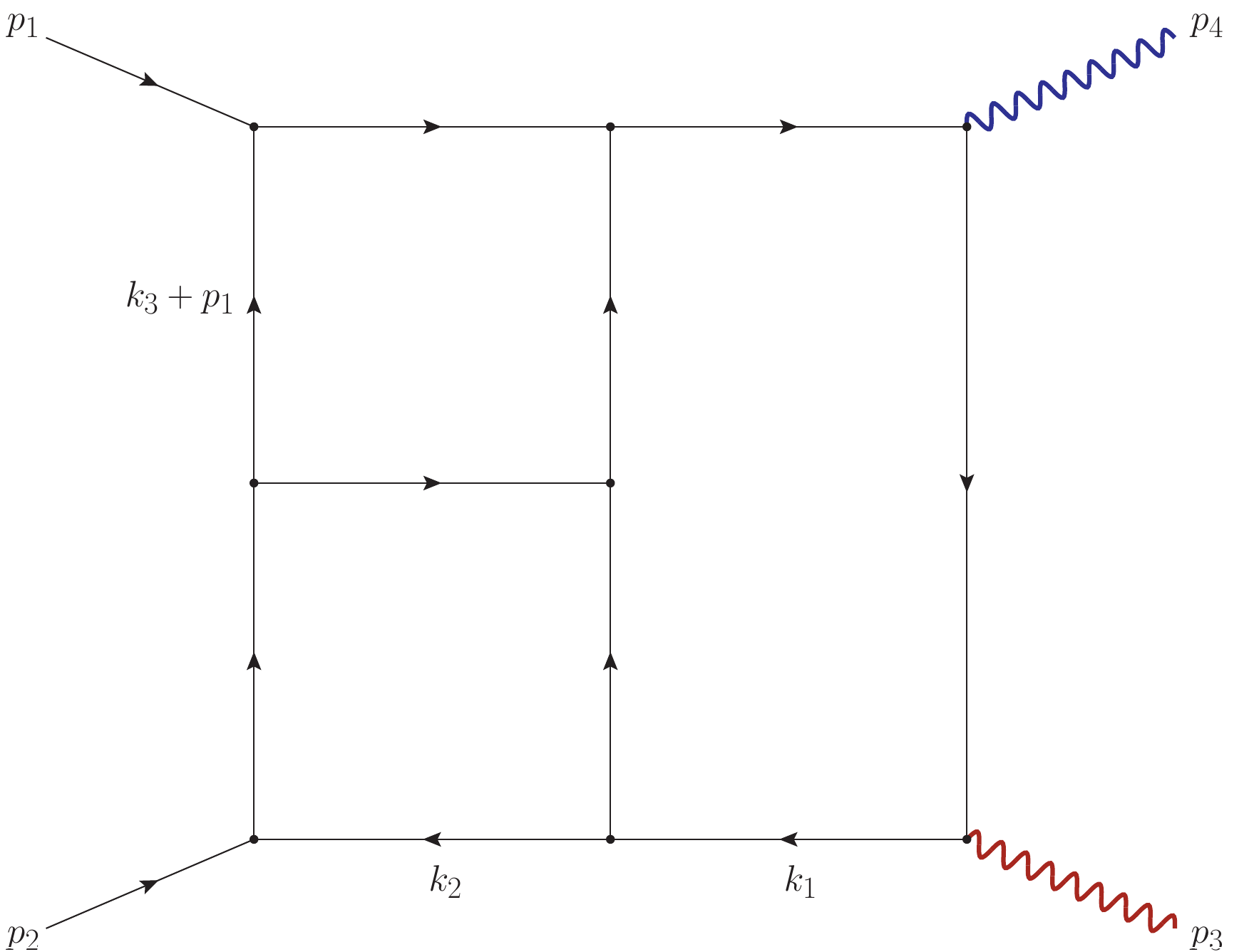}
    \caption{The two irreducible integral families. The upper one is $\mathrm{PL}_1$ and the lower one is $\mathrm{PT}_4$. The black lines represent massless particles, while the two curled color lines correspond to the two vector bosons. The particle with mass $m_4$ ($m_3$) is represented by a blue (red) color.}
    \label{fig:irreducible}
\end{figure}

All of the Feynman integrals that occur in a three-loop four-point scattering amplitude can be written in the following standard form
\begin{equation}
G_{a_1,\ldots,a_{15}}= e^{3 \gamma_E \varepsilon } \int \frac{\prod_{j=1}^3 \mathrm{d}^dk_j}{(i \pi)^{3 d/2}} \frac{1}{D_1^{a_1} \cdots D_{15}^{a_{15}}}, \qquad (a_1, \dots, a_{15}) \in \mathbb{Z}^{15},
\end{equation}
where $a_i \leq 0$ for the propagators corresponded to irreducible scalar products (ISPs).

The integral families treated in this paper can be combined for convenience into two propagator superfamilies, related by the transformation $p_2 \leftrightarrow p_3$, which we shall call $F_{123}$ and $F_{132}$. Superfamily $F_{123}$ is defined by the following set of propagators
\begin{equation}
\label{props F123}
\begin{split}
&D_1=k_1^2, \qquad D_2=(k_1 + p_1)^2, \qquad D_3=(k_1+p_{12})^2, \qquad D_{4}=(k_1 + p_{123})^2, \\
&  D_5=k_2^2, \qquad D_{6}=(k_2 + p_{1})^2, \qquad  D_7=(k_2+p_{12})^2,  \qquad D_{8}=(k_2+p_{123})^2, \\
&D_9=k_3^2, \qquad D_{10}=(k_3+p_1)^2, \qquad D_{11}=(k_3 + p_{12})^2, \qquad D_{12}=(k_3+p_{123})^2, \\
& D_{13}=(k_1-k_2)^2, \qquad  D_{14}=(k_1-k_3)^2, \qquad  D_{15}=(k_2-k_3)^2,
\end{split}
\end{equation}
while superfamily $F_{132}$ by
\begin{equation}
\label{props F132}
\begin{split}
&D_1=k_1^2, \qquad D_2=(k_1 + p_1)^2, \qquad D_3=(k_1+p_{13})^2, \qquad D_{4}=(k_1 + p_{123})^2, \\
&  D_5=k_2^2, \qquad D_{6}=(k_2 + p_{1})^2, \qquad  D_7=(k_2+p_{13})^2,  \qquad D_{8}=(k_2+p_{123})^2, \\
&D_9=k_3^2,\qquad D_{10}=(k_3+p_1)^2, \qquad D_{11}=(k_3 + p_{13})^2, \qquad D_{12}=(k_3+p_{123})^2, \\
& D_{13}=(k_1-k_2)^2, \qquad  D_{14}=(k_1-k_3)^2, \qquad  D_{15}=(k_2-k_3)^2,
\end{split}
\end{equation}
where we use the abbreviations $p_{12} = p_1+p_2$, $p_{13} = p_1+p_3$ and $p_{123}=p_1+p_2+p_3$. 

\begin{figure}[t]
    \centering
    \includegraphics[scale=0.252]{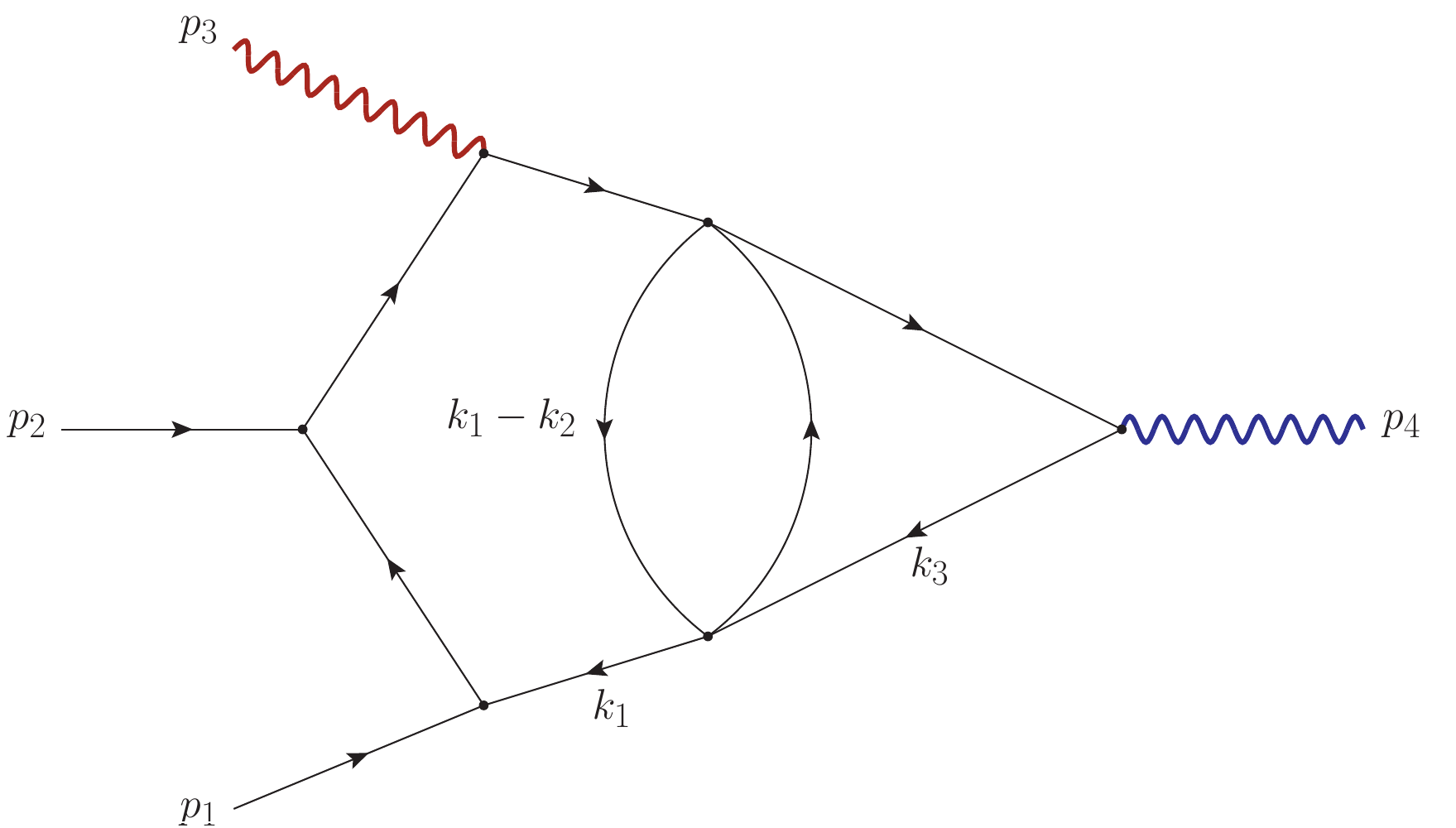}
    \includegraphics[scale=0.252]{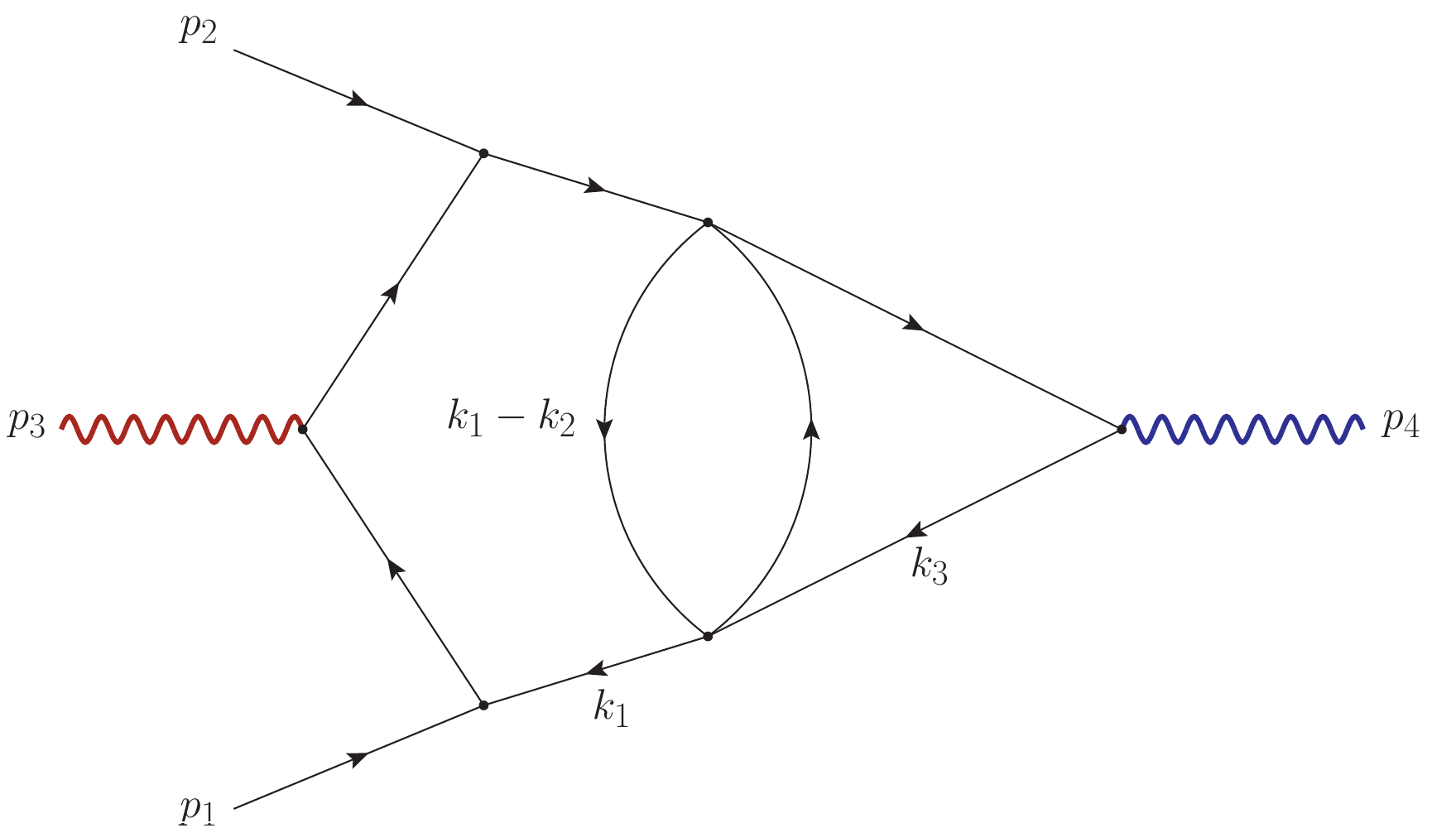}
    \caption{The two reducible integral families. The left one is $\mathrm{RL}_1$ and the right one is $\mathrm{RL}_2$. The black lines represent massless particles, while the two curled color lines correspond to the two vector bosons. The particle bearing mass $m_3$ ($m_4$) is drawn by a red (blue) color.}
    \label{fig:reducible}
\end{figure}


Concerning the irreducible families of the problem at hand, in general two classes of them appear: the so-called ladder-box and tennis-court families. In this paper we have chosen to study one representative of each class, both belonging to the superfamily $F_{123}$. Using the set of exponents $(a_1, \dots, a_{15})$ for their definition, the first one is the ladder-box family $\mathrm{PL}_1$, whose top-sector is described by the indices $(1,1,1,0,1,0,1,0,1,0,1,1,1,0,1)$ and is shown in the upper part of Fig.~\ref{fig:irreducible}. The second one is the tennis-court family $\mathrm{PT}_4$, whose top-sector is described by the propagators $(1,0,1,1,1,1,0,0,0,1,1,0,1,1,1)$ and it is pictured in the lower part of the Fig.~\ref{fig:irreducible}. Using \texttt{FFIntRed}\footnote{An in-house package written by Tiziano Peraro.} to generate a system of integration-by-parts identities (IBPs) and \texttt{FiniteFlow}~\cite{Peraro:2019svx} to solve them, we identified a set of 150 and 189 master integrals (MIs), for $\mathrm{PL}_1$ and $\mathrm{PT}_4$, respectively.

Apart from the irreducible families chosen above, in this work we further study two reducible ladder-box-like families. The first one, which we call $\mathrm{RL}_1$, contains 27 MIs and belongs to the superfamily $F_{123}$. Its top-sector is described by the set of exponents $(1,1,1,1,0,0,0,0,1,0,0,1,1,0,1)$ and is shown on the left in Fig.~\ref{fig:reducible}. The other reducible family, denoted by $\mathrm{RL}_2$, is related to $\mathrm{RL}_1$ by $p_2 \leftrightarrow p_3$, so it belongs to the superfamily $F_{132}$ and its top-sector is defined by the same set of exponents and is shown on the right in Fig.~\ref{fig:reducible}. We found $\mathrm{RL}_2$ to contain 25 MIs.

\section{Differential equations}
\label{Pure_Bases_and_DEs}

In this section, we describe the details of the choice of basis for the master integrals. We recall the theory about the method of differential equations and the existence of a pure basis. We constructed such a basis for the four families we studied and we wrote their differential equation in canonical \textit{dlog}-form. 

\subsection{Canonical form of the differential equations}
\label{Canonical_DEs}

In order to solve integral families under consideration we employ the method of differential equations (DEs)~\cite{Kotikov:1990kg, Bern:1993kr, Remiddi:1997ny, Gehrmann:1999as}. We use \texttt{FFIntRed} and \texttt{FiniteFlow}~\cite{Peraro:2019svx} to compute derivatives of the MIs with respect to the kinematic invariants, reduce them to MIs and reconstruct the differential equation over finite fields.

For a generic basis of MIs $\Vec{G}$ the DEs take the form\footnote{Henceforth, we shall employ the abbreviation $\partial_{\xi} X \equiv \partial X /\partial \xi$ in our notation.}
\begin{equation}
    \partial_\xi \Vec{G} (\Vec{s}) = B_\xi(\Vec{s};\epsilon) \ \Vec{G}(\Vec{s})  \qquad \forall \ \xi \in \Vec{s}= \begin{cases}
        (s_{12},s_{23},m_3^2,m_4^2)\\
        (x,y,z)
    \end{cases}.\\
\end{equation}
By construction, the connection matrices $B_\xi$ satisfy the integrability conditions
\begin{equation}
    (\partial_\xi \partial_\eta - \partial_\eta \partial_\xi) \Vec{G} = 0 \Rightarrow \partial_\xi B_\eta - \partial_\eta B_\xi -[B_\xi, B_\eta] = 0.
\end{equation}
In general, $B_\xi$ will depend on both the kinematic invariants and the dimensional regulator.
A possible strategy to solve these differential equations more efficiently is constructing a basis of master integrals such that the DEs are canonical~\cite{Henn:2013pwa}. Such a basis is composed of pure master integrals, i.e.~integrals with a uniform transcendental weight which is lowered by differentiation. The differential equation for such a basis is $\varepsilon$-factorised
\begin{equation}
\label{eq:canonical_DE}
\partial_\xi \Vec{I}(\Vec{s}) = \varepsilon \ A_\xi(\Vec{s}) \ \Vec{I}(\Vec{s})
\end{equation}
and can be cast in \textit{dlog}-form
\begin{equation}
\label{eq:dlog Form}
\mathrm{d} \Vec{I}(\Vec{s}) = \varepsilon \ \mathrm{d} \tilde{A} (\Vec{s}) \ \Vec{I} (\Vec{s}), \qquad
\mathrm{d} \tilde{A}(\Vec{s}) = \sum_i a_i \ \mathrm{d} \log w_i (\Vec{s}),
\end{equation}
where $\mathrm{d}$ is the total differential, $a_i$ are constant matrices and $w_i (\Vec{s})$ are functions of the kinematic invariants alone, called letters. The set of all letters is called alphabet.
For such a differential equation the integrability conditions take the simpler form
\begin{equation}
\partial_\xi A_\eta - \partial_\eta A_\xi = 0, \qquad [A_\xi, A_\eta] = 0.
\end{equation}

The $\varepsilon-$factorized form of Eq.~\eqref{eq:canonical_DE} 
implies that the differential equation can be solved iteratively by expanding the basis $\Vec{I}$ in a power series in $\varepsilon$
\begin{equation}
    \Vec{I}(\Vec{s}; \varepsilon) = \sum_{k=0}^\infty \varepsilon^k \ \Vec{I}^{(k)}(\Vec{s}),
\end{equation}
and solving for the coefficients $\Vec{I}^{(k)}(\Vec{s})$ order by order, starting at $k=0$, since the differential equations for $\Vec{I}^{(k)}$ depend only on the solution at lower weights. Moreover, Eq.~\eqref{eq:dlog Form} implies that $\Vec{I}^{(k)}(\Vec{s})$ can be formally expressed as a $k-$fold iterated integral over the \textit{dlog}-forms in $\Tilde{A}$. In particular $\Vec{I}^{(k)}(\Vec{s})$ is a pure function of transcendental weight $k$, as expected.

In our case, we were able to achieve a canonical \textit{dlog}-form for the DEs by employing the techniques outlined in the subsection~\ref{Pure_Bases}. In order to determine the matrix $\Tilde{A}$ we first constructed an \textit{ansatz} for the alphabet using the denominator factors of the differential equation for the rational letters and employing the package \texttt{BaikovLetter}~\cite{Jiang:2024eaj} for the algebraic letters. We determined the entries of $\Tilde{A}$ by fitting the logarithmic derivative of a generic linear combination of letters to the entries of the differential equation. The total number of letters appearing in the alphabet of all the families is 16 in the Mandelstam invariants
\begin{equation}
\label{total_alpha_Man}
\begin{split}
\bar{W} \equiv \{\bar{w}_i\}_{i=1}^{15} =\{&m_3^2, \, m_4^2, \, s_{12}, \, s_{23}, \, m_3^2 - s_{23}, \, m_4^2 - s_{23}, \, s_{12} + s_{23} -m_3^2,\\ 
&s_{12} + s_{23} - m_4^2, s_{12} + s_{23} - m_3^2 - m_4^2, \,  m_3^2 (m_4^2 - s_{23}) + s_{12} s_{23},\\
&m_3^2 m_4^2 + (s_{12}-m_4^2) s_{23},m_3^4 + (m_4^2 - s_{12})^2 - 2 m_3^2 (m_4^2 + s_{12}),\\ 
&m_3^2 (m_4^2 - s_{23}) + s_{23} (s_{12} + s_{23} - m_4^2),\frac{m_3^2 + m_4^2 - s_{12} - R}{m_3^2 + m_4^2 - s_{12} + R},\\ 
&\frac{m_3^2 - m_4^2 + s_{12} - R}{m_3^2 - m_4^2 + s_{12} + R}, \, \frac{m_3^2 + m_4^2 - s_{12} - 2 s_{23} - R}{m_3^2 + m_4^2 - s_{12} - 2 s_{23} + R} \},
\end{split}
\end{equation}
where $R$ is the square root defined in Eq.~\eqref{R_invs}, and 15 in the rationalized variables, since we set $m_3^2 = 1$
\begin{equation}
\label{total_alpha}
\begin{split}
W \equiv \{w_i\}_{i=1}^{15} =\{&x, \, y, \, z, \, 1 + x, \, 1 - y, \, 1 - z, \, z - y, \, 1 + y - z, \, 1 + x y, \, 1 + x z,\\ 
&z+ x y, \, 1 - z+ y(1 + x), \, 1 + x (1 + y - z),\\
&z - y (1 - z - x z), \, z - x (y - z - y z)\} .
\end{split}
\end{equation}
It is worth noting that the above alphabet is the same as that of the corresponding two-loop planar families~\cite{Henn:2014lfa}. The sets of letters and the dimensions of the alphabet  appearing in the differential equations of the individual families are  
\begin{equation*}
\begin{array}{lll}
   \bar{W}_{\mathrm{RL}_1}=\bar{W} \backslash \{\bar{w}_8,\, \bar{w}_9, \, \bar{w}_{10}\},  & W_{\mathrm{RL}_1}= W \backslash \{w_8, \, w_{10}, \, w_{11} \}, & \mathrm{dim}(W_{\mathrm{RL}_1})=13, \\
    \bar{W}_{\mathrm{RL}_2}= \bar{W} \backslash \{ \bar{w}_{10}, \, \bar{w}_{11} \}, & W_{\mathrm{RL}_2}=W \backslash \{w_{14}, \, w_{15} \}, & \mathrm{dim}(W_{\mathrm{RL}_2})=14,\\
    \bar{W}_{\mathrm{PL}_1}=\bar{W}_{\mathrm{RL}_2}, & W_{\mathrm{PL}_1}=W_{\mathrm{RL}_2}, & \mathrm{dim}(W_{\mathrm{PL}_1})= \mathrm{dim}(W_{\mathrm{RL}_2}),\\
    \bar{W}_{\mathrm{PT}_4}= W, & W_{\mathrm{PT}_4}= W, & \mathrm{dim}(W_{\mathrm{PL}_1})= \mathrm{dim}(W)=15.
\end{array}
\end{equation*}
The complete alphabet in both variables, as well as the differential equations for the families under consideration, can be found in the ancillary files~\cite{canko_2024_14284044} attached to this paper and described in Appendix~\ref{Ancillary}.


\subsection{Construction of pure bases}
\label{Pure_Bases}

In general, the construction of a pure basis for a family, assuming that such a form exists for the family under consideration, is not a process that can be carried out in the same way each time via a systematic approach. A variety of techniques can be employed, and a combination of them is typically required. This was also the case in the present study. 

Some of the MIs of our families, for instance, some two- and three-point integrals, are already known from previous works~\cite{DiVita:2014pza, Henn:2020lye, Canko:2021xmn}, and for them, we used pure candidates proposed therein. For the genuinely new MIs, we employed a bottom-up approach, identifying first the pure candidates for the lower sectors and then moving to higher sectors. We started by working on the maximal cut (i.e.~setting to zero all the integrals that do not belong to the sector under investigation) until we achieved a canonical form for the block of the differential equations corresponding to the sector at hand. We then fixed the entries of the DEs corresponding to the coupling to lower sectors by appropriately rotating the pure candidates.

In most cases, the technique proposed in reference~\cite{Gehrmann:2014bfa} was employed, which is similar to the method of Magnus exponential~\cite{Argeri:2014qva}. The starting point of this method is the selection of appropriate candidates, for the sector under study, chosen in a way that renders the DEs in the maximal cut linear on $\varepsilon$
\begin{equation}
\label{iniDEs}
    \partial_{\xi} \Vec{G}^{\text{\tiny MC}}= (H_{0,\xi} + \varepsilon H_{1,\xi}) \Vec{G}^{\text{\tiny MC}}\qquad \forall \ \xi \in \Vec{s} .
\end{equation}
Then the $H_0$ term can be removed from the MIs by rescaling them with a matrix that satisfies the following system of DEs
\begin{equation}
    \partial_{\xi} \tilde{T}^{\text{\tiny MC}}=- \tilde{T}^{\text{\tiny MC}} H_{0,\xi}.
\end{equation}
The differential equation is in general difficult to solve. In practice, one determines each entry separately, by observing that the diagonal ones correspond to the normalisation of the MIs and the off-diagonal ones to the contribution of the MIs to the derivatives of the others. Therefore, one has to build linear combinations of appropriately normalised master integrals. The new candidates defined as $\Vec{I}^{\text{\tiny MC}}=\tilde{T}^{\text{\tiny MC}}\Vec{G}^{\text{\tiny MC}}$ satisfy canonical DEs in the maximal cut\footnote{The notation MC (LS) here and later on stands for symbolizing quantities in the maximal cut (lower sectors).}
\begin{equation}
\partial_{\xi} \Vec{I}^{\text{\tiny MC}}= \varepsilon A_{\xi}^{\text{\tiny MC}} \Vec{I}^{\text{\tiny MC}}, \qquad A_{\xi}^{\text{\tiny MC}}=\tilde{T}^{\text{\tiny MC}} H_{1,\xi} (\tilde{T}^{\text{\tiny MC}})^{-1}.
\end{equation}

In order to construct MIs that are pure beyond the maximal cut too, one needs to relax the cut conditions and derive DEs for the sector at hand including sub-maximal sectors or without cuts at all. In most cases that we encountered in this study, the sub-sector entries were fine and no extra care needed to be taken. In some of them, the sub-sector entries were again linear on the dimensional regulator
\begin{equation}
\partial_{\xi} \Vec{I}^{\text{\tiny}}= \varepsilon A_{\xi}^{\text{\tiny MC}} \Vec{I}^{\text{\tiny MC}}+ (h_{0,\xi}+\varepsilon h_{1,\xi})\Vec{I}^{\text{\tiny LS}}
\end{equation}
and to set them in canonical form we needed to rotate properly the lower sector $\varepsilon^0$ contributions by integrating out $h_{0,\xi}$ through the transformation
\begin{equation}
\Vec{I}=\Vec{I}^{\text{\tiny MC}}+\tilde{T}^{\text{\tiny LS}}\Vec{I}^{\text{\tiny LS}}, \qquad \partial_{\xi} \tilde{T}^{\text{\tiny LS}}=-h_{0,\xi}
\end{equation}

It is evident that the identification of a suitable set of candidates represents a fundamental aspect of the aforementioned methodology. In order to achieve this, educated guesses were made and subsequently verified by reconstructing the DEs on the maximal cut only in $\varepsilon$, keeping numerical values for the kinematic invariants. For sectors with up to seven propagators, it is often enough to dot some propagators of the MIs provided by Laporta's algorithm. For higher sectors, the scalar integrals and MIs with ISPs are usually a good starting point. Except for dotting propagators and multiplying with ISPs, two methods that helped us in our quest of searching for good candidates are the building-blocks method and the choice of candidates whose integrands can be put in \textit{dlog}-form with constant leading singularities~\cite{Wasser:2018qvj,Henn:2020lye}.

In the former, the constant leading singularities of one-loop and two-loop pure candidates are used as building blocks, which combined create the leading singularity of a three-loop candidate, in a graphical and heuristic approach. This method has been shown to work well with massless propagators, and in our study it helped us to identify pure candidates including also the necessary lower-sector contributions. For the latter method, we used two approaches. Firstly, we used the \texttt{DLogBasis} package~\cite{Henn:2020lye}, which brings the 4-dimensional integrand of a Feynman integral into $d \log$ form using the spinor helicity formalism. In order to employ the latter for the massive momenta, we used a two-variable simplified differential equations parametrization~\cite{Papadopoulos:2014lla} to re-express them in terms of massless momenta, as proposed in~\cite{Canko:2021hvh}.
Secondly, we exploited the loop-by-loop Baikov representation of Feynman integrals~\cite{Baikov:1996iu, Baikov:1996rk} for some sectors, using the package \texttt{Baikov.m}~\cite{Frellesvig:2017aai}, trying to put the resulting d-dimensional integrand into a \textit{dlog}-form for the Baikov variables, first in the maximal cut and later by investigating sub-maximal contributions.

Employing these various methods, we constructed a basis of pure master integrals satisfying canonical differential equations for all the four families we study in this paper. We refer to the ancillary files~\cite{canko_2024_14284044} for the complete list of pure master integrals, and to section~\ref{Appendix} for the list of genuinely new sectors of integrals.
\section{Results}
\label{Results}
In this section, we discuss in detail the two strategies we followed to solve the DEs and compute the master integrals. Firstly we computed the solution analytically in terms of multiple polylogarithms, employing the PSLQ algorithm to numerically fix the boundary values. Secondly, we set up a semi-numerical implementation of the generalized series expansion method using \texttt{DiffExp} and the differential equations we derived in the previous section. We compare the performance of both methods.

\subsection{Analytic results in terms of multiple polylogarithms}
\label{Analytic_Results_GPLs}


Multiple polylogarithms (MPLs)~\cite{Goncharov:1998kja, Goncharov:2010jf}, also known as Goncharov polylogarithms, are a well-studied class of iterated integrals suitable to express the solution of the DEs whenever the alphabet is rational. MPLs emerge in numerous computations of multi-loop Feynman integral families and can be defined recursively via the relation
\begin{equation}
\label{GPLs}
\mathcal{G}(a_1,\dots,a_n;X)=\int_0^x \frac{dt}{t-a_1}\mathcal{G}(a_2,\dots,a_n;t) \qquad \text{with} \qquad \mathcal{G}(;X) \equiv 1,
\end{equation}
where $a_i$ are the indices and $X$ the argument of the MPL. The number of logarithmic integrations is referred to as the transcendental weight of the MPL. For $a_n=0$, when the integration in Eq.~\eqref{GPLs} diverges, GPLs are defined through the relation
\begin{equation}
\label{gplzero}
\mathcal{G}(\Vec{0}_n;X)=\frac{1}{n!} \text{log}^n(X) \qquad \text{with} \qquad \Vec{0}_n=(0,\ldots,0).
\end{equation}
An important property of MPLs is that they satisfy a shuffle algebra~\cite{Duhr:2011zq}:
\begin{equation}
\label{shuffle algebra}
\mathcal{G}(a_1,\dots,a_{n_1};X) \mathcal{G}(a_{n_1+1},\dots,a_{n_1+n_2};X)=\sum_{\sigma \in \Sigma (n_1,n_2)} \mathcal{G}(a_{\sigma (1)},\dots, a_{\sigma (n_1+n_2)};X), 
\end{equation}
where $\Sigma (n_1,n_2)$ denotes the set of all the shuffles of $(a_1,...,a_{n_1})$ and $(a_{n_1+1}, ..., a_{n_1+n_2})$. The shuffle product of Eq.~\eqref{shuffle algebra} preserves both the weight of the MPLs and the ordering of the indices inside the vectors $(a_1, \dots, a_{n_1})$ and $(a_{n_1+1}, \dots, a_{n_1+n_2})$. In particular, Eq.~\eqref{shuffle algebra} implies that a product of MPLs with weight $n_1$ and $n_2$ can be expressed as a sum of MPLs with weight $n_1+n_2$.

\begin{table}[t!]
\centering
\begin{tabular}{|c|c|c|c|c|c|c|c|}
\hline
\textbf{Order} & $\varepsilon^0$ & $\varepsilon^1$ & $\varepsilon^2$ & $\varepsilon^3$ & $\varepsilon^4$ & $\varepsilon^5$ & $\varepsilon^6$ \\
\hline
\textbf{Constants} & $1$ & $\mathrm{i} \pi $ & $\pi^2$ & $\mathrm{i} \pi^3, \zeta_3$  & $\pi^4, \mathrm{i} \pi\zeta_3$ & $\mathrm{i} \pi^5,\pi^2 \zeta_3, \zeta_5$ & $\pi^6, \mathrm{i} \pi^3 \zeta_3, \zeta_3^2, \mathrm{i} \pi \zeta_5$ \\
 \hline
\end{tabular}
\caption{Table consisting of the transcendental constants expected to appear in the boundary conditions of pure candidates at each order of the expansion on $\varepsilon$.}
\label{tab:transcendental}
\end{table}

Concerning the analytic properties of MPLs, as is apparent from their definition, these functions develop branch cuts whenever any of the indices lies along the integration path connecting 0 and $X$, as in such cases the integral is not well-defined. For the sake of a fast evaluation, it is desirable to express the solution in terms of MPLs without discontinuities in the physical region. Then MPLs are real-valued functions and the imaginary part of the MIs comes solely from the boundary terms. For further insight into the properties of MPLs and their numerical evaluation, we direct the reader to~\cite{Vollinga:2004sn}.


To obtain an analytic solution in terms of MPLs for the problem at hand, we solve the canonical \textit{dlog}-form DEs of the variables $(x,y,z)$, where the alphabet is rationalized. We do so by applying the methodology outlined in~\cite{Badger:2023xtl}, which involves integrating Eq.~\eqref{eq:dlog Form} along piecewise continuous paths, with each sub-path being linearly parametrized. In order to minimize, as much as possible, the number of MPLs in the solutions we follow two different integration paths for each superfamily
\begin{equation}
\begin{split}
&\text{$F_{123}$: } (0,0,0) \xrightarrow{\gamma_1} (x,0,0) \xrightarrow{\gamma_2} (x,0,z) \xrightarrow{\gamma_3} (x,y,z) \\
&\text{$F_{132}$: } (0,0,0) \xrightarrow{\gamma_1'} (0,0,z) \xrightarrow{\gamma_2'} (0,y,z) \xrightarrow{\gamma_3'} (x,y,z)
\end{split} \, \, .
\end{equation}
The two integration paths result in MPLs with the same arguments but different indices and are chosen so that we obtain real-valued MPLs directly in the physical region~\cite{Henn:2014lfa}. In particular, for the MPLs resulting from each sub-path we have
\begin{itemize}
    \item $\gamma_1$: argument $x$ and indices $\{-1, 0\}$.
    \item $\gamma_2$: argument $z$ and indices $\{0, -\frac{1}{x}, 1, 1 + \frac{1}{x}\}$.
    \item $\gamma_3$: argument $y$ and indices $ \{ 0, - \frac{1}{x}, -\frac{z}{x}, z, \frac{z}{x}\frac{1 + x}{1 - z}, 1, \frac{z-1}{1 + x}, z -1 - \frac{1}{x}, z-1 , \frac{-z}{-1 + z + x z} \}$.
    \item $\gamma_1'$: argument $z$ and indices $\{ 0, 1\}$.
    \item $\gamma_2'$: argument $y$ and indices $\{ 0, z - 1, z \}$.
    \item $\gamma_3'$: argument $x$ and indices $\{ 0, - \frac{1}{z}, -\frac{z}{y}, \frac{1}{z - 1 - y}, \frac{z -1 - y}{y}, -1, -\frac{1}{y} \}$.
\end{itemize}

We write the expansion of the solution of the DEs up to order six in the dimensional regulator in the general form
\begin{align}
\label{Analytic_Solution}
\vec{I}&=\varepsilon^0 \left( \vec{b}_0 \right) \nonumber\\
&+ \varepsilon^1 \left(c^i_1 {\cal G}_i \vec{b}_0+\vec{b}_1\right)\nonumber\\
&+ \varepsilon^2 \left(c^{ij}_2{\cal G}_{ij} \vec{b}_0+c^{i}_2{\cal G}_{i} \vec{b}_1+ \vec{b}_2\right)\nonumber\\
&+ \varepsilon^3 \left(c^{ijl}_3{\cal G}_{ijl} \vec{b}_0+c^{ij}_3{\cal G}_{ij} \vec{b}_1+ c^i_3 {\cal G}_{i}\vec{b}_2 +\vec{b}_3\right)\\
&+ \varepsilon^4 \left(c^{ijlk}_4{\cal G}_{ijlk} \vec{b}_0+c^{ijl}_4{\cal G}_{ijl} \vec{b}_1+ c^{ij}_4 {\cal G}_{ij}\vec{b}_2 +c^{i}_4 {\cal G}_{i} \vec{b}_3 + \vec{b}_4 \right) \nonumber\\
&+ \varepsilon^5 \left(c^{ijlkn}_5{\cal G}_{ijlkn} \vec{b}_0+c^{ijlk}_5{\cal G}_{ijlk} \vec{b}_1+ c^{ijl}_5 {\cal G}_{ijl}\vec{b}_2 +c^{ij}_5 {\cal G}_{ij} \vec{b}_3 + c^{i}_5 {\cal G}_{i} \vec{b}_4 +\vec{b}_5 \right) \nonumber\\
&+ \varepsilon^6 \left(c^{ijlkno}_6{\cal G}_{ijlkno} \vec{b}_0 + c^{ijlkn}_6 {\cal G}_{ijlkn} \vec{b}_1 + c^{ijlk}_6 {\cal G}_{ijlk} \vec{b}_2 + c^{ijl}_6 {\cal G}_{ijl} \vec{b}_3+c^{ij}_6 {\cal G}_{ij} \vec{b}_4+c^i_6 {\cal G}_i \vec{b}_5+ \vec{b}_6 \right) \nonumber
\end{align}
where $c_{\alpha}^{i \dots o}$ are constant matrices and $\vec{b}_{\alpha}$ are the boundary constants at order $\varepsilon^{\alpha}$. In Eq.~\eqref{Analytic_Solution} summation over repeated indices is assumed, and we used the shorthand notation ${\cal G}_{i \dots o}\equiv\mathcal{G}(a_i,\dots,a_o;X)$ or $\mathcal{G}(a_i,\dots,a_j;X) \ldots \mathcal{G}(a_l,\dots,a_o;X')$ for MPLs and their products. 

In lieu of using conventional methodologies for the computation of the boundary conditions that appear in Eq.~\eqref{Analytic_Solution}, such as studying the solution of the DEs to all the unphysical (regularity conditions) and physical limits~\cite{Henn:2020lye} or using the expansion by regions method~\cite{Jantzen:2012mw}, we employed the procedure described in~\cite{Badger:2023xtl}. Because of the choice of the integration base point $(x,y,z)=(0,0,0)$, the boundary constants are expected to contain only the transcendental constants listed in table~\ref{tab:transcendental}. In particular, each entry of the vector $\vec{b}_w$ is a linear combination of transcendental constants of weight $w$. We determined the coefficients recursively weight-by-weight, so that when dealing with $\vec{b}_w$ all the boundary constants $\vec{b}_{w'}, w'< w$ were already fixed. We did so numerically as follows: we used the \texttt{Mathematica} package \texttt{AMFlow}~\cite{Liu:2022chg}, which implements the auxiliary mass flow method for the numerical calculation of Feynman integrals via DEs~\cite{Liu:2017jxz}, to compute the values of the pure candidates $\vec{I}$ numerically with 70 digits precision at the physical point 
\begin{equation}
\label{ideal point}
\left(s_{12},s_{23},m_3^2,m_4^2\right) = \left(\frac{13}{4},-\frac{3}{4},1,\frac{9}{20}\right) \qquad \longleftrightarrow \qquad (x,y,z)=\left(\frac{3}{2},\frac{1}{5},\frac{1}{2}\right),
\end{equation} 
and the \texttt{Mathematica} package \texttt{DiffExp}~\cite{Hidding:2020ytt} for the evaluation of the GPLs at the same point with the same precision. After inserting these numerical values, the coefficient of $\varepsilon^w$ in Eq.~\eqref{Analytic_Solution} is composed of a linear combination of complex numbers with undetermined rational coefficients (the boundary constant $\vec{b}_w$) and a complex number obtained by summing all other terms at this weight (including $\vec{I}^{(w)}$). In this way, the problem of determining the unknown boundary conditions of the pure MIs is reduced to that of finding integer coefficients for a linear combination of real numbers\footnote{Formally, we multiplied the equation by the minimum common multiple of the rational coefficients and treated separately its real and the imaginary part, resulting in two equations.}. We solved this problem employing the PSLQ algorithm~\cite{pslq}, as implemented in \texttt{Mathematica}. 

In order to further reduce the number of MPLs appearing in the solution we also employed a partial application of the decomposition method in Lyndon words~\cite{Lyndon}. 
For each argument, we fixed an ordering of the allowed indices and we used the shuffle algebra of Eq.~\eqref{shuffle algebra} to express all the solutions in terms of MPLs with indices with this ordering. We tried all the possible orderings for the indices, selecting the one that minimizes the number of MPLs. The decomposition into Lyndon words is implemented in \texttt{PolyLogTools}~\cite{Duhr:2019tlz} for MPLs up to weight six and at most five different indices. 
But from the integration paths $\{\gamma_3,\gamma_3'\}$ we obtained MPLs with more than five different indices. For those cases, we applied the Lyndon decomposition to a selected subset of indices containing the letters mostly appearing in the MPLs. The final form of the analytic results for each family can be found in the ancillary files~\cite{canko_2024_14284044} described in the appendix~\ref{Ancillary}.

In table~\ref{tab:2} we present the number of MPLs included in the final results, classified by weight and grouped according to family, superfamily, and all categories combined. A review of the data in the table reveals that the weight 6 (respectively weight 5) MPLs make up approximately $46\%$ (respectively $38\%$) of all MPLs in the solution and their number is $\mathcal{O}(10^3)$. This observation, along with the fact that higher-weight MPLs are the most time-consuming to evaluate numerically, renders the current form of the analytic solution expensive to evaluate. For this reason, in the next subsection, we consider an alternative (semi-)numerical solution for our DEs and compare the computational times for both methods.



\begin{table}[t!]
\centering
\begin{tabular}{|c|c|c|c|c|c|c|c|}
\hline
  \textbf{Family} & \textbf{W1} & \textbf{W2} & \textbf{W3} & \textbf{W4} & \textbf{W5} & \textbf{W6} &  \textbf{Total} \\
\hline
  \textbf{RL}$_\mathbf{1}$ & $12$ & $27$ & $137$ & $492$  & $1320$ & $1631$ & $3619$ \\
 \hline
  \textbf{PL}$_\mathbf{1}$ & $14$ & $ 35 $ & $188$ & $690$  & $1935$ & $2554$ & $5416$  \\
 \hline
  \textbf{PT}$_\mathbf{4}$ & $16$ & $51 $ & $312$ & $1170$  & $3032$ & $3709$ & $8290$ \\
 \hline
  \textbf{$\mathbf{F_{123}}$} & $16$ & $51$ & $312$ & $1176$  & $3066$ & $3739$ & $8360$  \\
 \hline
  \textbf{$\mathbf{F_{132}}$} & $12$ & $29$ & $168$ & $996$  & $4549$ & $5219$ & $10973$  \\
 \hline
  \textbf{ALL} & $21$ & $78$ & $478$ & $2169$  & $7609$ & $8951$ & $19306$ \\
 \hline
\end{tabular}
\caption{Table presenting the number of MPLs, at each weight (denoted W) and in total, as observed in the solution of each family, superfamily, and across all of them collectively. Since $\mathrm{RL}_2$ is the only family belonging to the superfamily $F_{132}$, the number of MPLs for this family is precisely that of the superfamily.}
\label{tab:2}
\end{table}

\subsection{Numerical results using \texttt{DiffExp}}
\label{Numerical_Results}
As an alternative to the analytic solution, whose evaluation is computationally inefficient, we evaluate the master integrals using the \texttt{Mathematica} package \texttt{DiffExp}~\cite{Hidding:2020ytt}, which relies on the generalized series expansion method~\cite{Moriello:2019yhu}. The method allows us to integrate the DEs from a base point, where the values of the integrals are known, to a target point in kinematic space. Although in principle one could choose any point as a base point, physical or not, we choose to work directly in the physical region as defined in section~\ref{Kinematics}. Then, assuming the path connecting the base- and target-point lies inside of the physical region as well, no physical singularity is crossed during the computation. If that were the case one would have to provide \texttt{DiffExp} with rules for the analytic continuation along these singularities. Such rules are hard to obtain for the complete multi-dimensional phase-space and they slow down the evaluation. As boundaries, we use the same values of the pure candidates we computed for the analytic solution using \texttt{AMFlow} at the physical point in Eq.~\eqref{ideal point}. We implemented the generalized series expansion method both for the variables $\left(s_{12},s_{23},m_3^2,m_4^2\right)$ and for the variables $(x,y,z)$.

We performed a first check of the implementation by computing the integrals at the following points
\begin{equation}
    \begin{split}
        \left(s_{12},s_{23},m_3^2,m_4^2\right) &= \left(\frac{304}{33},-\frac{5}{7},3,\frac{25}{33}\right) \qquad \longleftrightarrow \qquad (x,y,z)=\left(\frac{5}{3},\frac{1}{11},\frac{1}{7}\right),\\
        \left(s_{12},s_{23},m_3^2,m_4^2\right) &= \left(\frac{88}{5},-\frac{35}{6},5,\frac{49}{15}\right) \qquad \longleftrightarrow \qquad (x,y,z)=\left(\frac{7}{5},\frac{1}{3},\frac{5}{6}\right),\\
        \left(s_{12},s_{23},m_3^2,m_4^2\right) &= \left(25,-\frac{36}{5},9,\frac{16}{7}\right) \qquad \longleftrightarrow \qquad (x,y,z)=\left(\frac{12}{9},\frac{1}{7},\frac{3}{5}\right),
    \end{split}
    \label{eq:checkpoints}
\end{equation}
both with \texttt{DiffExp} and with \texttt{AMFlow}, finding perfect agreement up to the requested precision.

\begin{table}[t!]
    \centering
    \begin{tabular}{|c|c|c|}
    \hline
        Family & $(s_{12},s_{23},m_3^2,m_4^2)$ & $(x,y,z)$ \\
        \hline
         $\mathrm{RL}_1$ & 175 $s$ & 55 $s$\\
         \hline
         $\mathrm{RL}_2$ & 169 $s$ & 64 $s$\\
         \hline
         $\mathrm{PL}_1$ & 2765 $s$ & 818 $s$\\
         \hline
         $\mathrm{PT}_4$ & 4987  $s$ & 1478 $s$\\
         \hline
    \end{tabular}
    \caption{Average time needed by \texttt{DiffExp} to integrate a kinematic point with 40 digits precision, per family, and per set of kinematic variables used for the DEs.}
    \label{tab:timings_points}
\end{table}

The time needed to compute the integrals at one kinematic point with \texttt{AMFlow} is of the order of days for $\mathrm{PL}_1$ and $\mathrm{PT}_4$. Moreover, we only managed to get results for kinematic points involving small primes as numerators and denominators, like those in Eq.~\eqref{eq:checkpoints}. In order to test the implementation with \texttt{DiffExp} at random physical points, we exploit the analytic solution from section~\ref{Analytic_Results_GPLs}. For the evaluation of the MPLs in the analytic solution we also used \texttt{DiffExp}, which proved faster than \texttt{Ginac}~\cite{ginac} interfaced with \texttt{Ginsh} and \texttt{PolyLogTools}~\cite{Duhr:2019tlz}. We compared the analytic solution and the results obtained with \texttt{DiffExp} at ten random physical points with 40 digits precision, finding perfect agreement up to the requested precision. We included in the ancillaries~\cite{canko_2024_14284044} the code that we used to run \texttt{DiffExp} (\texttt{DiffExpRun.wl}).

\begin{table}[t!]
    \centering
    \begin{tabular}{|c|c|c|}
    \hline
        Family & $(s_{12},s_{23},m_3^2,m_4^2)$ & $(x,y,z)$ \\
        \hline
         $\mathrm{RL}_1$ & 8 $s$ & 9  $s$\\
         \hline
         $\mathrm{RL}_2$ & 9 $s$ & 11 $s$\\
         \hline
         $\mathrm{PL}_1$ & 126 $s$ & 136 $s$\\
         \hline
         $\mathrm{PT}_4$ & 226 $s$ & 246 $s$\\
         \hline
    \end{tabular}
    \caption{Average time needed by \texttt{DiffExp} to integrate a segment with 40 digits precision, per family, and per set of kinematic variables used for the DEs.}
    \label{tab:segments}
\end{table}
We conclude this section by commenting on the performance of \texttt{DiffExp} for the two sets of kinematic invariants. In the table~\ref{tab:timings_points} we show the average time needed to compute a single point for the two sets of kinematic invariants. As can be seen from the results, evaluating the integrals in the variables $(x,y,z)$ is considerably faster. However, the time per kinematic point is not the only criterion on which the performance of \texttt{DiffExp} should be judged. Since \texttt{DiffExp} internally divides the path from the base-point to the target point into segments, which are then integrated one after the other, one should also look at the time needed to integrate a single segment. We computed the average time per segment for all the families. The results, shown in table~\ref{tab:segments}, paint the opposite picture compared to table~\ref{tab:timings_points}. This is due to the fact that the DEs in the variables $(s_{12},s_{23},m_3^2,m_4^2)$ are less complicated than those in $(x,y,z)$, as can be deduced from the size of the files in the ancillaries. Therefore, the integration of the differential equation is faster. However, the phase-space has a simpler topology when the square root is rationalized, hence \texttt{DiffExp} splits the integration path into fewer sub-segments. Finally, with an eye towards phenomenology, we remark that constructing a grid in a smart way one could integrate each point using only one segment, in which case the timings in table~\ref{tab:segments} would approximately be the time needed to compute one point.

\section{Conclusions}
\label{Conclusions}

In this paper, we undertook a study of a ladder-box, a tennis-court, and two reducible ladder-box-like three-loop integral families with massless propagators and four external particles, where two of the external particles are massive but bear different masses. We constructed pure bases and derived canonical DEs, which we subsequently solved on physical kinematics up to order six in the dimensional regulator. This was done analytically in terms of MPLs and numerically utilizing \texttt{DiffExp}, verifying the validity of our results by comparisons with \texttt{AMFlow}, finding perfect agreement in all cases. These families are of significant phenomenological interest, as they are relevant for the study of the production of two massive vector bosons at N3LO precision in QCD.

We expect that the analytic solution can still be optimized, for example by identifying additional relations between the MPLs that appear in it. For this reason, future work will involve the complete decomposition into Lyndon words and the application of symbol algebra to reveal relations between the existing MPLs, so that solutions can be expressed in terms of a minimal set of independent MPLs. Moreover, we plan to implement the evaluation of the analytic solution using more conventional tools, such as a \texttt{C++} (\texttt{FORTRAN}) code suitable for fast and parallel numerical evaluations through \texttt{GiNaC}~\cite{ginac} (\texttt{handyG}~\cite{Naterop:2019xaf}). Alternatively, the efficiency of the numerical solution of the \texttt{DiffExp} solution can be enhanced by creating an optimized grid of numerical solutions in multiple phase-space points, as demonstrated in \cite{Abreu:2020jxa}. Whether pursuing one or the other path, we expect that our solutions will prove efficient for the future computation of scattering amplitudes involving these integrals.

The study of the remaining planar (and non-planar) three-loop four-point integral families with two off-shell legs represents a logical subsequent step in the same direction of this work. Based on the existing studies of analogous problems \cite{Canko:2021hvh, Long:2024bmi}, it is reasonable to expect that the remaining planar families can be expressed in terms of MPLs too. Consequently, a comparable approach to the one employed in this study can be effectively applied to address these families too. Moreover, once all the planar families have been computed, it would be worthwhile to investigate the adjacency conditions between the letters of their alphabets and the possibility of finding a cluster algebra describing them, as has already been done for the corresponding planar families with one off-shell leg \cite{Chicherin:2020umh, Aliaj:2024zgp}.



\acknowledgments

We would like to thank Matteo Becchetti, Vsevolod Chestnov, Tiziano Peraro, and Simone Zoia for useful discussions and comments on the draft. This work was supported by the European Research Council (ERC) under the European Union’s Horizon Europe research and innovation program grant agreement 101040760, \textit{High-precision multi-leg Higgs and top physics with finite fields} (ERC Starting Grant FFHiggsTop).

\appendix

\section{Description of the ancillary files}
\label{Ancillary}
The ancillary files can be downloaded from~\cite{canko_2024_14284044}. The notation in the files translates to that of the article as follows:
\begin{table}[ht!]
    \centering
    \begin{tabular}{ll}
        \texttt{sij} = $s_{ij}$, & \texttt{FFG[Fam,\{a1,...,a15\}]} = $G^{Fam}_{a_1,\dots,a_{15}}$,\\
        \texttt{mi2} = $m_i^2$, & \texttt{W[i]},\texttt{W[r[1],i]} = $w_i$ (respectively $\Bar{w}_i$),\\
        \texttt{eps} = $\epsilon$, & \texttt{dlog[W[i]]} = $\mathrm{d}\log(w_i)$ (respectively $\mathrm{d}\log(\Bar{w}_i)$),\\
         \texttt{r[1]} = $R$, & \texttt{G[a1,\dots,an,x]} = $\mathcal{G}(a_1, \dots, a_n; X)$,\\
         \texttt{Pi} = $\pi$, & \texttt{Zeta[3]} = $\zeta_3$,\\
         \texttt{Zeta[5]} = $\zeta_5$, & \texttt{I} = $\mathrm{i}$.
    \end{tabular}
    \label{tab:ancillary_notation}
\end{table}

The folder \texttt{GlobalDefinitions} contains the definitions of the alphabet for both choices of variables, in the format
\begin{equation*}
    \texttt{\{\dots, W[i] -> expression, \dots\}}.
\end{equation*}
We remark that in general $w_i \neq \Bar{w}_i$, even though both are called \texttt{W[i]} in the ancillary files.

For each family, we provide two folders:
\begin{itemize}
    \item \texttt{MIs\_DEs}.
    \item \texttt{AnalyticSolution}.
\end{itemize}
The first one contains five files:
\begin{itemize}
    \item \texttt{masters.m}: the basis of pure candidates, in terms of \texttt{FFG[Fam,\{a1,...,a15\}]}, \texttt{eps} and external invariants \texttt{(s12, s23, m32, m42)}.
    \item \texttt{boundaries.m} and \texttt{dAtilde.m}: the values of the pure candidates at the boundary point of Eq.~\eqref{ideal point} and the matrix $\Tilde{A}$ from Eq.~\eqref{eq:dlog Form}, both in the Mandelstam variables $\left(s_{12},s_{23},m_3^2,m_4^2\right)$.
    \item \texttt{boundariesxyz.m} and \texttt{dAtildexyz.m}: the same as the previous two files, but in the rationalized variables $(x,y,z)$.
\end{itemize}
The second folder contains the zip file with the analytic solution, in terms of the MPLs \texttt{G[a1,\dots,an,x]}, the transcendental constants \texttt{Pi}, \texttt{Zeta[3]}, \texttt{Zeta[5]} and the dimensional regulator $\texttt{eps}$.

Lastly, we provide the codes \texttt{DiffExpRun.wl}, which computes numerically the pure basis elements using \texttt{DiffExp}. In order to use it, one should first modify the variable \texttt{PathToDiffExp} in the file \texttt{DiffExpRun.wl} appropriately, with the path to the \texttt{DiffExp.m} file on their machine. Then the only input needed is a file with the physical point where one wants to evaluate the integrals, in the same format as \texttt{testpoint.m} (which is provided in the ancillaries). The command to run the computation is then
\begin{align*}
    \texttt{math -script DiffExpRun.wl } &\texttt{-family 'fam' -vars 'vars'/ }\\
    &\texttt{-filein "point.m"/}\\
    &\texttt{-fileout "output.m"},
\end{align*}
where \texttt{'fam'} is the integral family one wants to compute, \texttt{'vars'} is either \texttt{'sij'} or \texttt{'xyz'} and the results are stored as a list in \texttt{output.m}. On the first use, the file will compute the partial derivatives of the pure basis from the matrix $\Tilde{A}$ and store them in the \texttt{fam/MIs\_DEs} folder.
\section{Genuinely new sectors}
\label{Appendix}

In this appendix we collect the pure candidates for the genuinely new sectors, referring to their position on the basis of the corresponding integral family (see ancillary files~\cite{canko_2024_14284044}). In particular, concerning the sector definition, we are using the notation
\begin{equation*}
\text{SFam}[a_1,\ldots,a_{15}],
\end{equation*}
with SFam$=$$\{\text{$F_{123}$}, \text{$F_{132}$}\}$ symbolizing the superfamily, and $a_i=\{0,1\}$ being the exponents of the propagators of the corresponding superfamily, which propagators are defined in Eqs. \eqref{props F123} and \eqref{props F132}. Pure candidates are denoted as
\begin{equation*}
I_{\text{PS}}^{\text{Fam}},
\end{equation*}
where Fam$=$$\{\text{RL1}, \text{RL2}, \text{PL1}, \text{PT4}\}$ the family index, and $\text{PS}=\{1,\ldots,189\}$ designating the position of the candidate in the corresponding pure basis.


In total, 50 new sectors were identified, consisting of 153 MIs. In the following we present a classification of the sectors based on the number of propagators they contain, commencing with those having five propagators and concluding with those comprising ten. 



\vspace{0.15cm}
\begin{center}
\textbf{\textit{Five-Propagator Pure Candidates}}\\
\end{center}
\vspace{0.2cm}

\textbf{Sector $\mathbf{F_{123}}$[0,0,1,0,0,0,0,0,1,0,0,1,1,0,1]}

\begin{multicols}{2}
\includegraphics[scale=0.20]{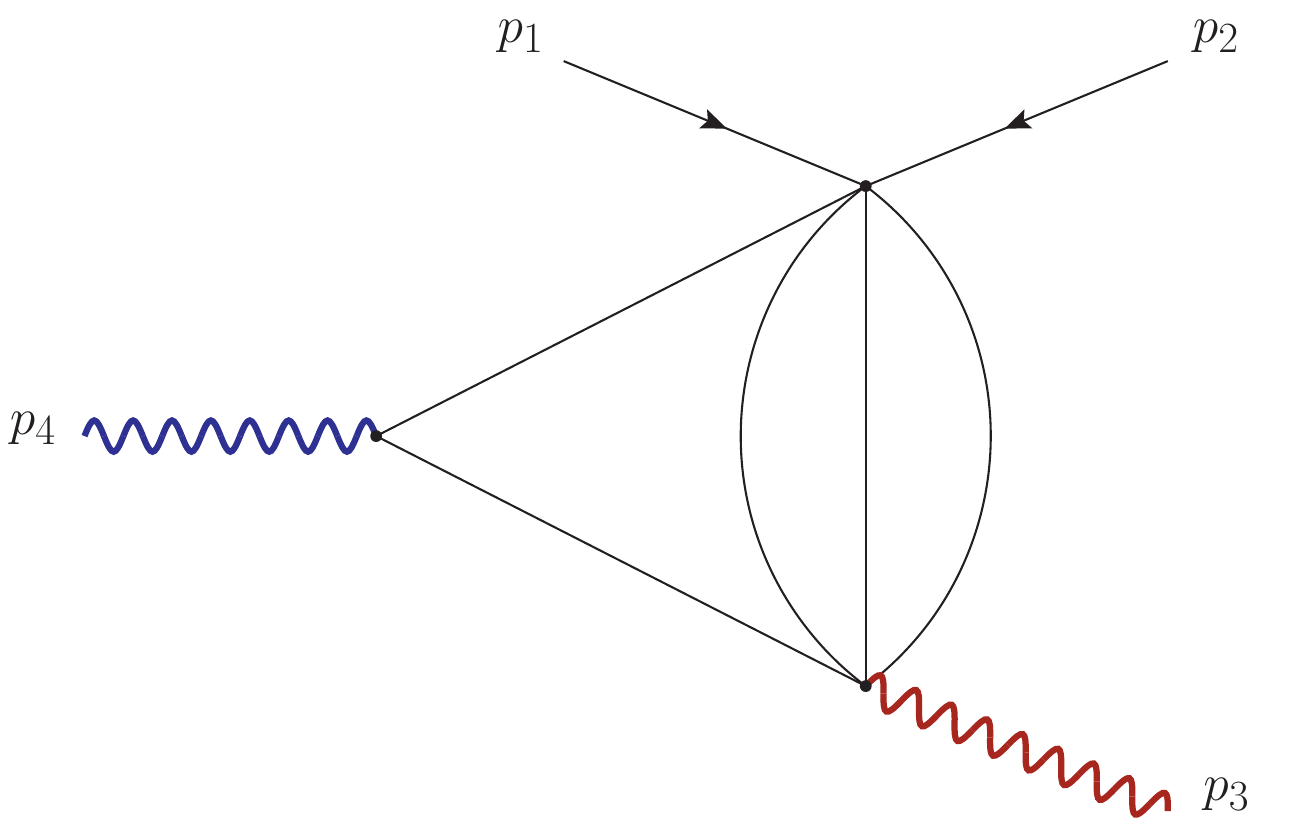}\\
\newline
\begin{equation*}
\{I_{133}^{\text{PL1}}, \, I_{134}^{\text{PL1}}\}
\end{equation*}
\end{multicols}

\textbf{Sector $\mathbf{F_{123}}$[0,0,1,0,1,0,0,0,0,0,0,1,1,0,1]}

\begin{multicols}{2}
\includegraphics[scale=0.20]{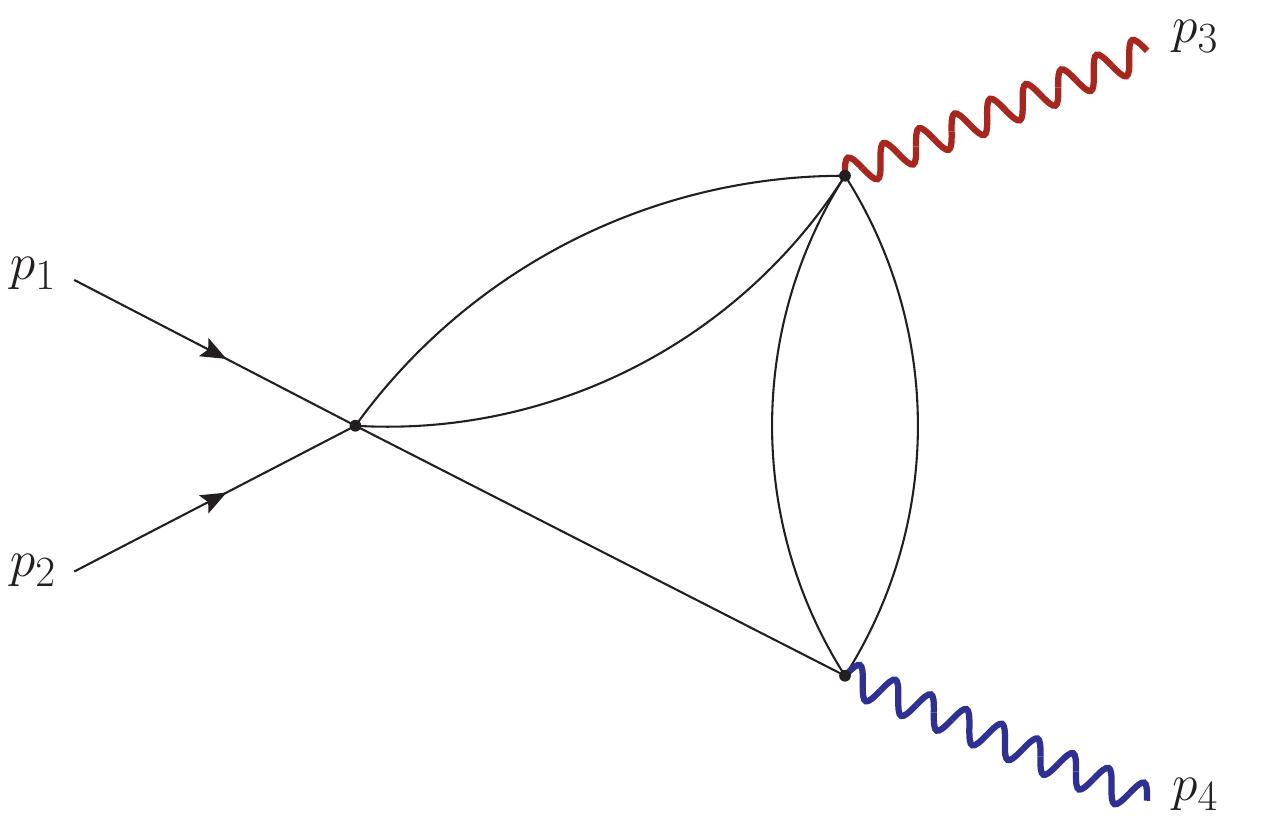}\\
\newline
\begin{equation*}
\{I_{126}^{\text{PL1}}, \, I_{127}^{\text{PL1}}, \, I_{128}^{\text{PL1}}\}
\end{equation*}
\end{multicols}



\begin{center}
\textbf{\textit{Six-Propagator Pure Candidates}}\\
\end{center}
\vspace{0.2cm}

\textbf{Sector $\mathbf{F_{123}}$[0,0,1,0,1,0,0,0,1,0,0,1,1,0,1]}

\begin{multicols}{2}
\includegraphics[scale=0.161]{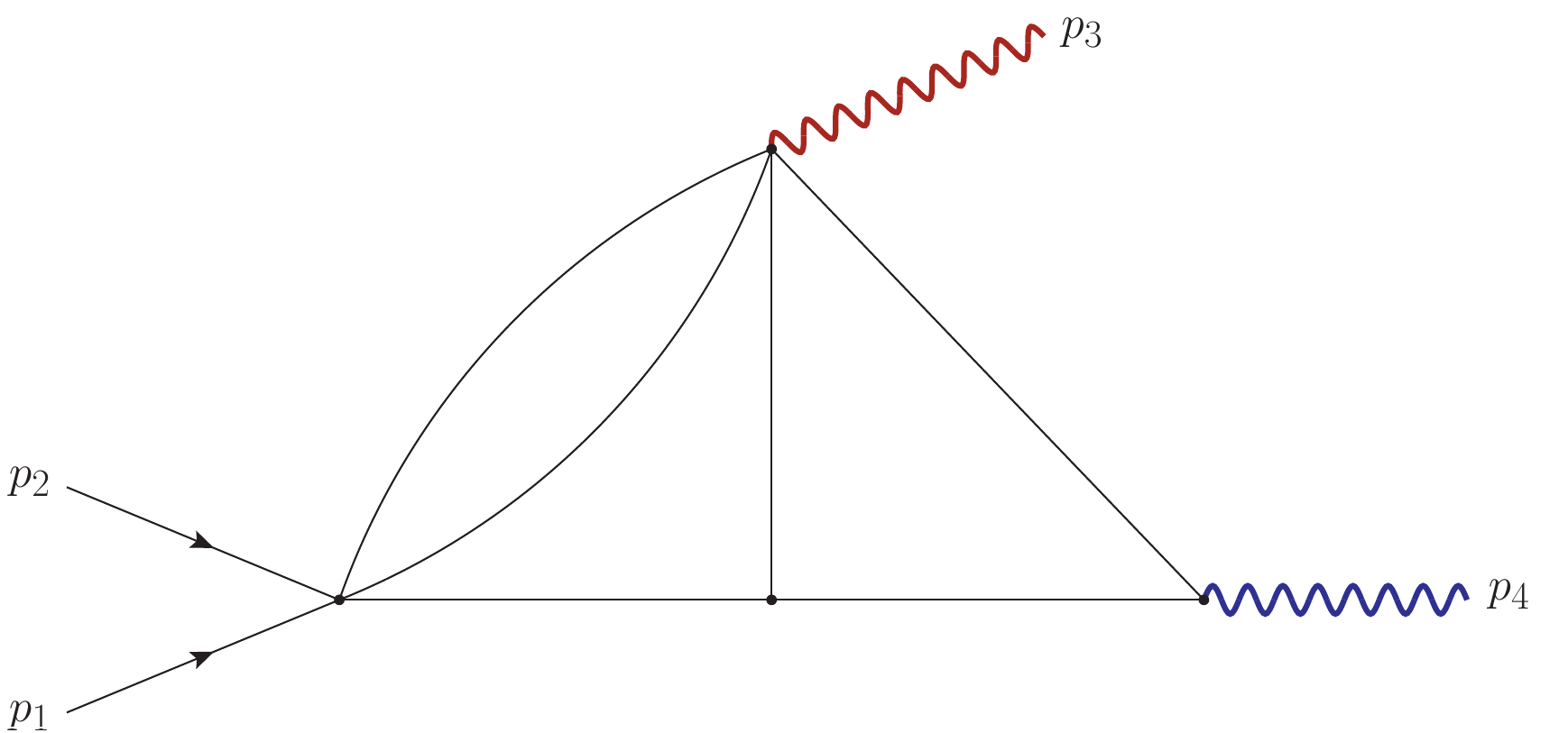}\\
\begin{equation*}
\{I_{122}^{\text{PL1}}\}
\end{equation*}
\end{multicols}

\textbf{Sector $\mathbf{F_{123}}$[1,0,1,0,0,0,0,0,0,0,1,1,1,0,1]}

\begin{multicols}{2}
\includegraphics[scale=0.165]{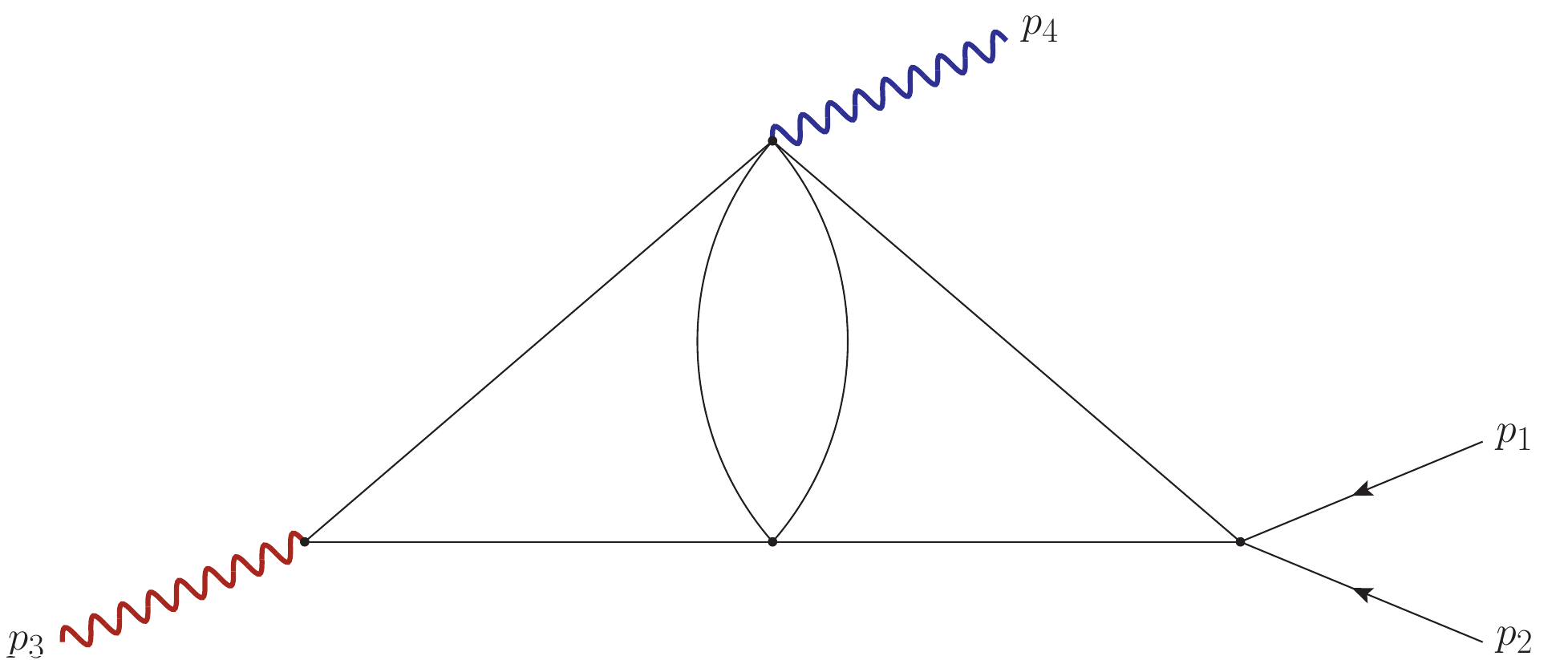}\\
\begin{equation*}
\{ I_{102}^{\text{PL1}}, \, I_{103}^{\text{PL1}} \}
\end{equation*}
\end{multicols}

\textbf{Sector $\mathbf{F_{123}}$[0,0,1,0,1,0,0,0,0,0,1,1,1,0,1]}

\begin{multicols}{2}
\includegraphics[scale=0.155]{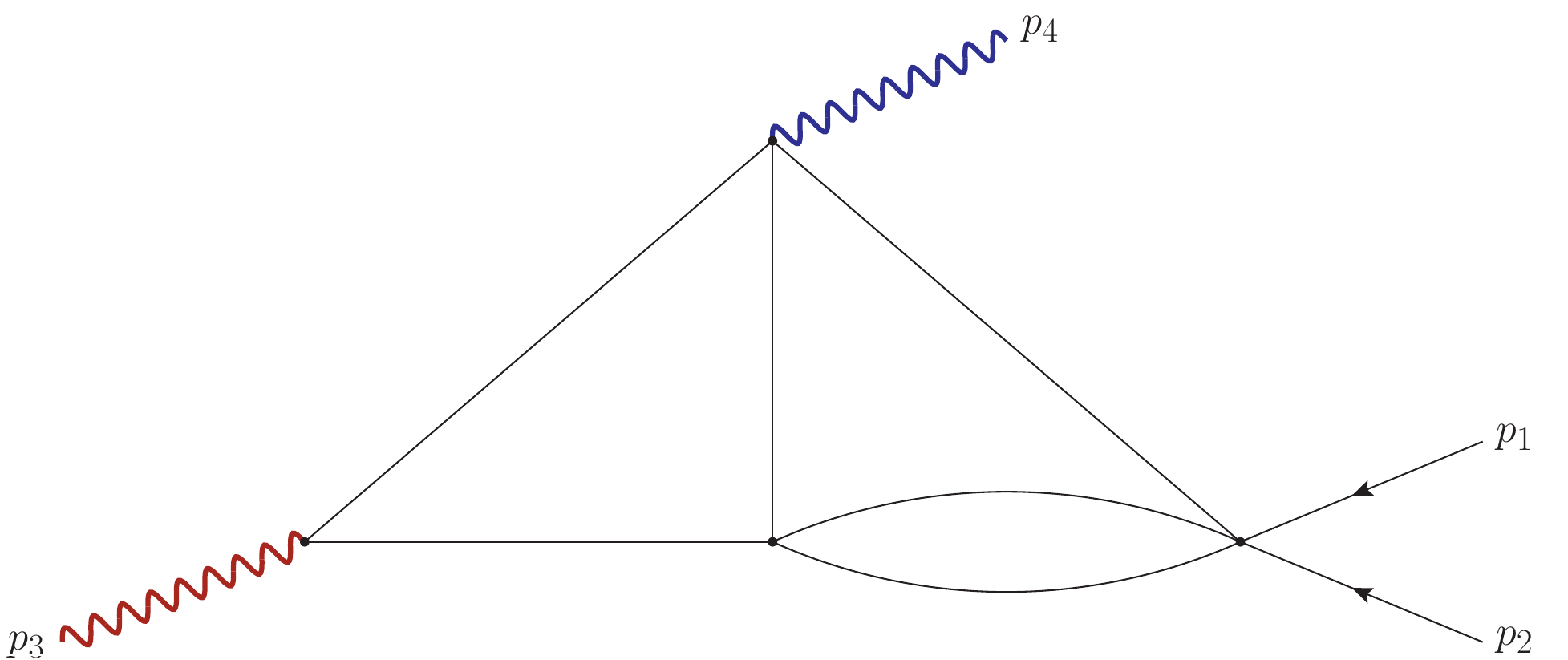}\\
\begin{equation*}
\{ I_{87}^{\text{PL1}}, \, I_{88}^{\text{PL1}}, \, I_{89}^{\text{PL1}} \}
\end{equation*}
\end{multicols}

\textbf{Sector $\mathbf{F_{123}}$[0,1,1,0,0,0,0,0,1,0,0,1,1,0,1]}

\begin{multicols}{2}
\includegraphics[scale=0.185]{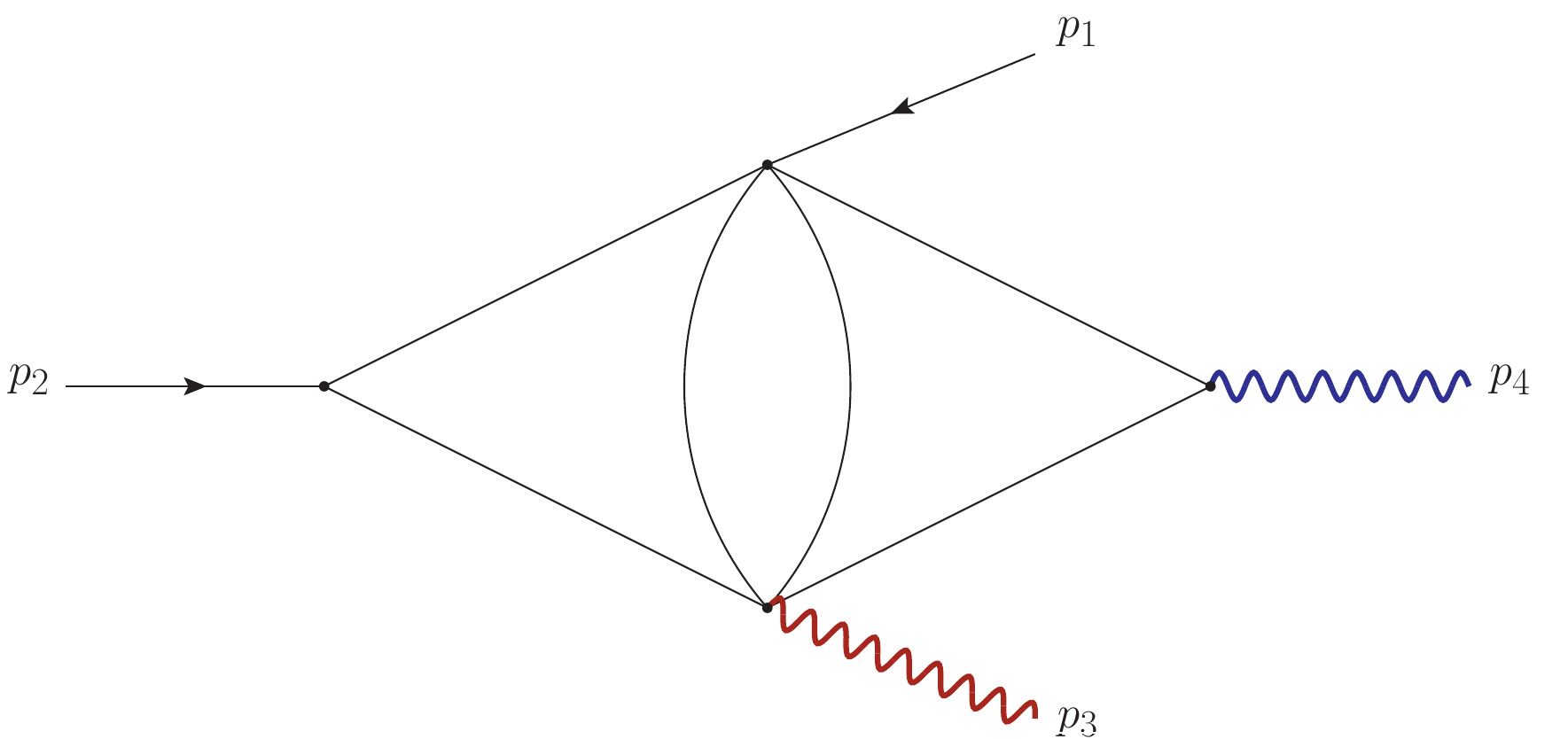}\\
\begin{equation*}
\{ I_{106}^{\text{PL1}}, \, I_{107}^{\text{PL1}} \}
\end{equation*}
\end{multicols}

\textbf{Sector $\mathbf{F_{132}}$[0,1,1,0,0,0,0,0,1,0,0,1,1,0,1]}

\begin{multicols}{2}
\includegraphics[scale=0.185]{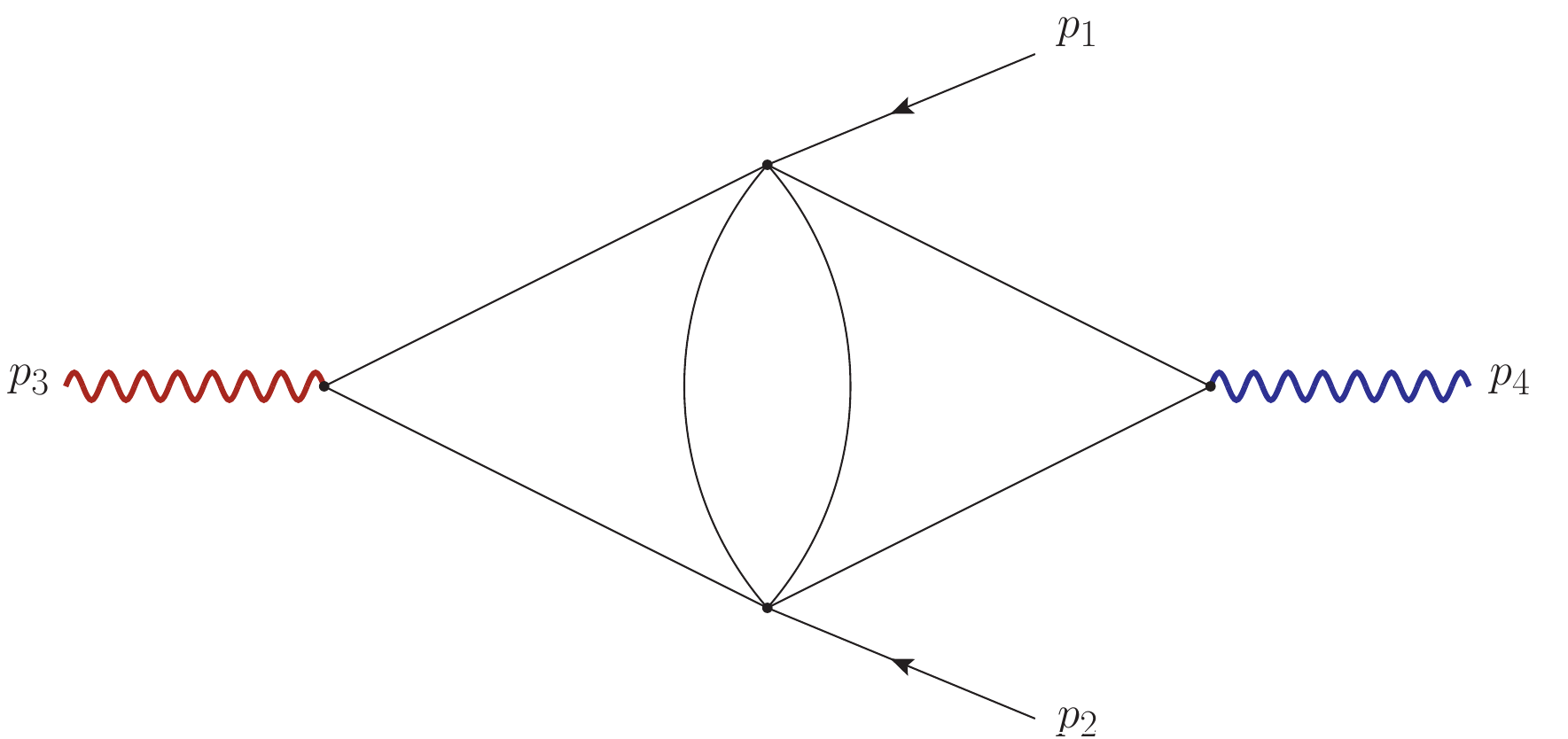}\\
\begin{equation*}
\begin{split}
\{ I_{4}^{\text{RL2}}, \, I_{5}^{\text{RL2}}, \, I_{6}^{\text{RL2}}, \, I_{7}^{\text{RL2}}, \, I_{8}^{\text{RL2}}, \, I_{9}^{\text{RL2}} \}
\end{split}
\end{equation*}
\end{multicols}

\textbf{Sector $\mathbf{F_{123}}$[0,1,1,0,1,0,0,0,0,0,0,1,1,0,1]}

\begin{multicols}{2}
\includegraphics[scale=0.187]{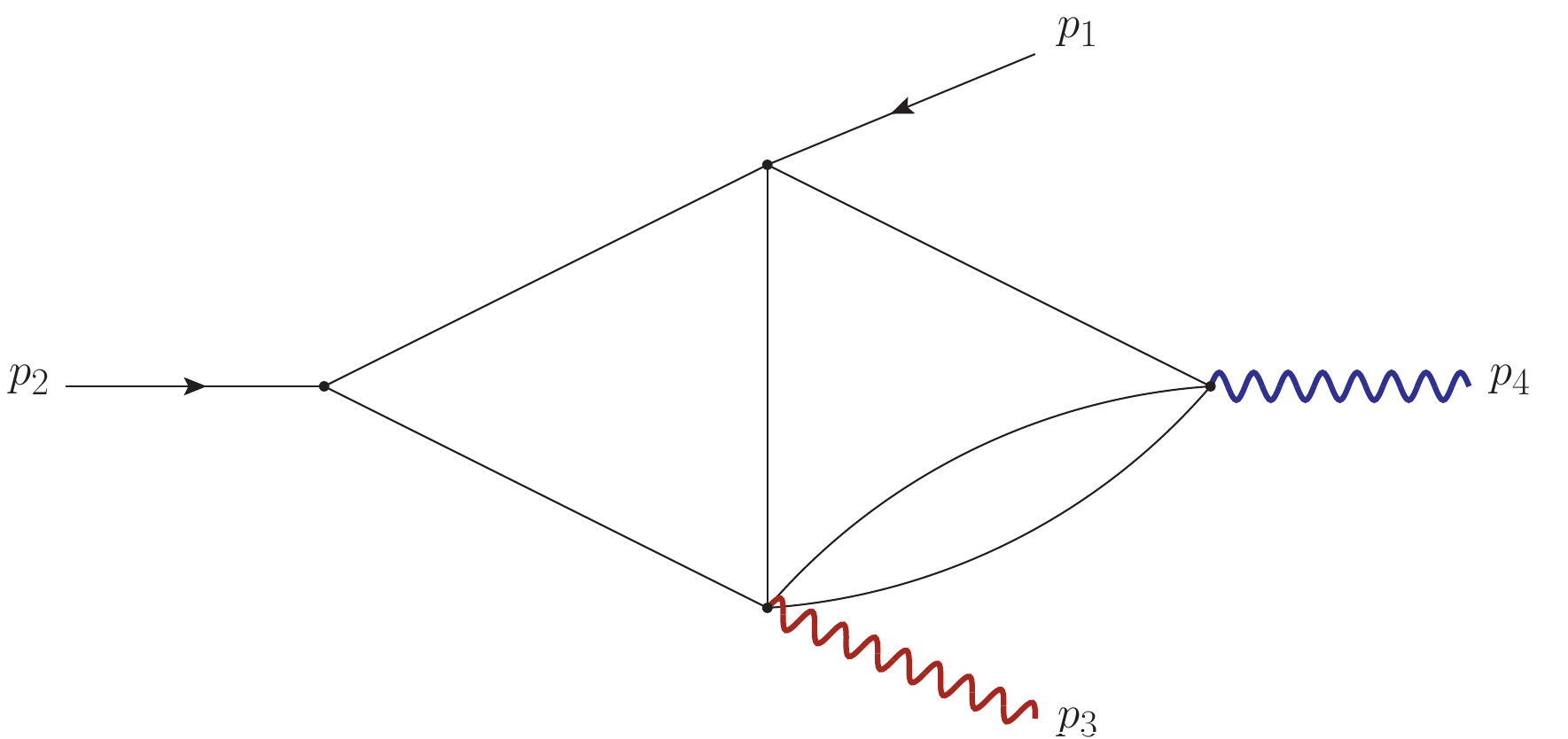}\\
\newline
\begin{equation*}
\{ I_{104}^{\text{PL1}}, \, I_{105}^{\text{PL1}} \}
\end{equation*}
\end{multicols}

\textbf{Sector $\mathbf{F_{123}}$[0,0,0,1,1,0,0,0,0,1,1,0,1,1,0]}

\begin{multicols}{2}
\includegraphics[scale=0.187]{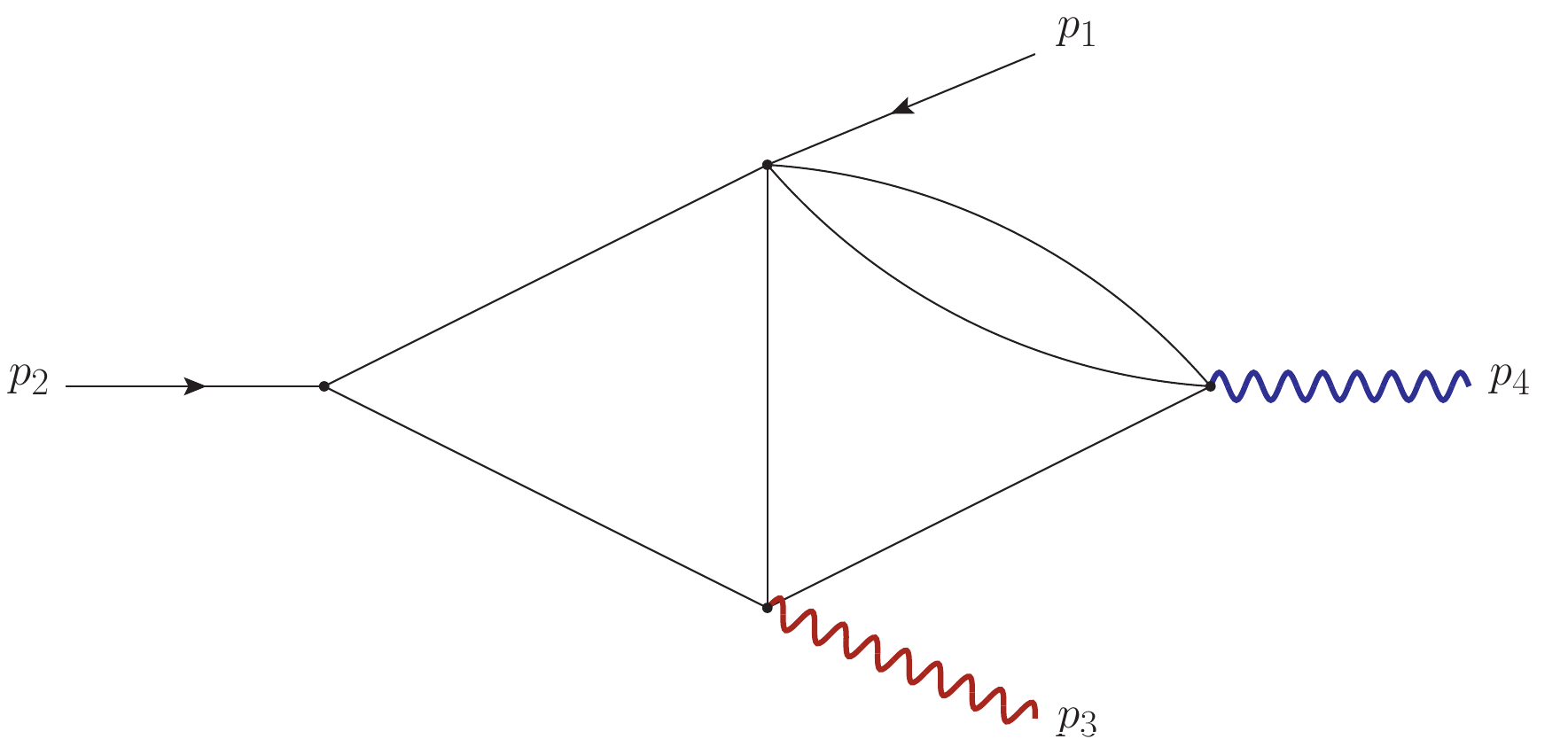}\\
\newline
\begin{equation*}
\{ I_{136}^{\text{PT4}}, \, I_{137}^{\text{PT4}} \}
\end{equation*}
\end{multicols}

\textbf{Sector $\mathbf{F_{123}}$[0,1,0,0,0,0,1,0,1,0,0,1,1,0,1]}

\begin{multicols}{2}
\includegraphics[scale=0.187]{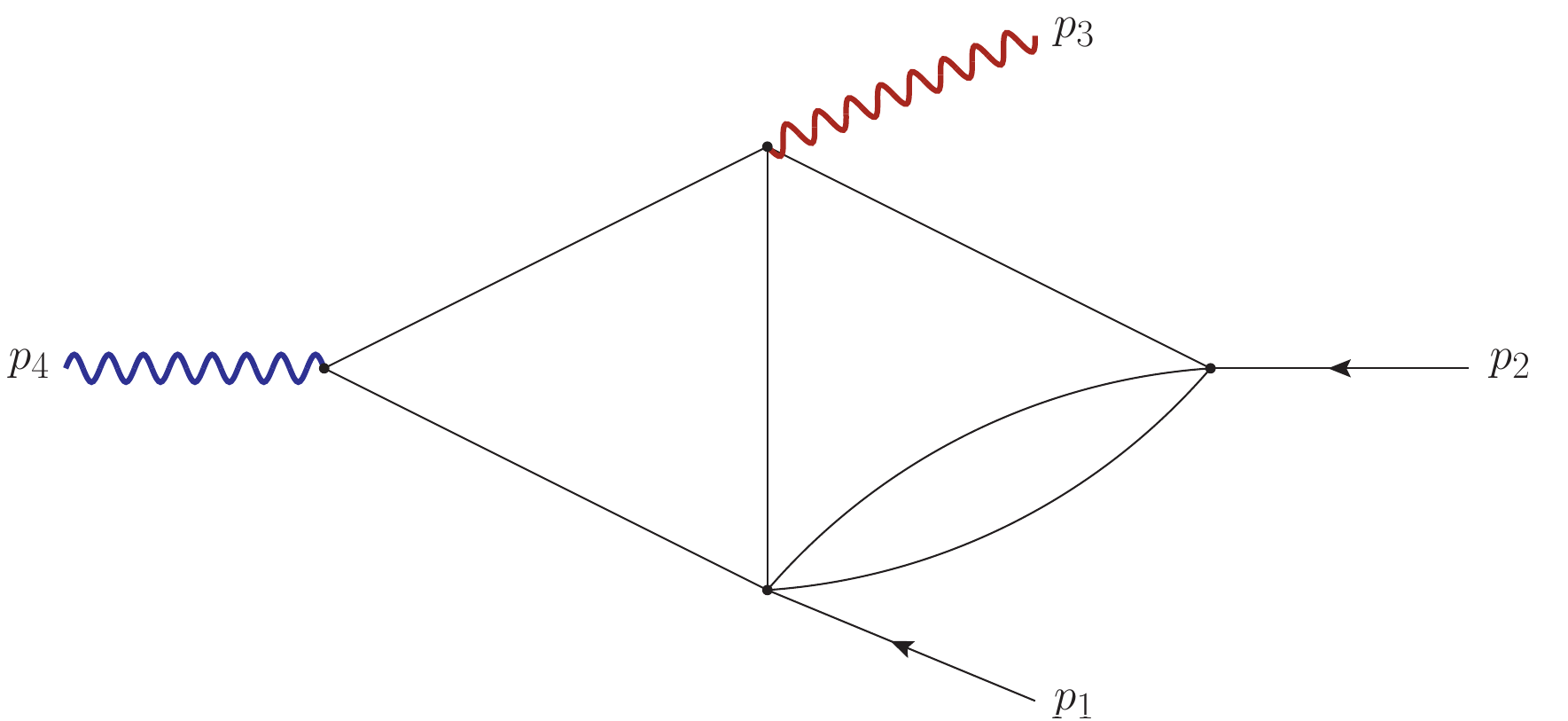}\\
\begin{equation*}
\{ I_{77}^{\text{PL1}}, \, I_{78}^{\text{PL1}}, \, I_{79}^{\text{PL1}}, \, I_{80}^{\text{PL1}} \}
\end{equation*}
\end{multicols}

\textbf{Sector $\mathbf{F_{123}}$[0,0,1,1,1,0,0,0,0,1,0,0,0,1,1]}

\begin{multicols}{2}
\includegraphics[scale=0.187]{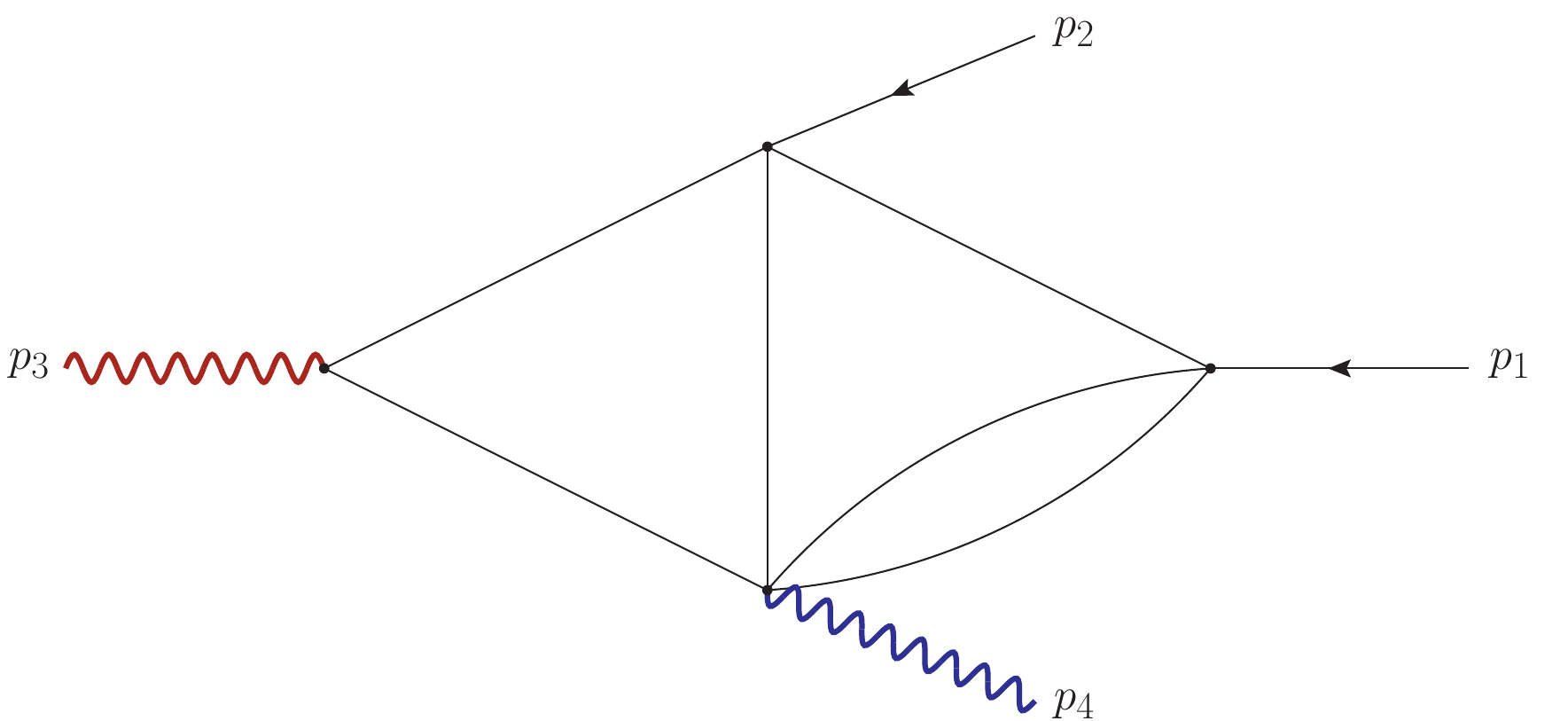}\\
\begin{equation*}
\{ I_{123}^{\text{PT4}}, \, I_{124}^{\text{PT4}}, \, I_{125}^{\text{PT4}} \}
\end{equation*}
\end{multicols}

\textbf{Sector $\mathbf{F_{123}}$[1,1,1,0,0,0,0,0,0,0,0,1,1,0,1]}

\begin{multicols}{2}
\includegraphics[scale=0.19]{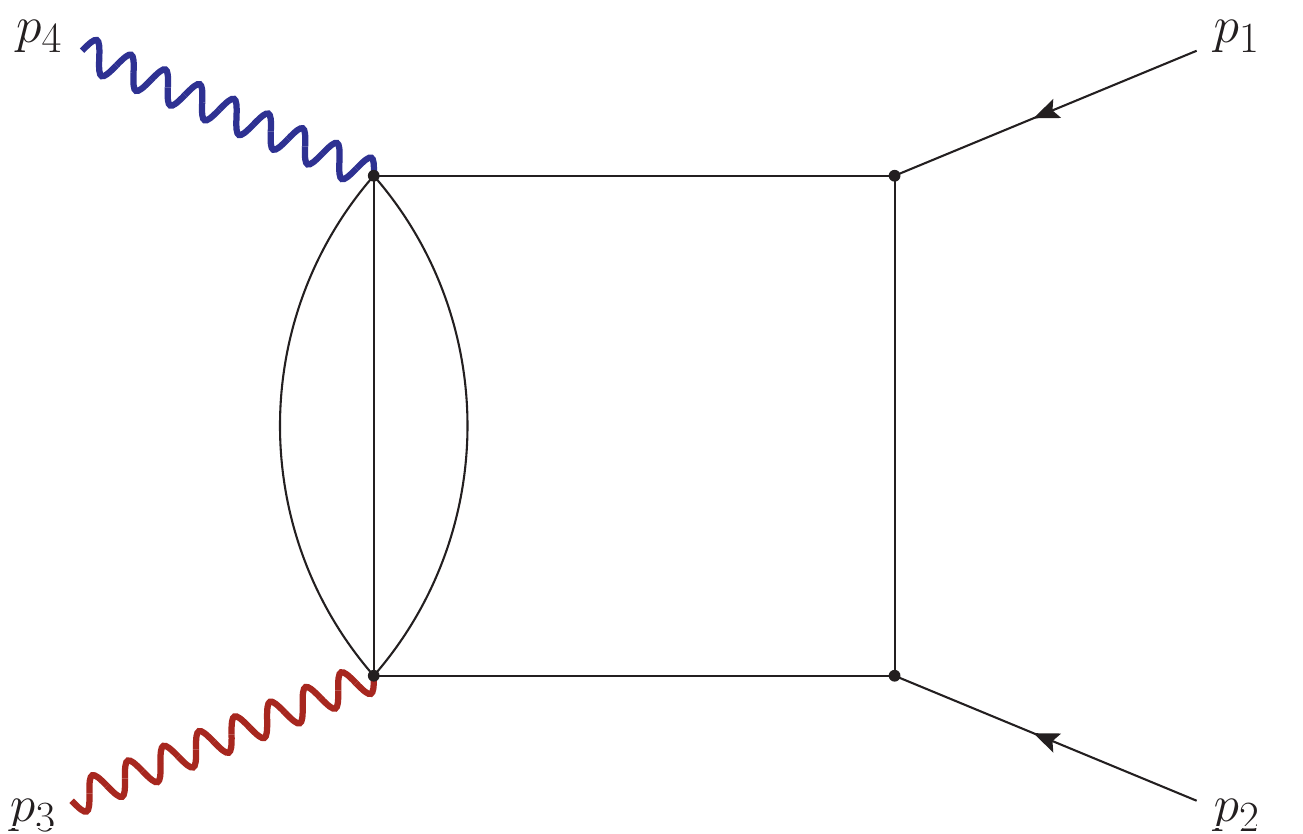}\\
\begin{equation*}
\{ I_{110}^{\text{PL1}} \}
\end{equation*}
\end{multicols}

\textbf{Sector $\mathbf{F_{123}}$[0,1,0,0,0,0,0,0,1,0,1,1,1,0,1]}

\begin{multicols}{2}
\includegraphics[scale=0.19]{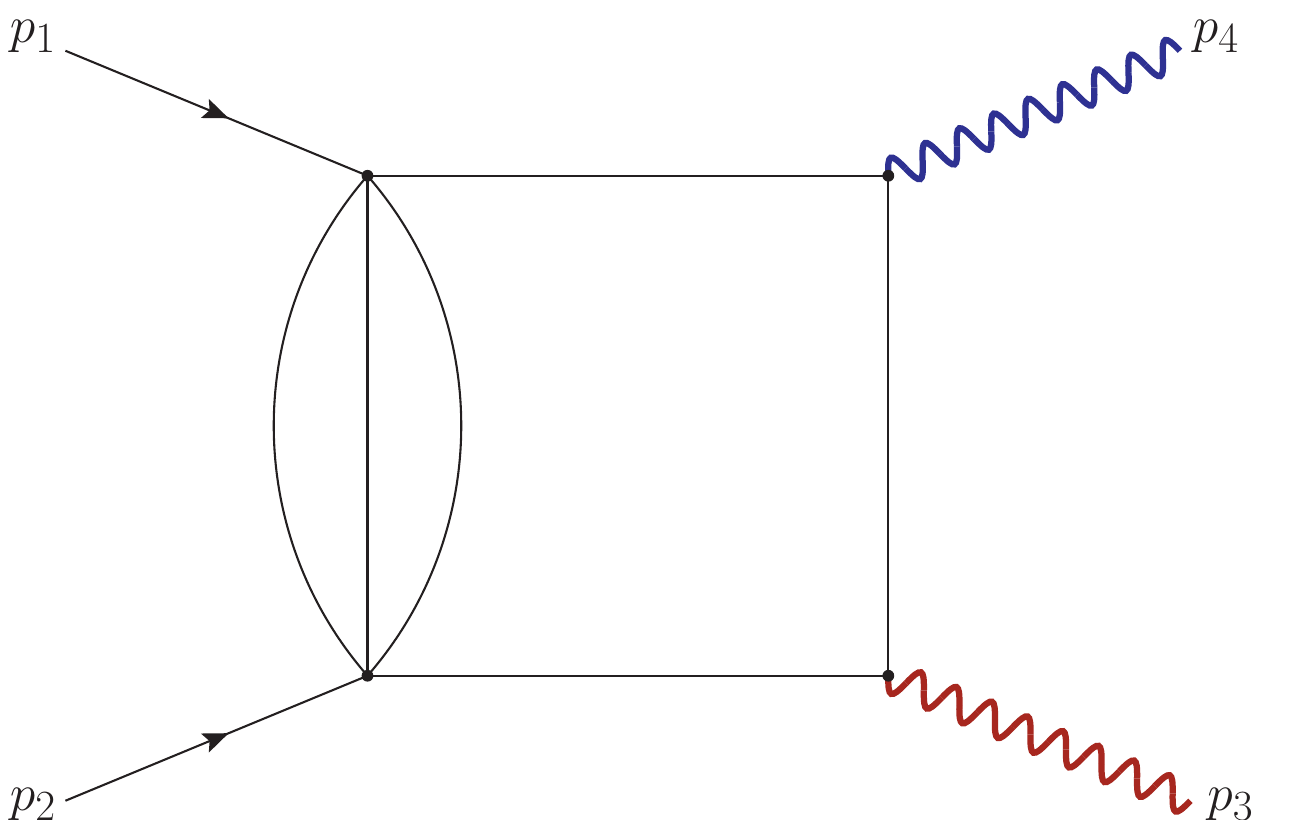}\\
\begin{equation*}
\{ I_{108}^{\text{PL1}}, \, I_{109}^{\text{PL1}} \}
\end{equation*}
\end{multicols}

\textbf{Sector $\mathbf{F_{123}}$[0,1,1,1,0,0,0,0,1,0,0,0,1,0,1]}

\begin{multicols}{2}
\includegraphics[scale=0.19]{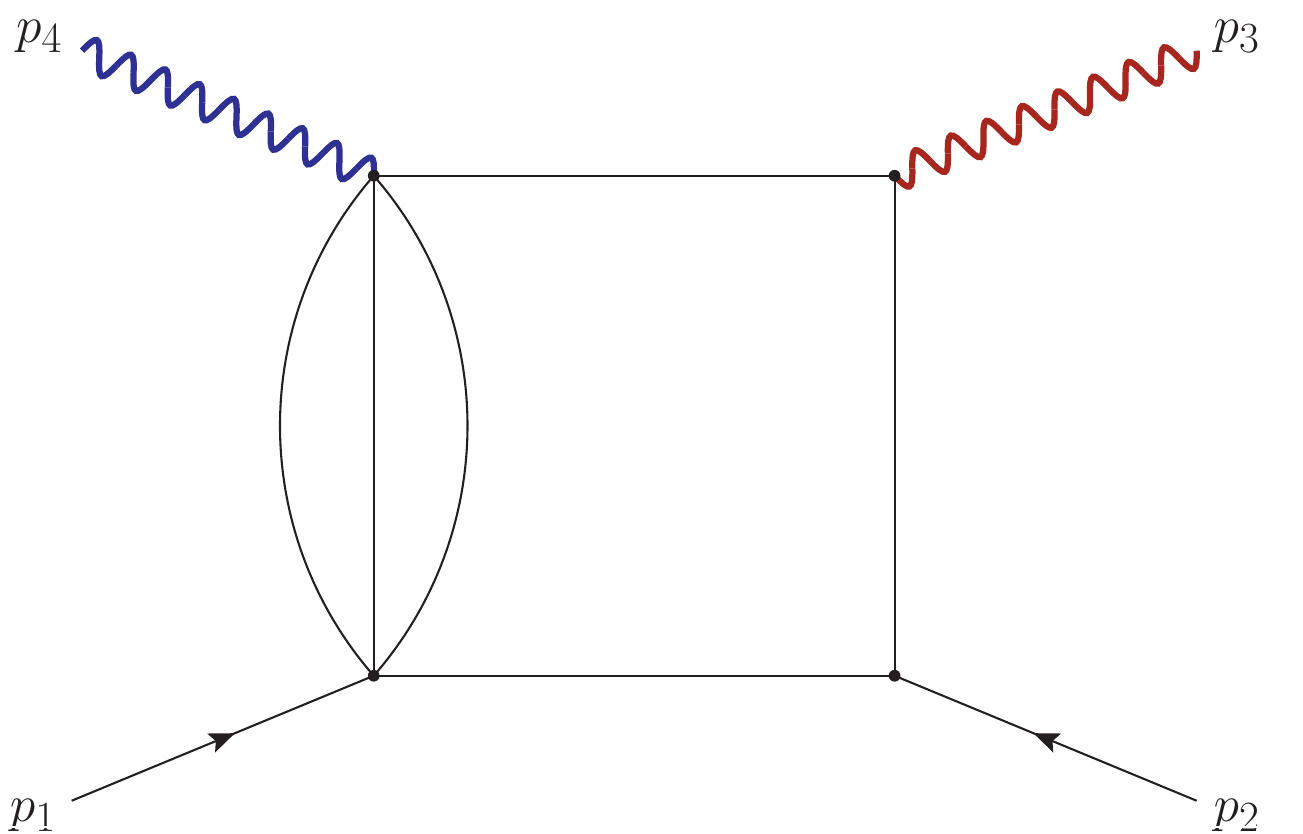}\\
\begin{equation*}
\{ I_{13}^{\text{RL1}} \}
\end{equation*}
\end{multicols}

\textbf{Sector $\mathbf{F_{132}}$[0,1,1,1,0,0,0,0,1,0,0,0,1,0,1]}

\begin{multicols}{2}
\includegraphics[scale=0.19]{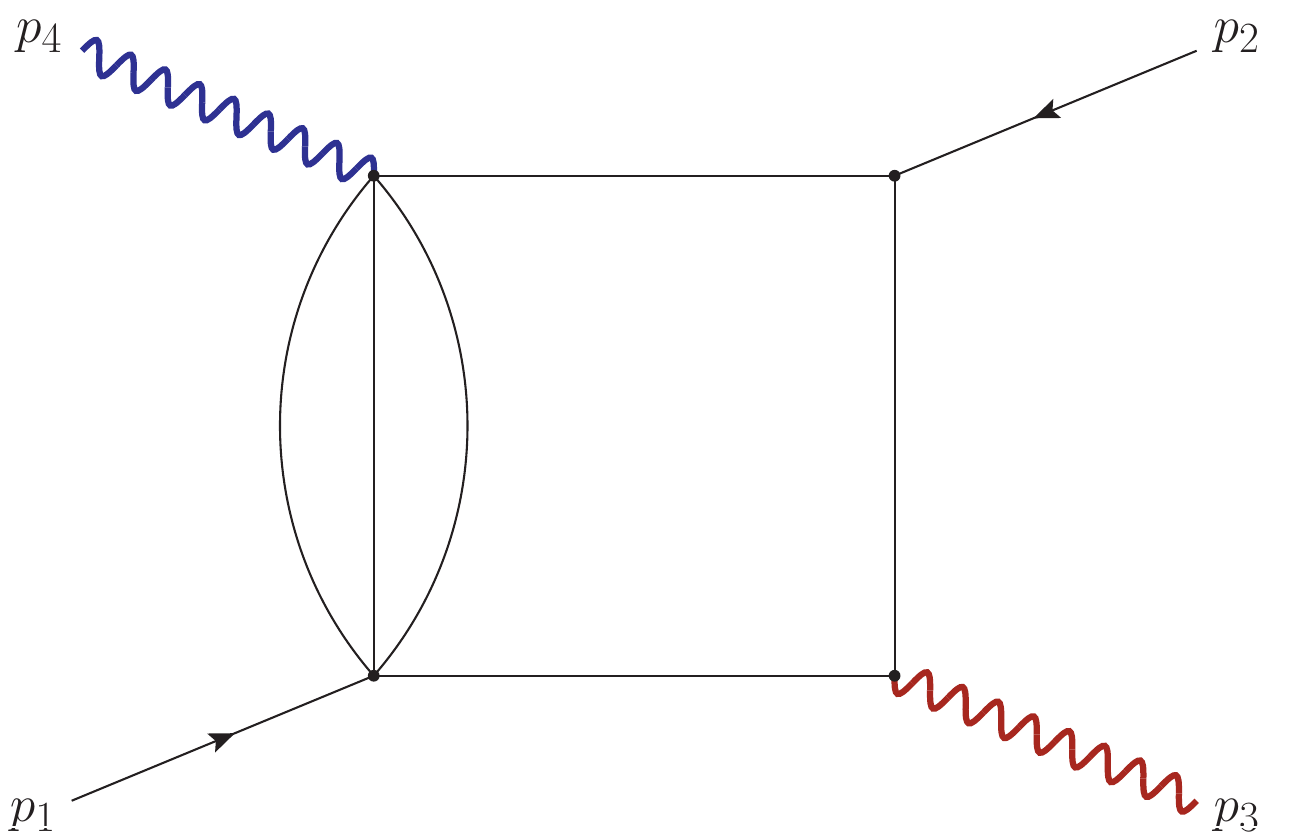}\\
\begin{equation*}
\{ I_{14}^{\text{RL2}} \}
\end{equation*}
\end{multicols}

\textbf{Sector $\mathbf{F_{123}}$[0,1,0,0,1,0,1,0,0,0,0,1,1,0,1]}

\begin{multicols}{2}
\includegraphics[scale=0.19]{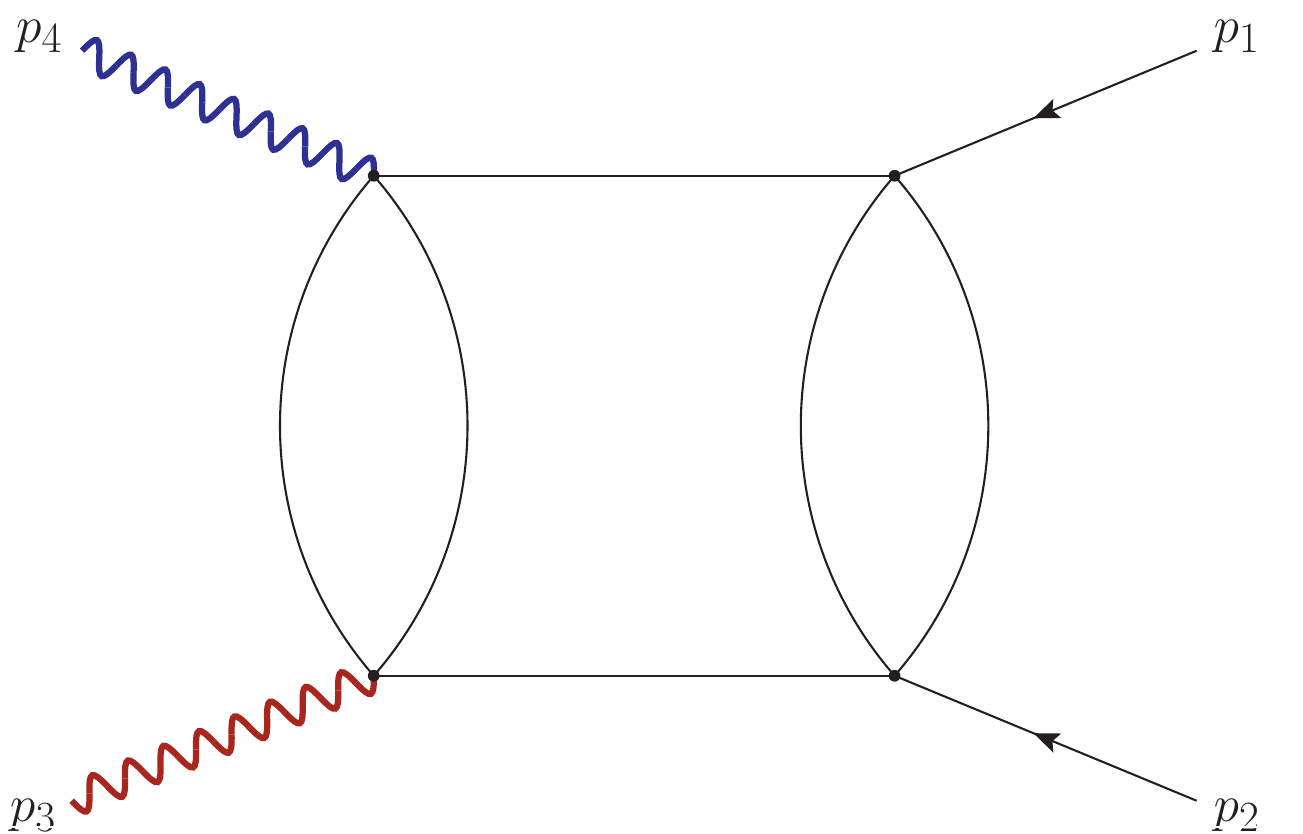}\\
\begin{equation*}
\{ I_{81}^{\text{PL1}}, \, I_{82}^{\text{PL1}}, I_{83}^{\text{PL1}} \}
\end{equation*}
\end{multicols}

\textbf{Sector $\mathbf{F_{123}}$[0,0,0,1,1,0,0,0,0,1,1,0,0,1,1]}

\begin{multicols}{2}
\includegraphics[scale=0.19]{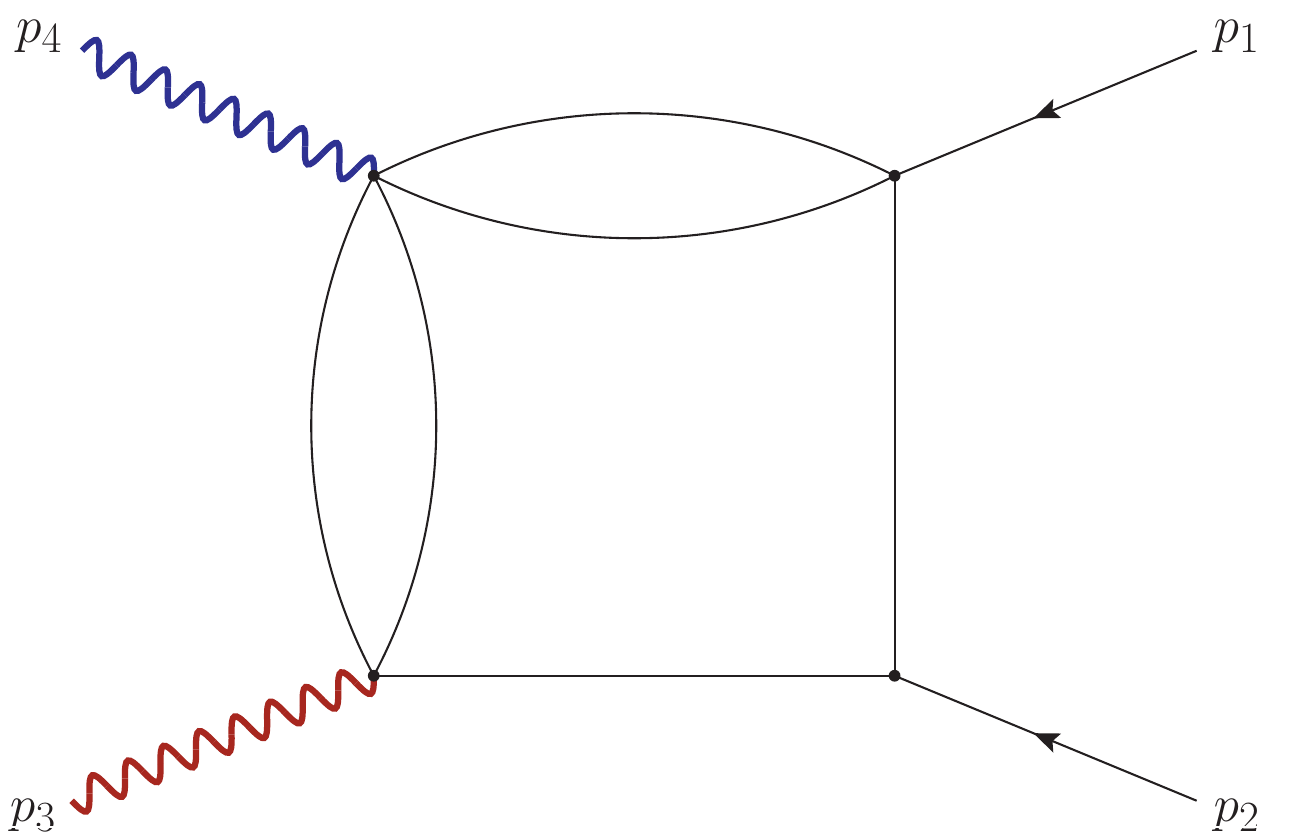}\\
\begin{equation*}
\{ I_{155}^{\text{PT4}} \}
\end{equation*}
\end{multicols}

\textbf{Sector $\mathbf{F_{123}}$[1,0,0,1,0,1,0,0,0,0,1,0,1,1,0]}

\begin{multicols}{2}
\includegraphics[scale=0.19]{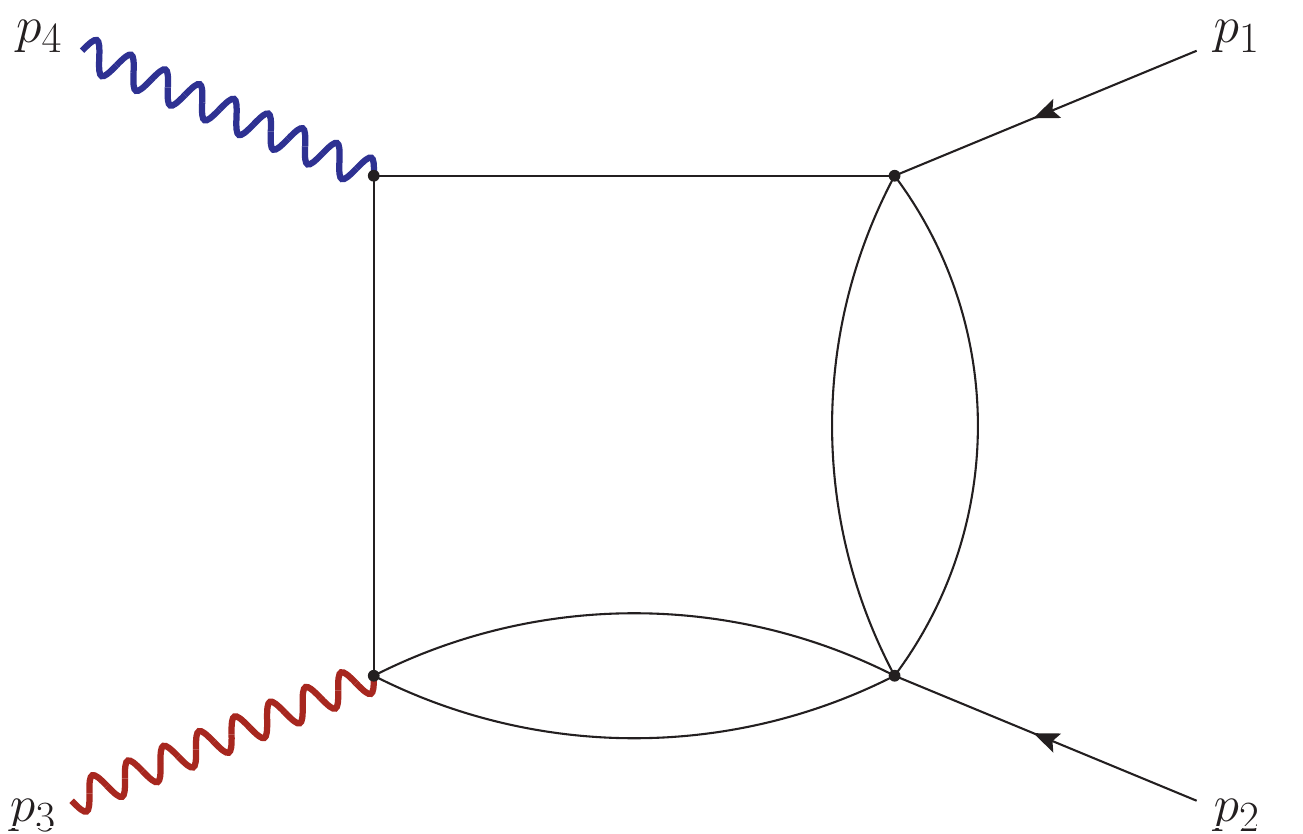}\\
\begin{equation*}
\{ I_{114}^{\text{PT4}}, I_{115}^{\text{PT4}}, I_{116}^{\text{PT4}} \}
\end{equation*}
\end{multicols}

\textbf{Sector $\mathbf{F_{123}}$[0,0,0,1,1,0,0,0,0,0,1,0,1,1,1]}

\begin{multicols}{2}
\includegraphics[scale=0.195]{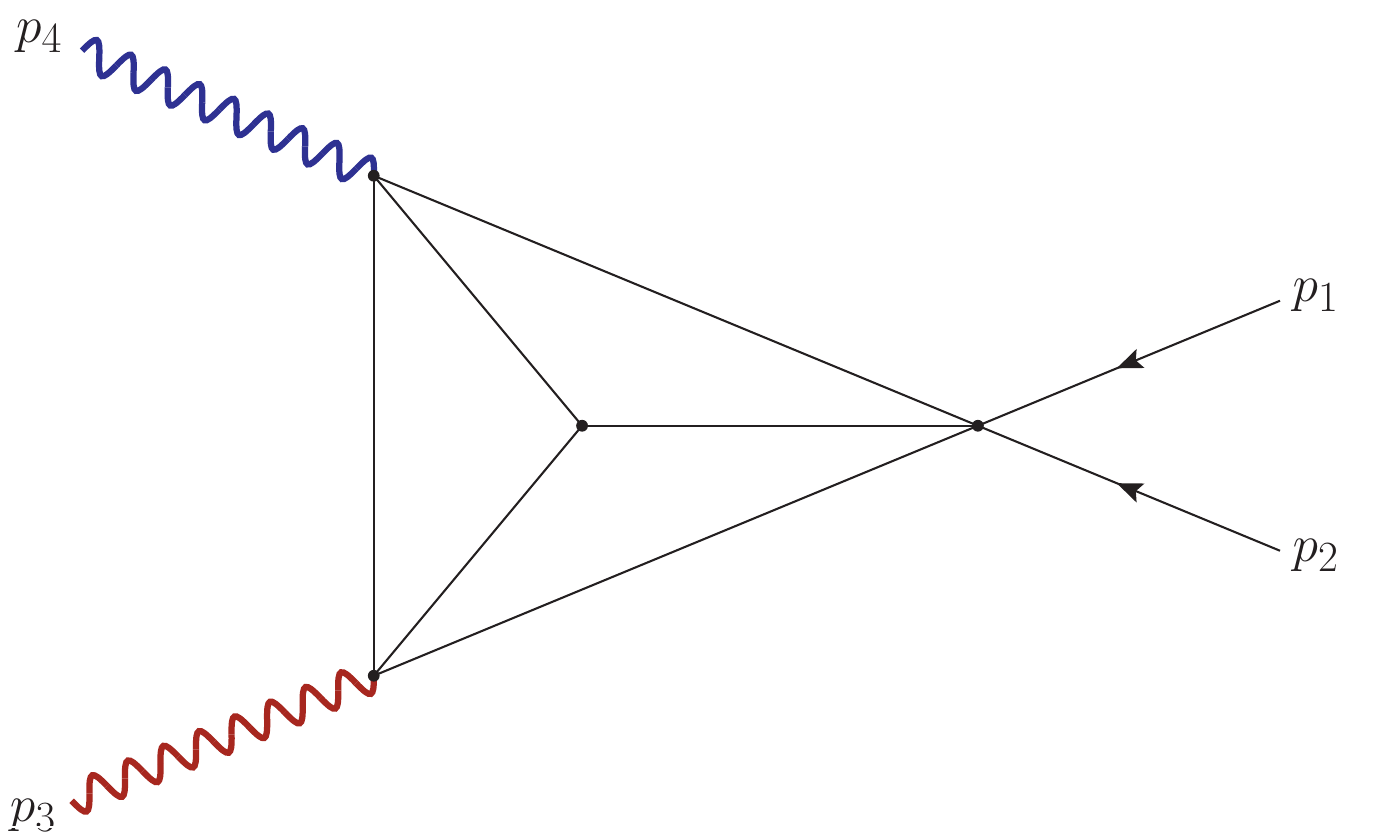}\\
\begin{equation*}
\{ I_{104}^{\text{PT4}}, I_{105}^{\text{PT4}}, I_{106}^{\text{PT4}}, I_{107}^{\text{PT4}} \}
\end{equation*}
\end{multicols}



\begin{center}
\textbf{\textit{Seven-Propagator Pure Candidates}}\\
\end{center}
\vspace{0.2cm}

\textbf{Sector $\mathbf{F_{123}}$[1,0,1,0,0,0,0,0,1,0,1,1,1,0,1]}

\begin{multicols}{2}
\includegraphics[scale=0.195]{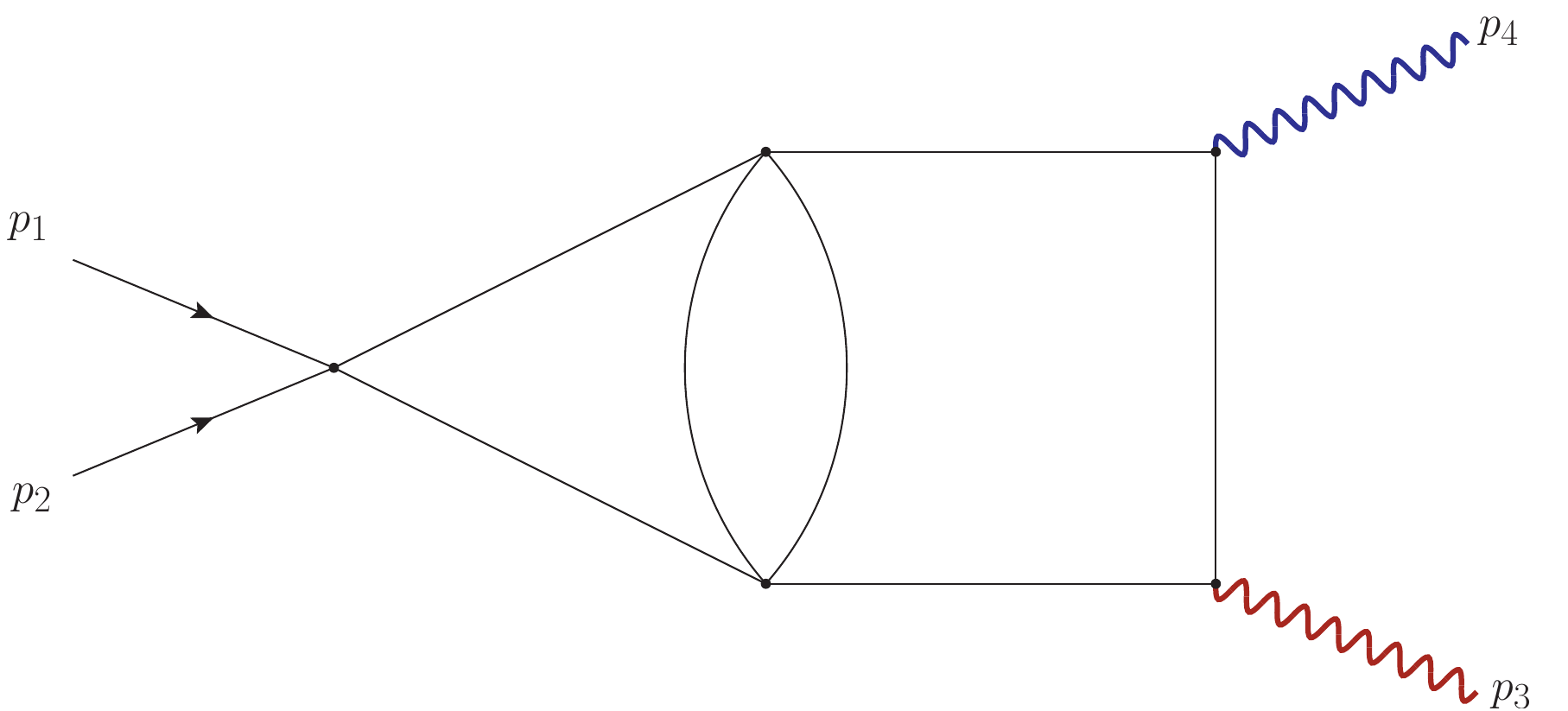}\\
\begin{equation*}
\{ I_{71}^{\text{PL1}} \}
\end{equation*}
\end{multicols}

\textbf{Sector $\mathbf{F_{123}}$[1,1,1,0,0,0,0,0,0,0,1,1,1,0,1]}

\begin{multicols}{2}
\includegraphics[scale=0.195]{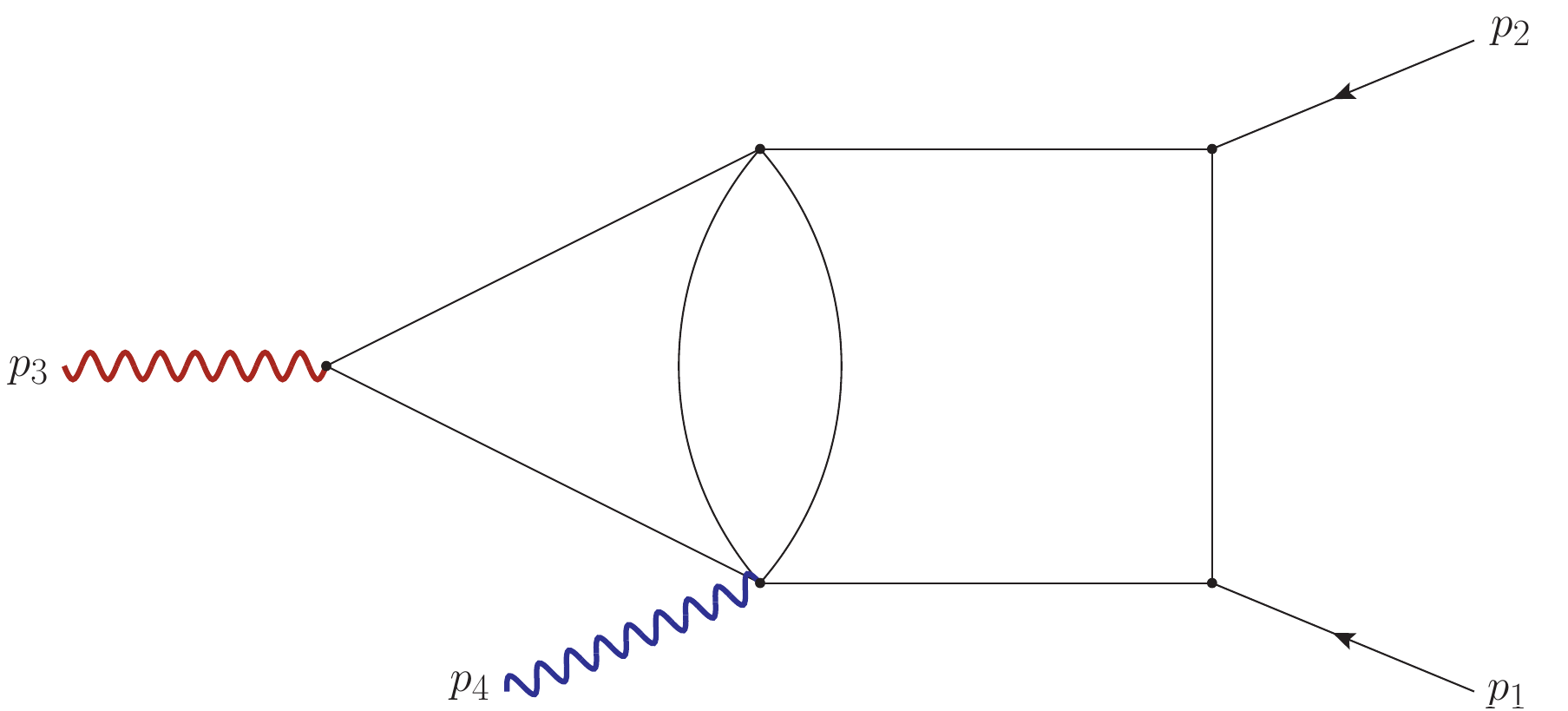}\\
\begin{equation*}
\{ I_{66}^{\text{PL1}} \}
\end{equation*}
\end{multicols}

\textbf{Sector $\mathbf{F_{123}}$[0,1,1,1,0,0,0,0,1,0,0,1,1,0,1]}

\begin{multicols}{2}
\includegraphics[scale=0.195]{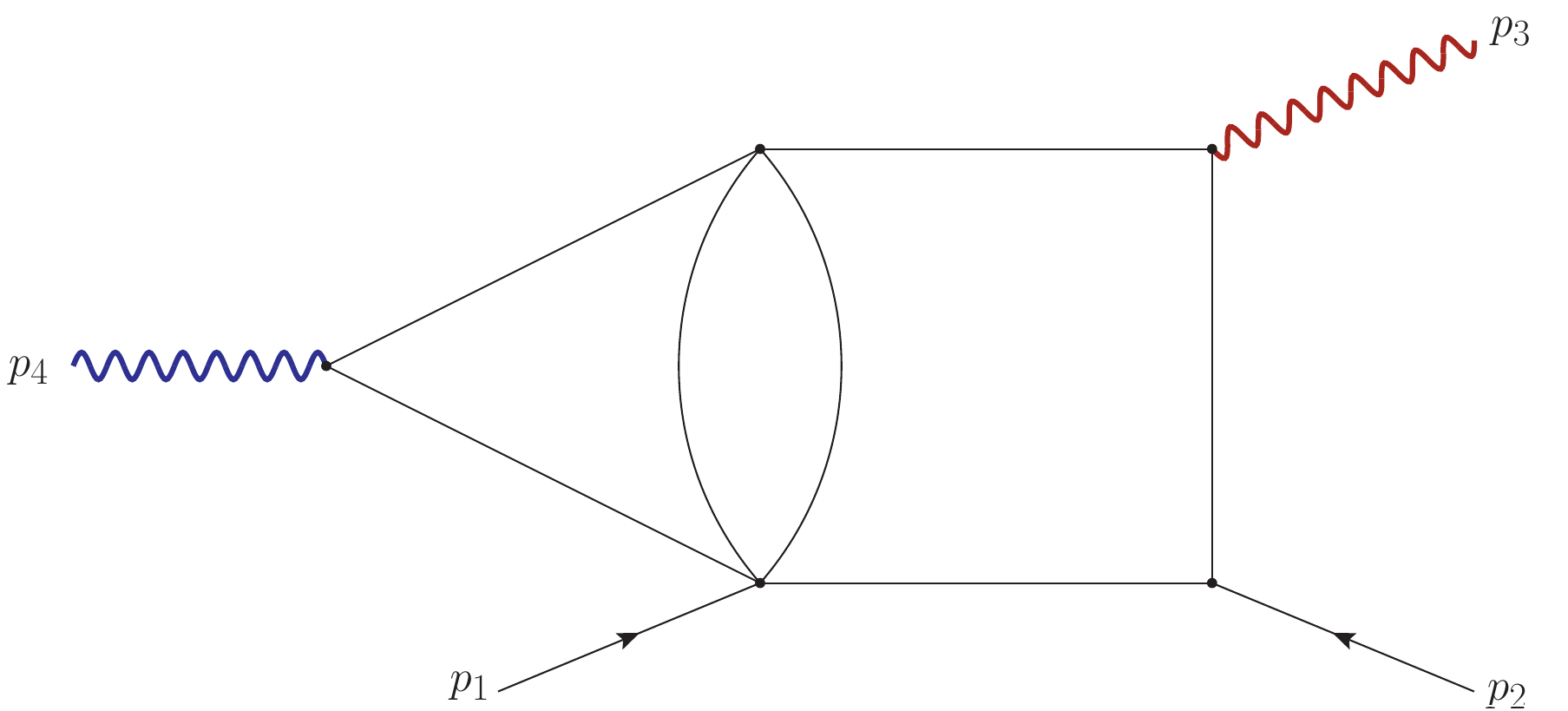}\\
\begin{equation*}
\{ I_{4}^{\text{RL1}} \}
\end{equation*}
\end{multicols}

\textbf{Sector $\mathbf{F_{132}}$[1,1,1,0,0,0,0,0,1,0,0,1,1,0,1]}

\begin{multicols}{2}
\includegraphics[scale=0.195]{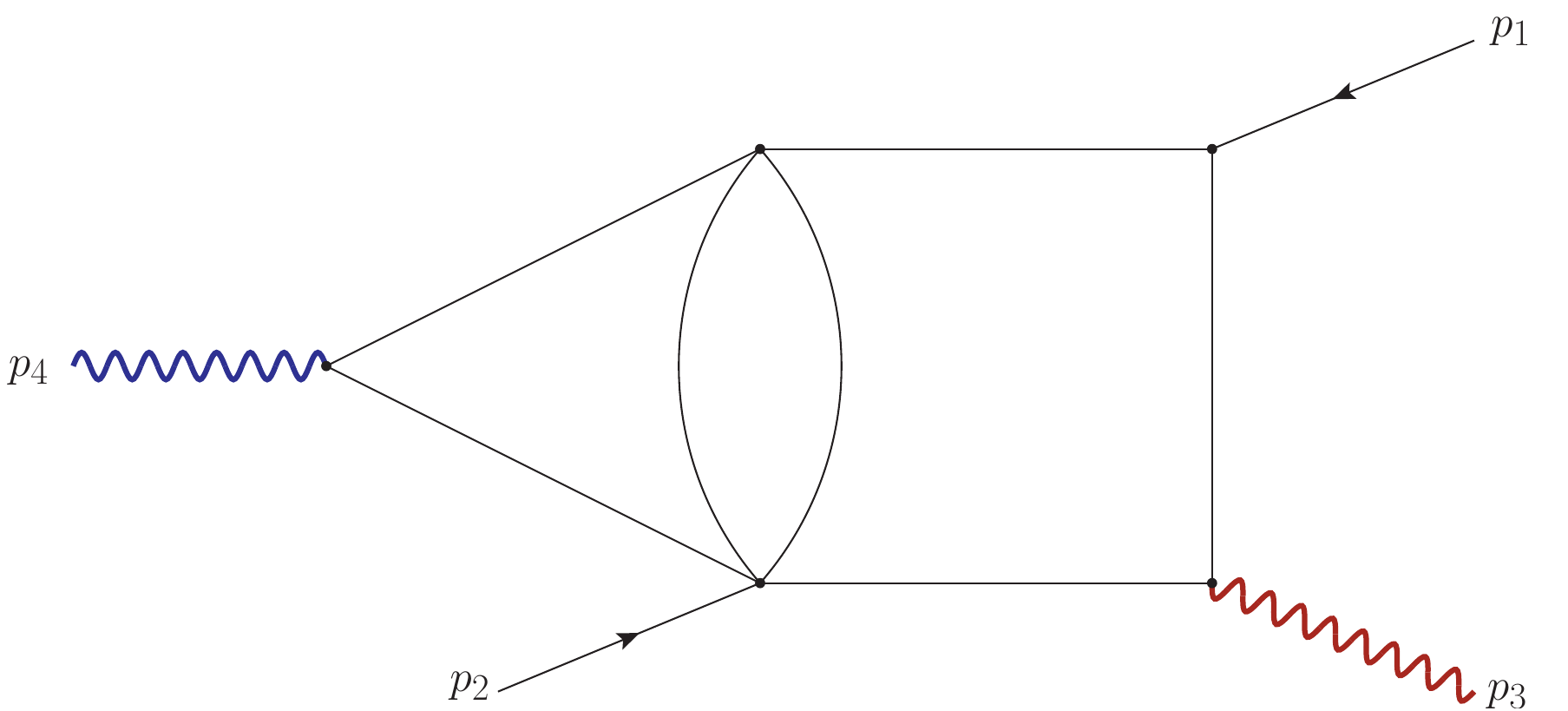}\\
\begin{equation*}
\{ I_{2}^{\text{RL2}} \}
\end{equation*}
\end{multicols}

\textbf{Sector $\mathbf{F_{123}}$[1,1,1,0,0,0,1,0,0,0,0,1,1,0,1]}

\begin{multicols}{2}
\includegraphics[scale=0.195]{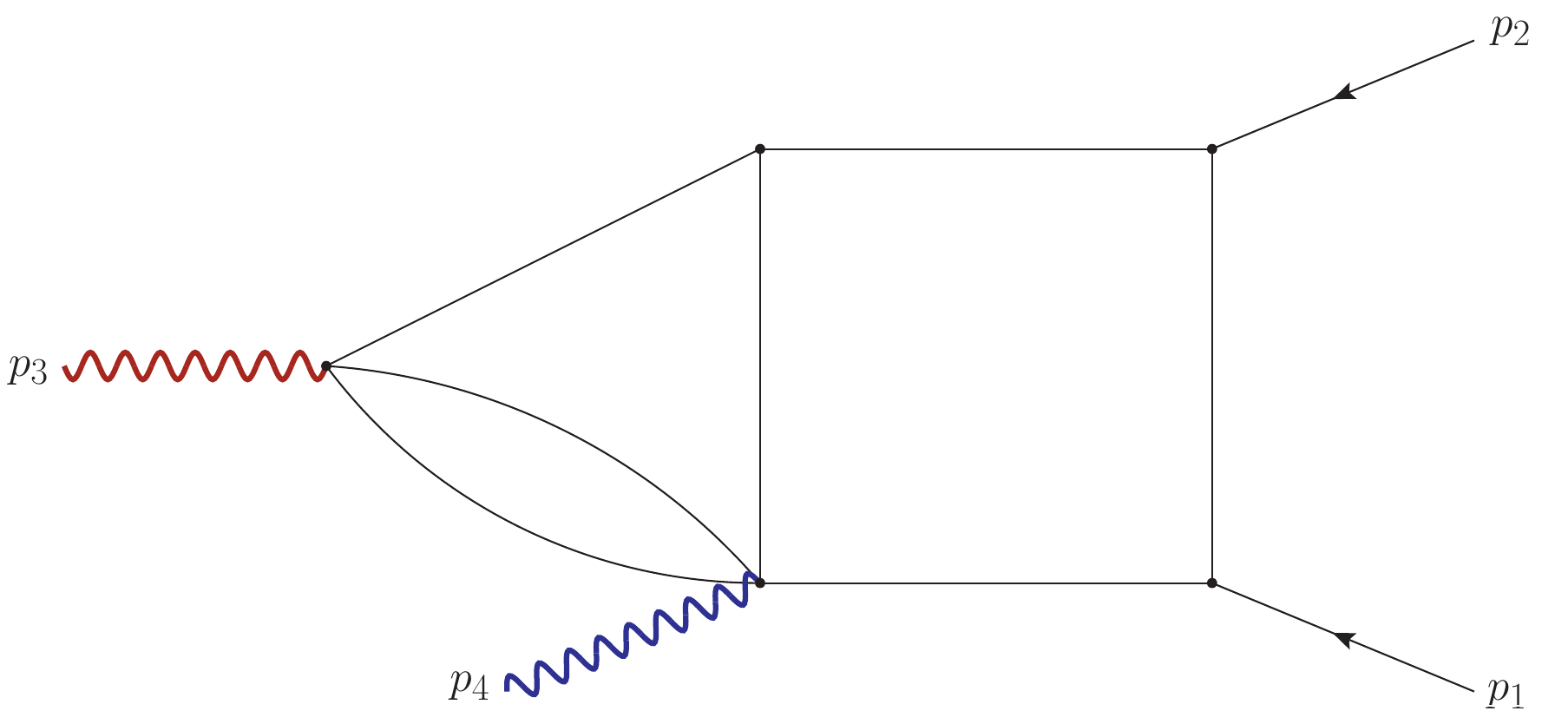}\\
\begin{equation*}
\{ I_{64}^{\text{PL1}} \}
\end{equation*}
\end{multicols}

\textbf{Sector $\mathbf{F_{123}}$[1,0,1,1,0,1,0,0,0,0,1,0,1,0,1]}

\begin{multicols}{2}
\includegraphics[scale=0.195]{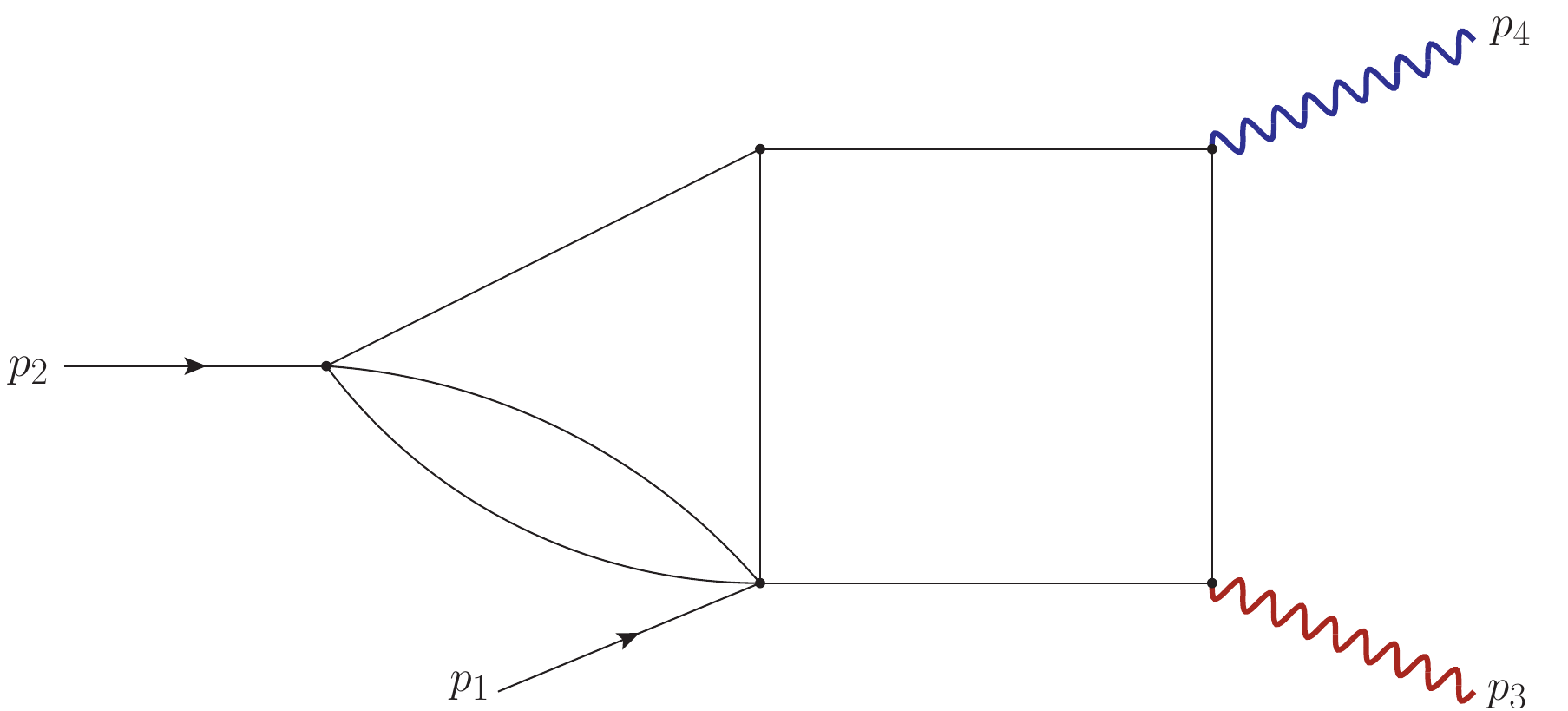}\\
\begin{equation*}
\{ I_{81}^{\text{PT4}}, \, I_{82}^{\text{PT4}} \}
\end{equation*}
\end{multicols}

\textbf{Sector $\mathbf{F_{123}}$[0,1,0,0,1,0,1,0,0,0,1,1,1,0,1]}

\begin{multicols}{2}
\includegraphics[scale=0.195]{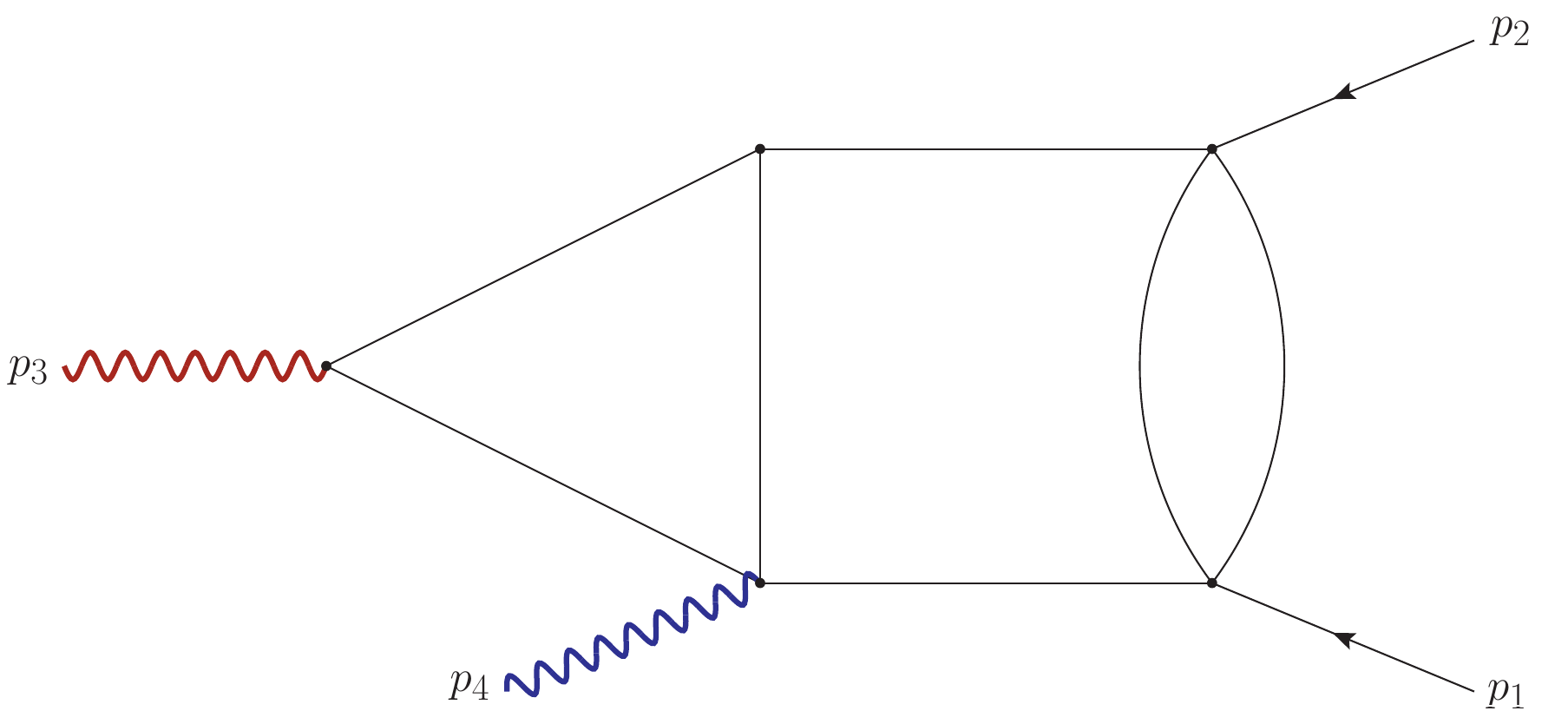}\\
\begin{equation*}
\{ I_{61}^{\text{PL1}}, \, I_{62}^{\text{PL1}} \}
\end{equation*}
\end{multicols}

\textbf{Sector $\mathbf{F_{123}}$[0,0,1,1,1,0,0,0,0,1,1,0,0,1,1]}

\begin{multicols}{2}
\includegraphics[scale=0.195]{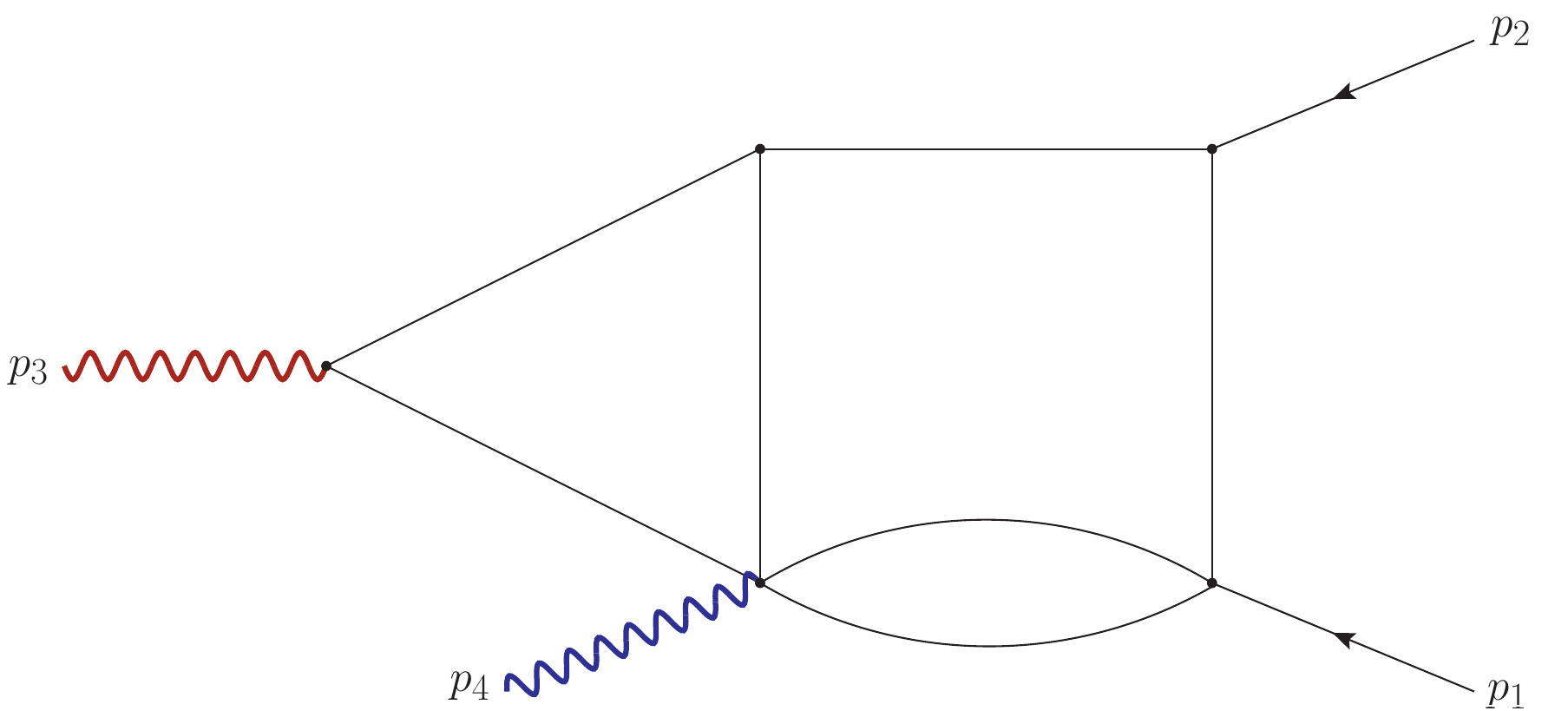}\\
\begin{equation*}
\{ I_{95}^{\text{PT4}} \}
\end{equation*}
\end{multicols}

\textbf{Sector $\mathbf{F_{123}}$[0,0,1,1,1,1,0,0,0,1,0,0,1,1,0]}

\begin{multicols}{2}
\includegraphics[scale=0.195]{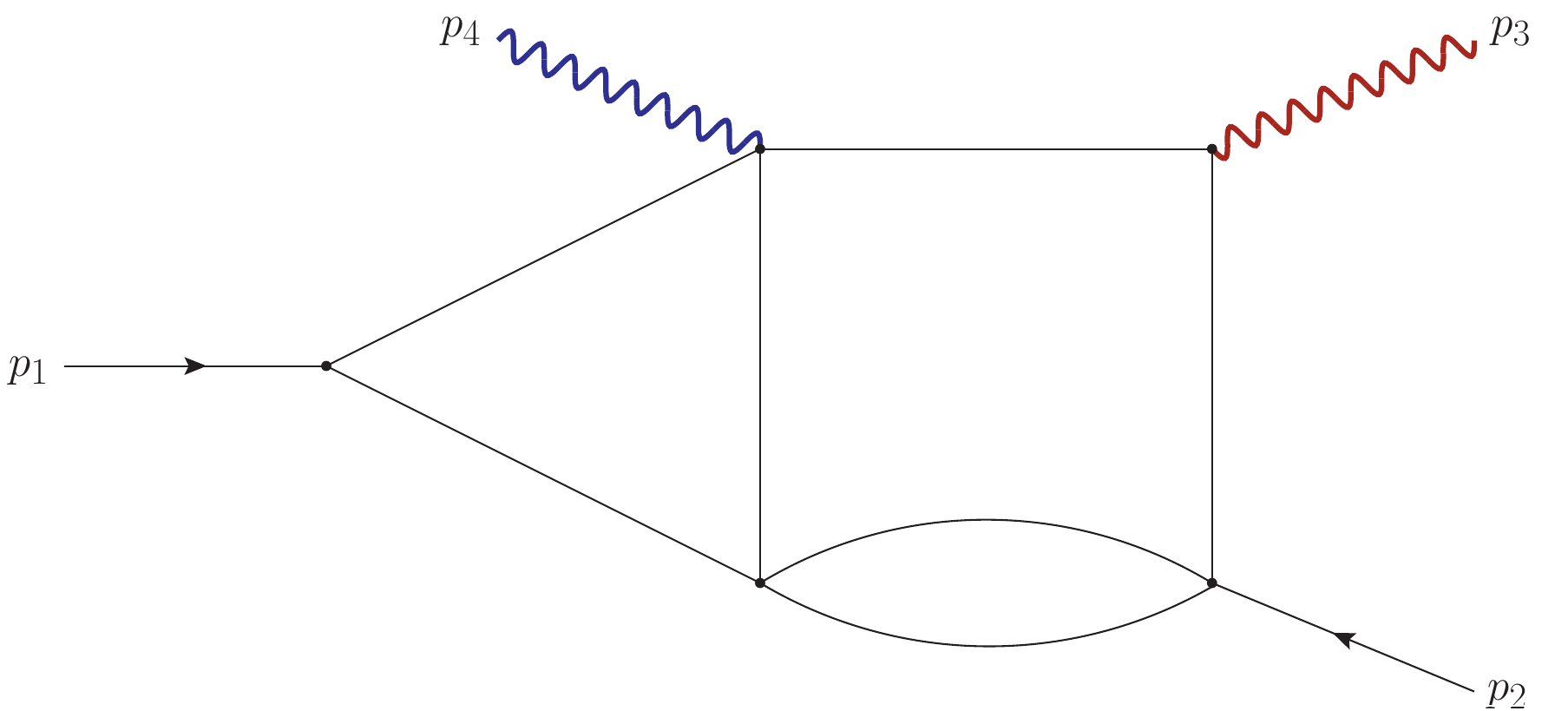}\\
\begin{equation*}
\{ I_{87}^{\text{PT4}}, \, I_{88}^{\text{PT4}} \}
\end{equation*}
\end{multicols}

\textbf{Sector $\mathbf{F_{123}}$[0,0,1,1,1,1,0,0,0,0,1,0,1,0,1]}

\begin{multicols}{2}
\includegraphics[scale=0.195]{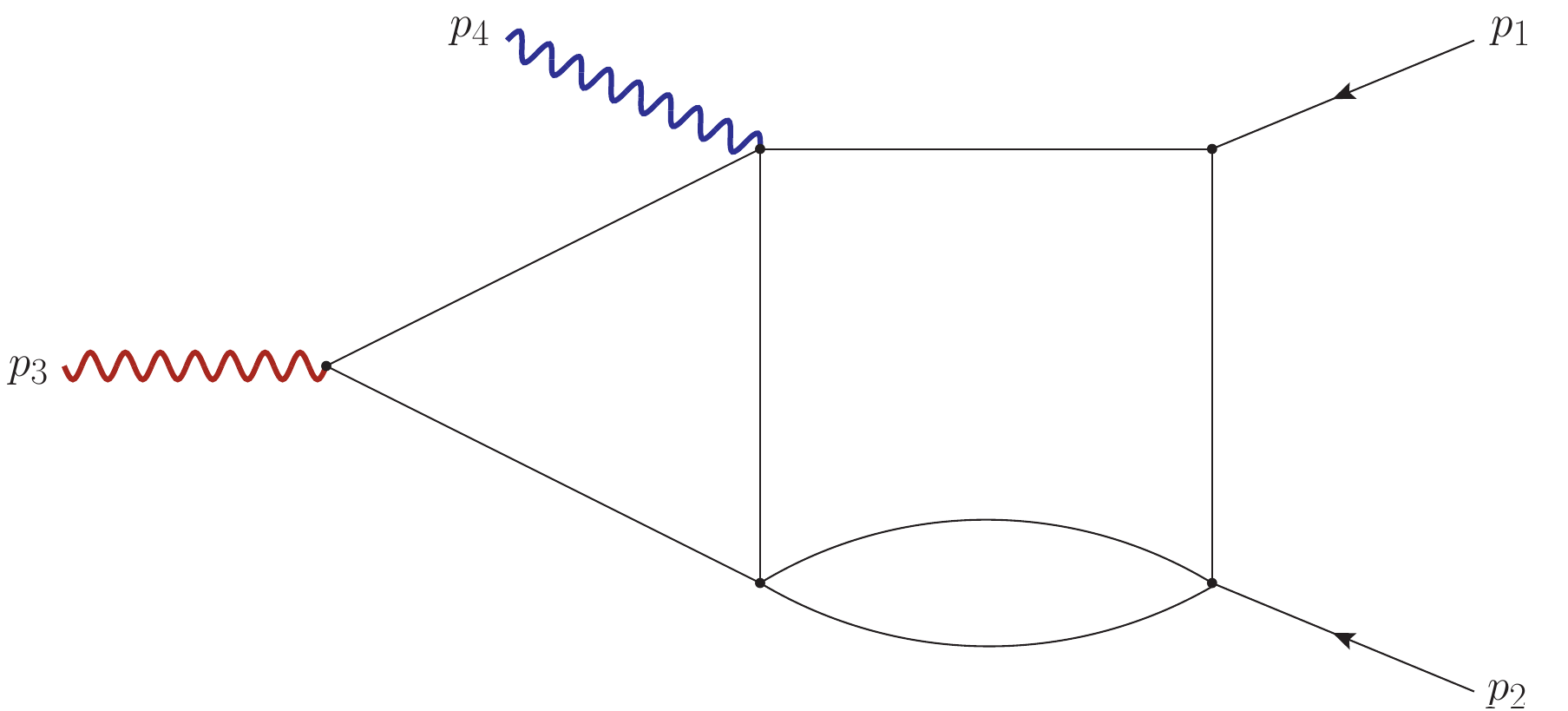}\\
\begin{equation*}
\{ I_{74}^{\text{PT4}}, \, I_{75}^{\text{PT4}}, \, I_{76}^{\text{PT4}} \}
\end{equation*}
\end{multicols}

\textbf{Sector $\mathbf{F_{123}}$[1,0,1,0,0,0,1,0,1,0,0,1,1,0,1]}

\begin{multicols}{2}
\includegraphics[scale=0.195]{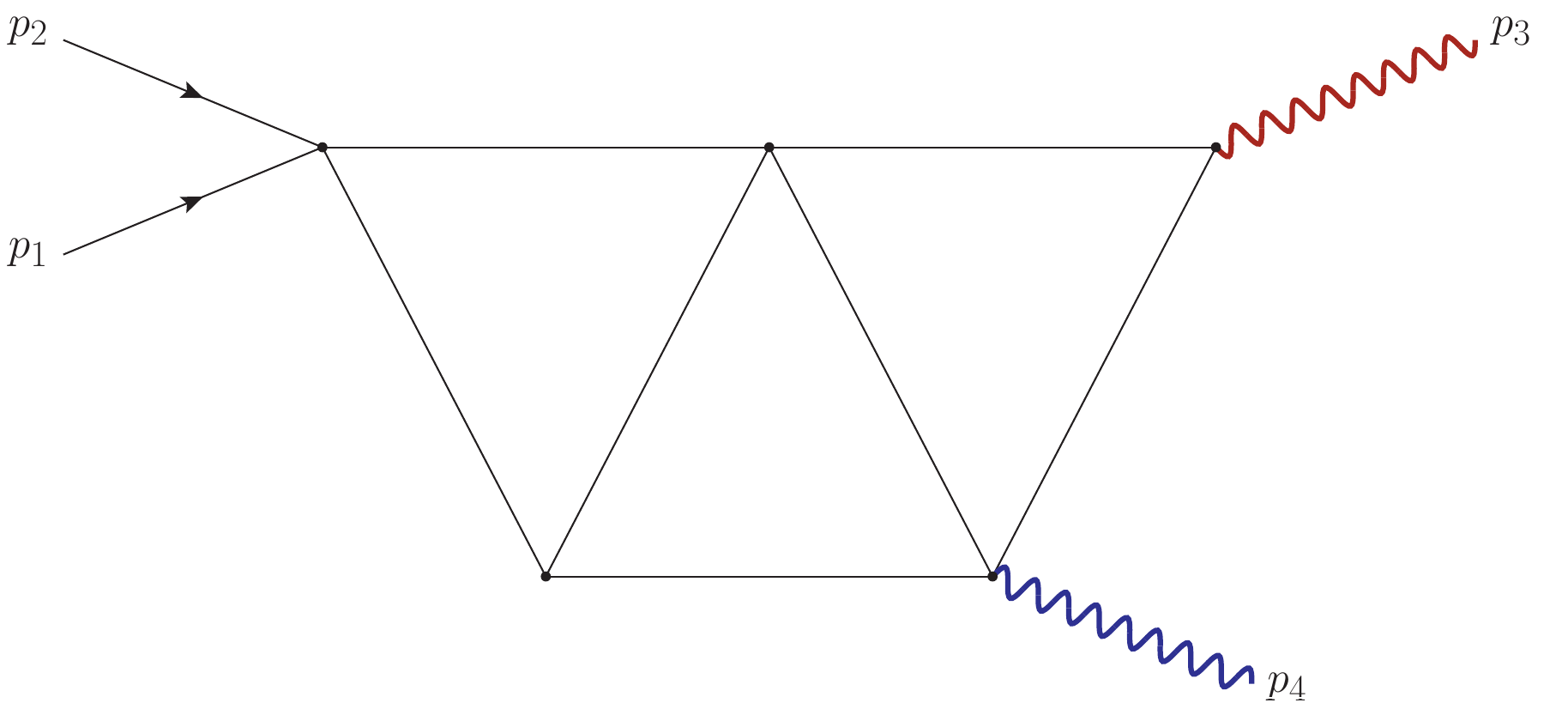}\\
\begin{equation*}
\{ I_{52}^{\text{PL1}}, \, I_{53}^{\text{PL1}}, \, I_{54}^{\text{PL1}} \}
\end{equation*}
\end{multicols}

\textbf{Sector $\mathbf{F_{123}}$[0,1,1,0,1,0,0,0,1,0,0,1,1,0,1]}

\begin{multicols}{2}
\includegraphics[scale=0.195]{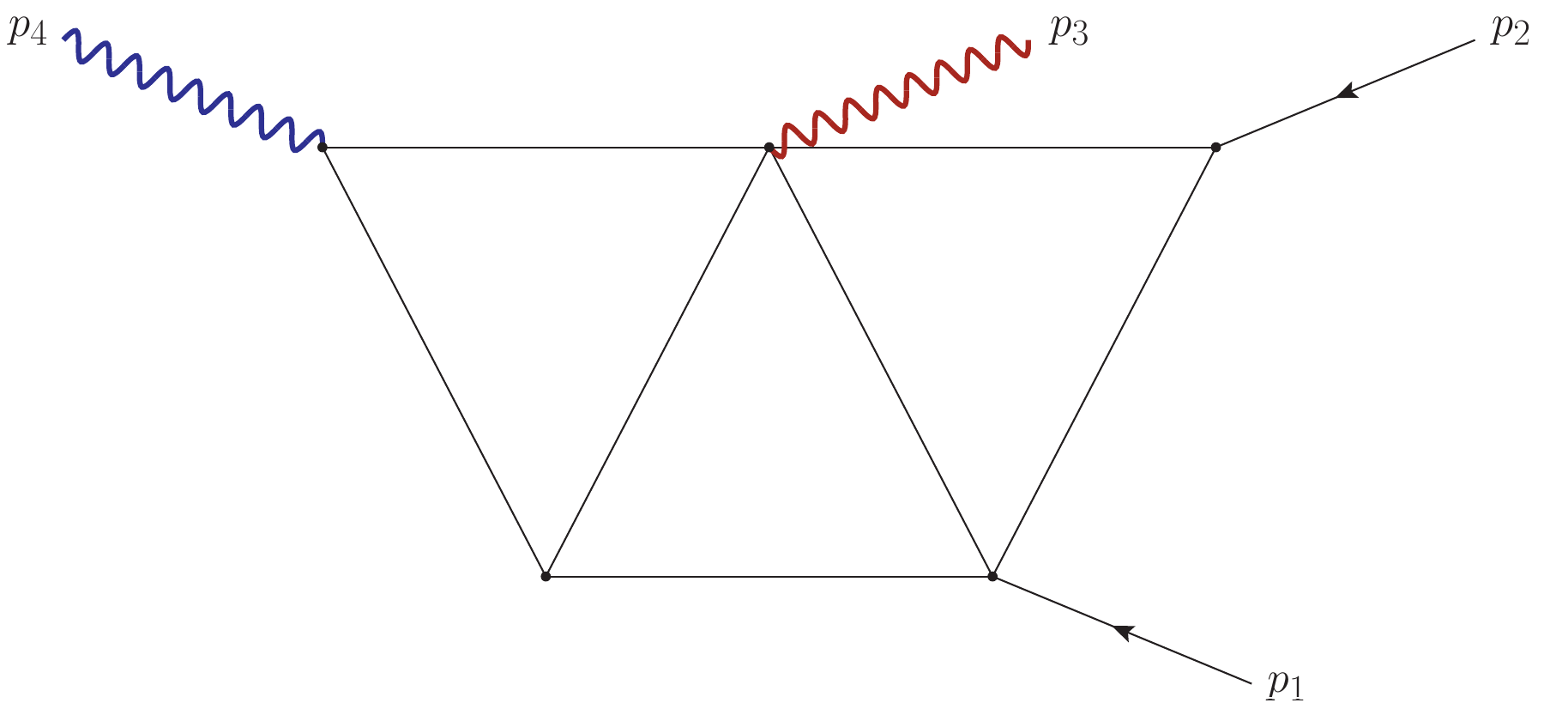}\\
\begin{equation*}
\{ I_{57}^{\text{PL1}}, \, I_{58}^{\text{PL1}} \}
\end{equation*}
\end{multicols}

\textbf{Sector $\mathbf{F_{123}}$[1,1,0,0,0,0,1,0,1,0,0,1,1,0,1]}

\begin{multicols}{2}
\includegraphics[scale=0.195]{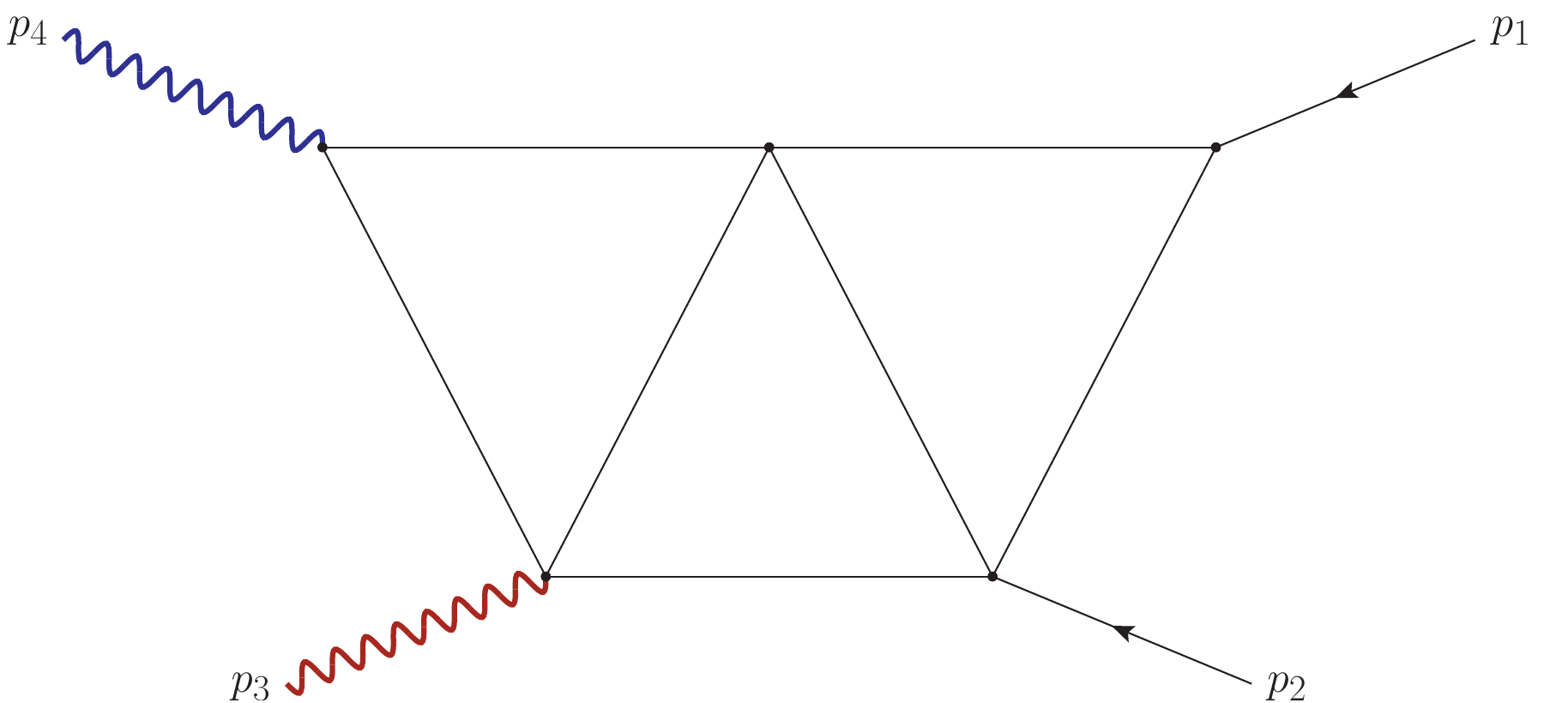}\\
\newline
\begin{equation*}
\{ I_{37}^{\text{PL1}}, \, I_{38}^{\text{PL1}}, \, I_{39}^{\text{PL1}}, \, I_{40}^{\text{PL1}}, \, I_{41}^{\text{PL1}}, \, I_{42}^{\text{PL1}}\}
\end{equation*}
\end{multicols}

\textbf{Sector $\mathbf{F_{123}}$[0,0,0,1,1,1,0,0,0,1,1,0,1,1,0]}

\begin{multicols}{2}
\includegraphics[scale=0.195]{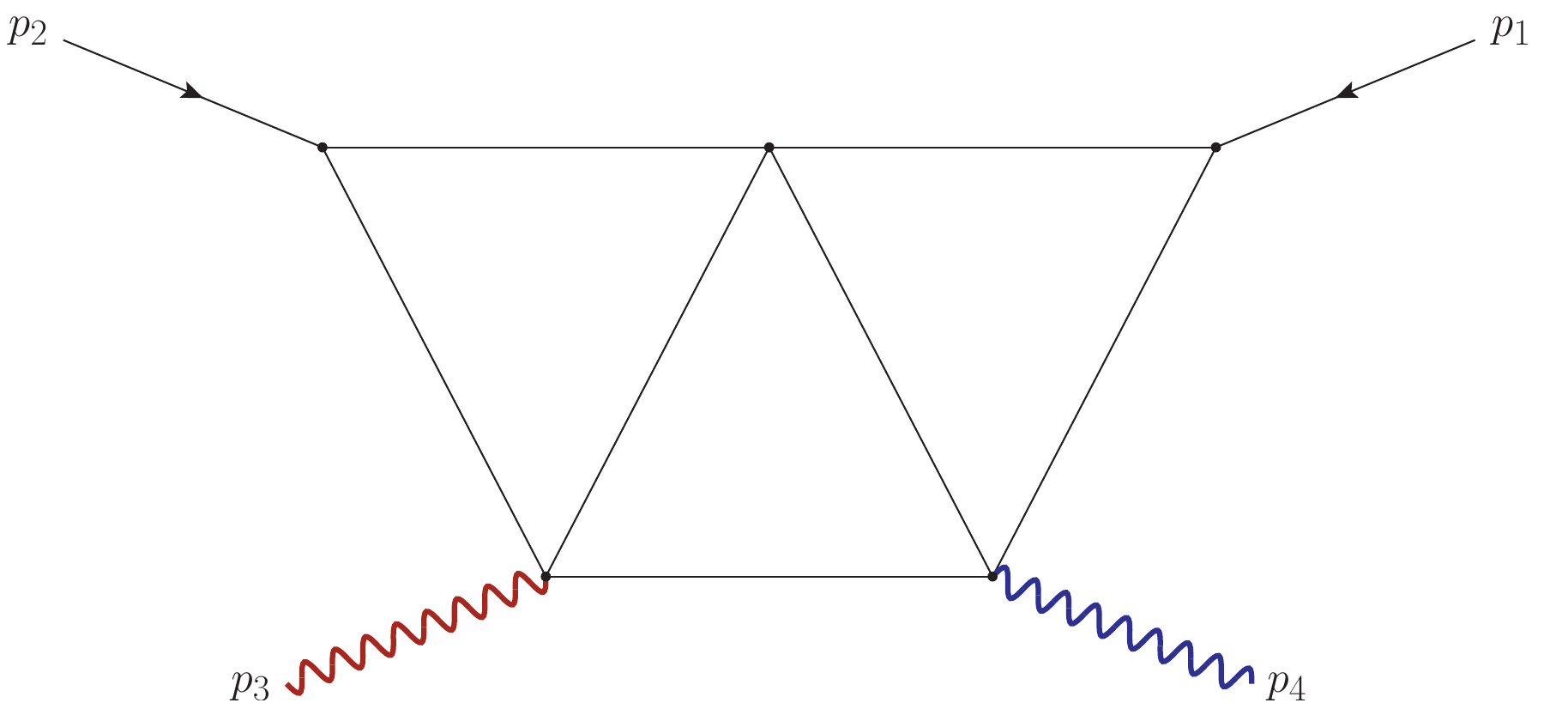}\\
\begin{equation*}
\{ I_{91}^{\text{PT4}}, \, I_{92}^{\text{PT4}} \}
\end{equation*}
\end{multicols}

\textbf{Sector $\mathbf{F_{123}}$[0,0,0,1,1,0,0,0,0,1,1,0,1,1,1]}

\begin{multicols}{2}
\includegraphics[scale=0.195]{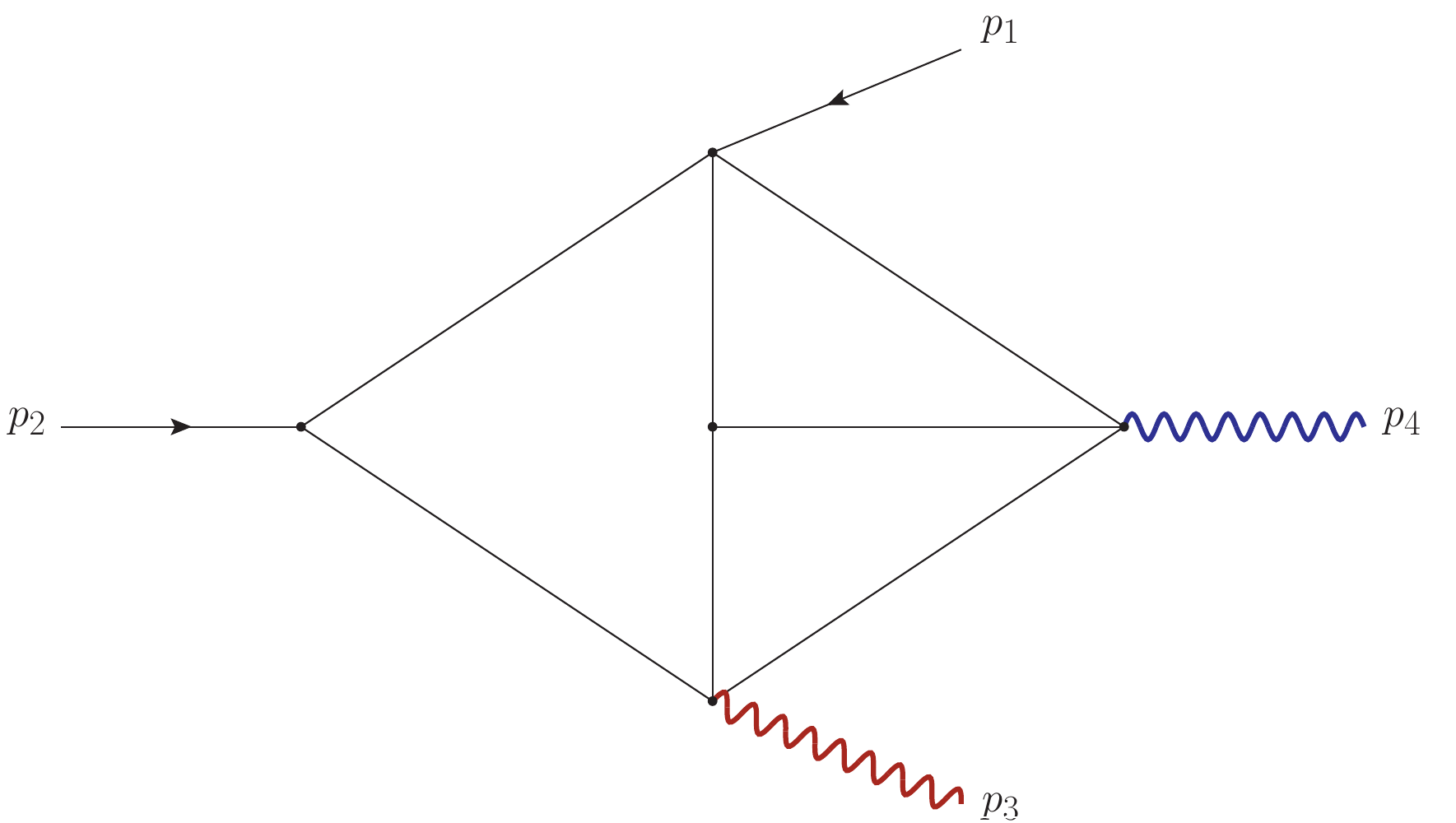}\\
\begin{equation*}
\{ I_{67}^{\text{PT4}}, \, I_{68}^{\text{PT4}}, \, I_{69}^{\text{PT4}}, \, I_{70}^{\text{PT4}} \}
\end{equation*}
\end{multicols}

\textbf{Sector $\mathbf{F_{123}}$[0,0,1,1,1,0,0,0,0,1,0,0,1,1,1]}

\begin{multicols}{2}
\includegraphics[scale=0.195]{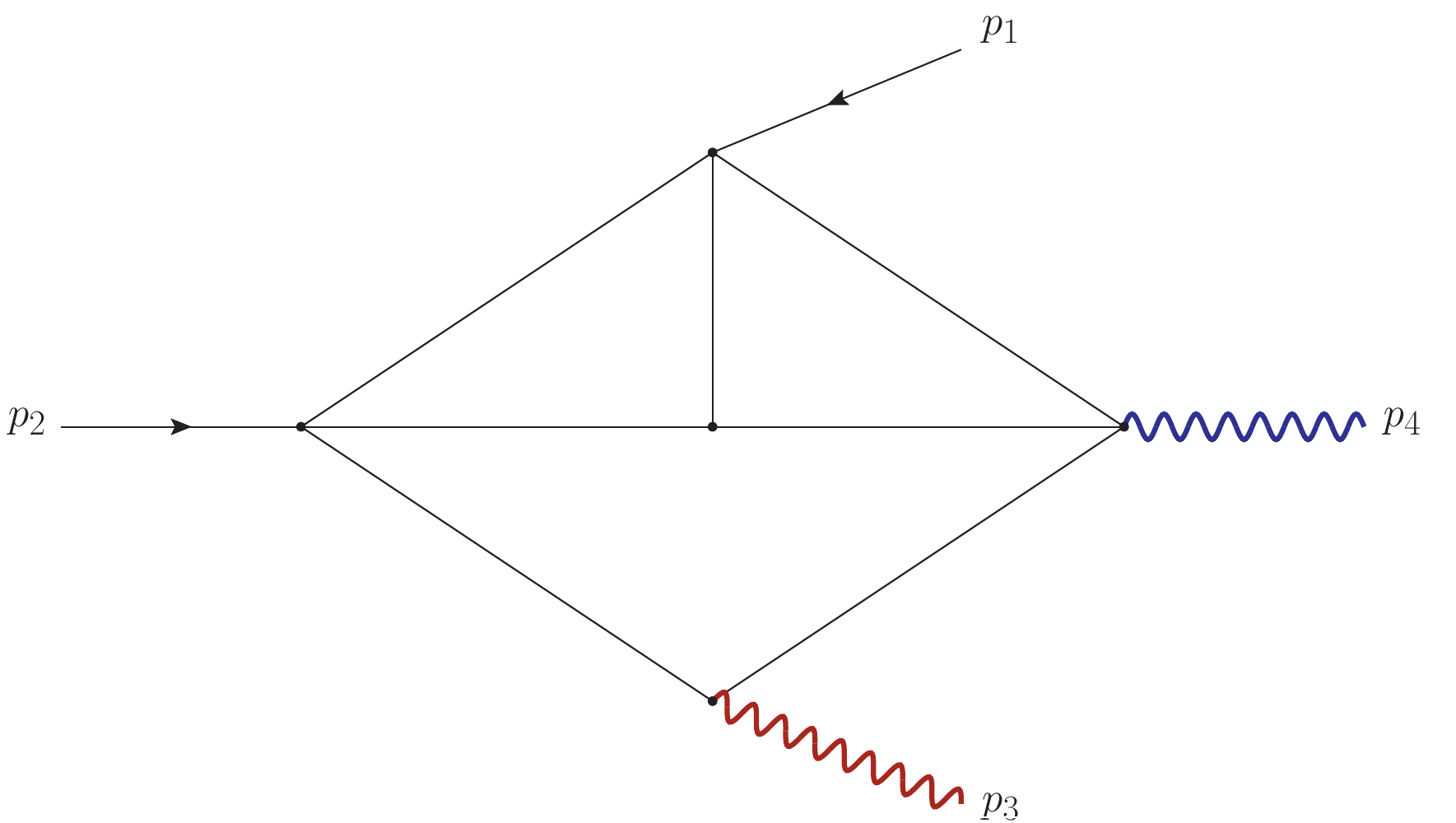}\\
\begin{equation*}
\begin{split}
\{ I_{58}^{\text{PT4}}, \, I_{59}^{\text{PT4}}, \, I_{60}^{\text{PT4}}, \, I_{61}^{\text{PT4}}, \, I_{62}^{\text{PT4}} \}
\end{split}
\end{equation*}
\end{multicols}

\textbf{Sector $\mathbf{F_{123}}$[1,0,1,1,0,1,0,0,0,1,0,0,1,1,0]}

\begin{multicols}{2}
\includegraphics[scale=0.195]{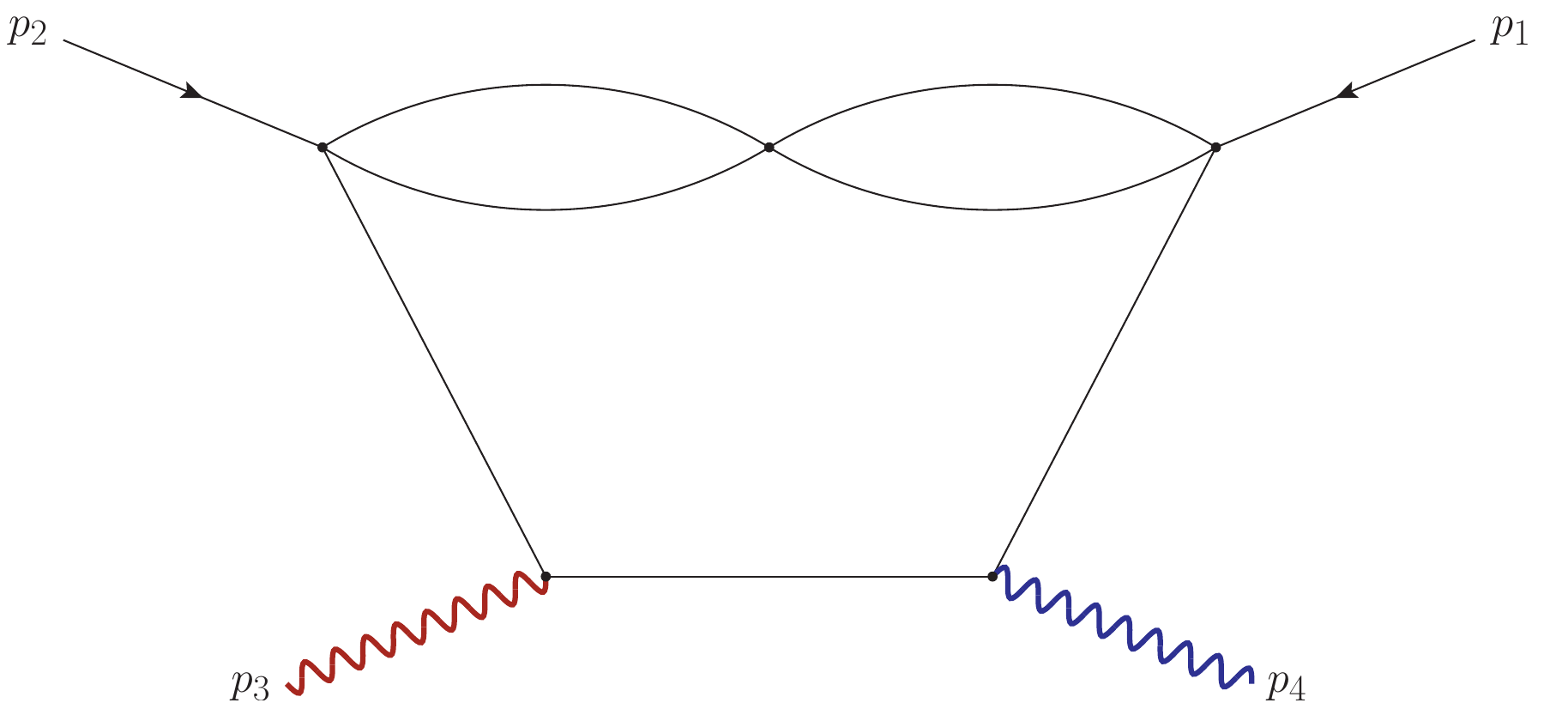}\\
\begin{equation*}
\begin{split}
\{ &I_{77}^{\text{PT4}}, \, I_{78}^{\text{PT4}} \}
\end{split}
\end{equation*}
\end{multicols}



\begin{center}
\textbf{\textit{Eight-Propagator Pure Candidates}}\\
\end{center}
\vspace{0.2cm}

\textbf{Sector $\mathbf{F_{123}}$[0,1,0,0,1,0,1,0,1,0,1,1,1,0,1]}

\begin{multicols}{2}
\includegraphics[scale=0.195]{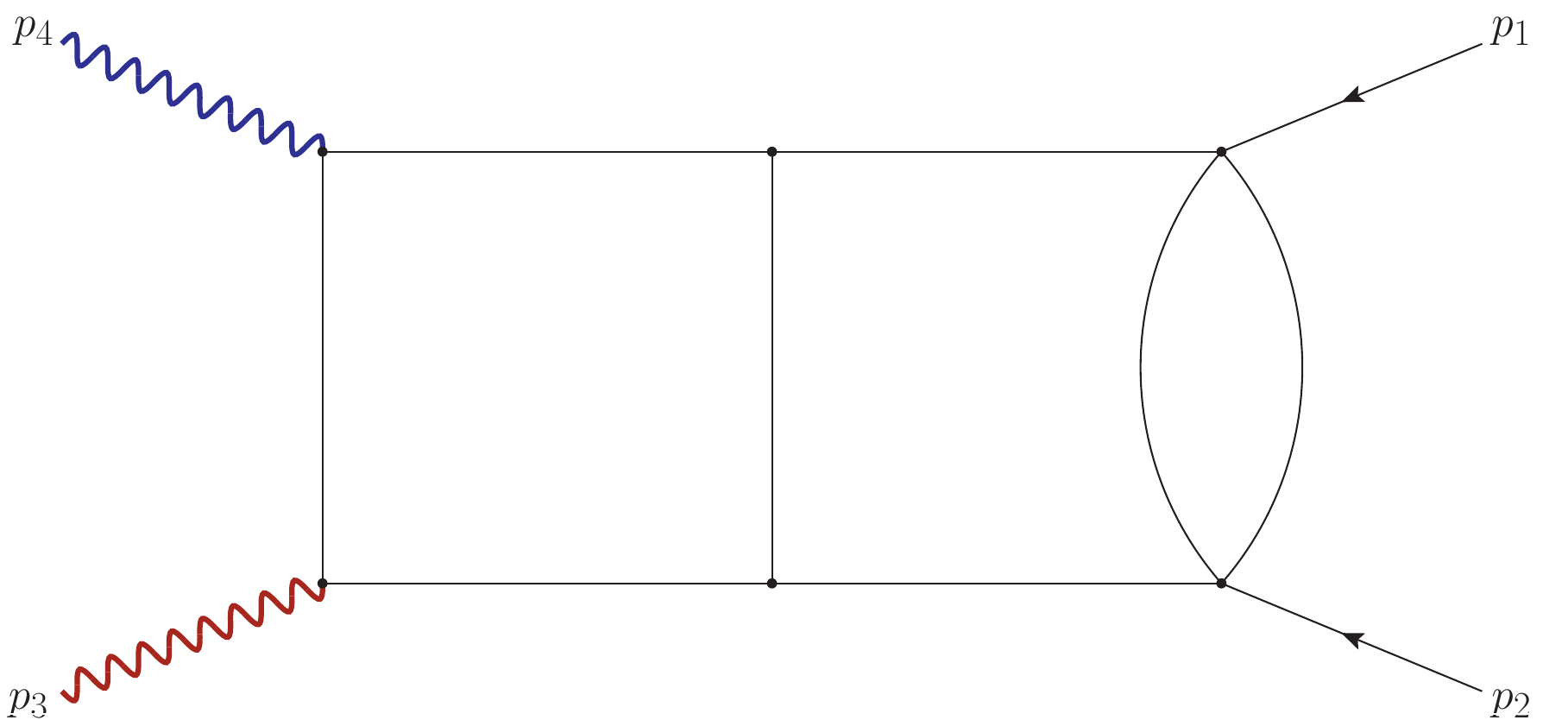}\\
\begin{equation*}
\{ I_{34}^{\text{PL1}}, \, I_{35}^{\text{PL1}}, \, I_{36}^{\text{PL1}} \}
\end{equation*}
\end{multicols}

\textbf{Sector $\mathbf{F_{123}}$[1,1,1,0,1,0,1,0,0,0,0,1,1,0,1]}

\begin{multicols}{2}
\includegraphics[scale=0.195]{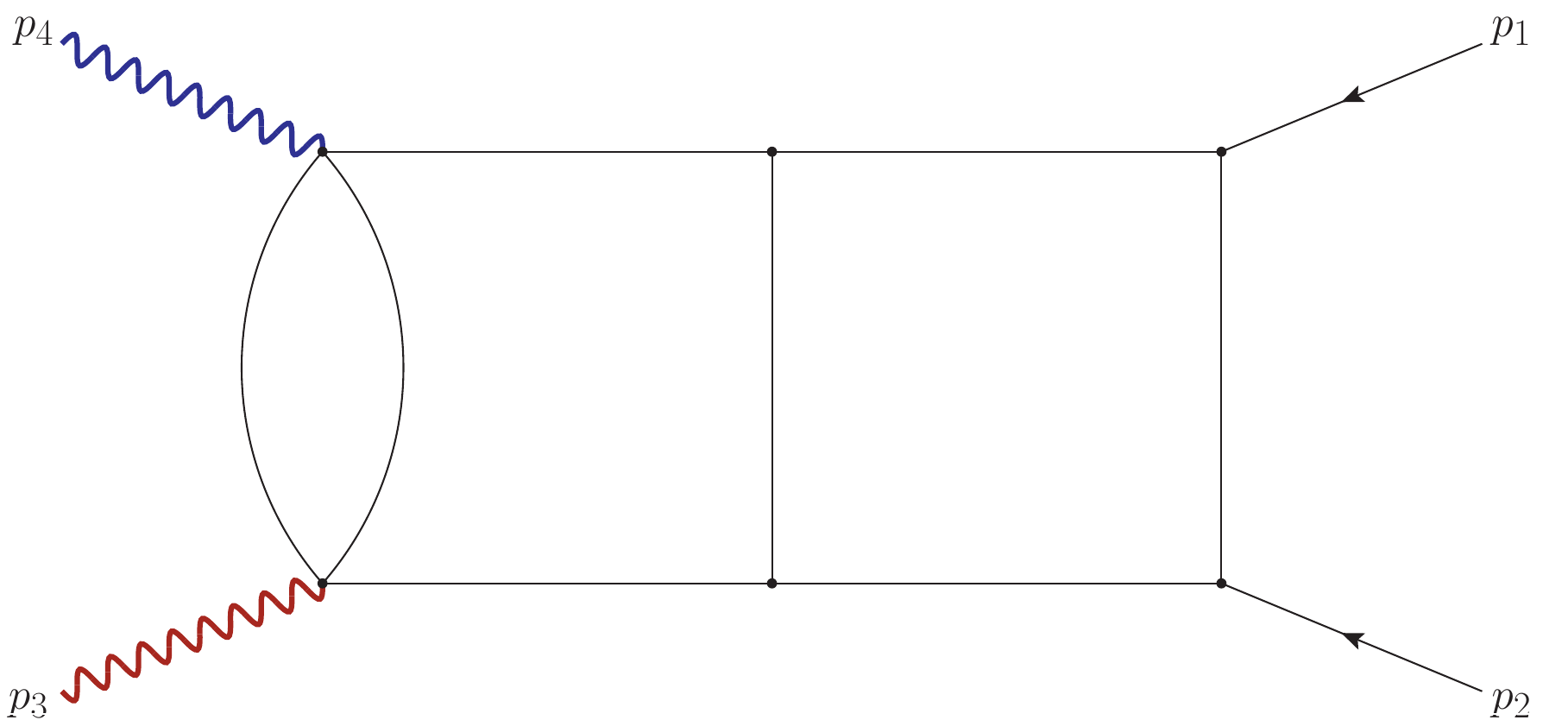}\\
\begin{equation*}
\{ I_{22}^{\text{PL1}}, \, I_{23}^{\text{PL1}}, \, I_{24}^{\text{PL1}} \}
\end{equation*}
\end{multicols}

\textbf{Sector $\mathbf{F_{123}}$[1,0,1,1,1,0,0,0,0,1,1,0,0,1,1]}

\begin{multicols}{2}
\includegraphics[scale=0.195]{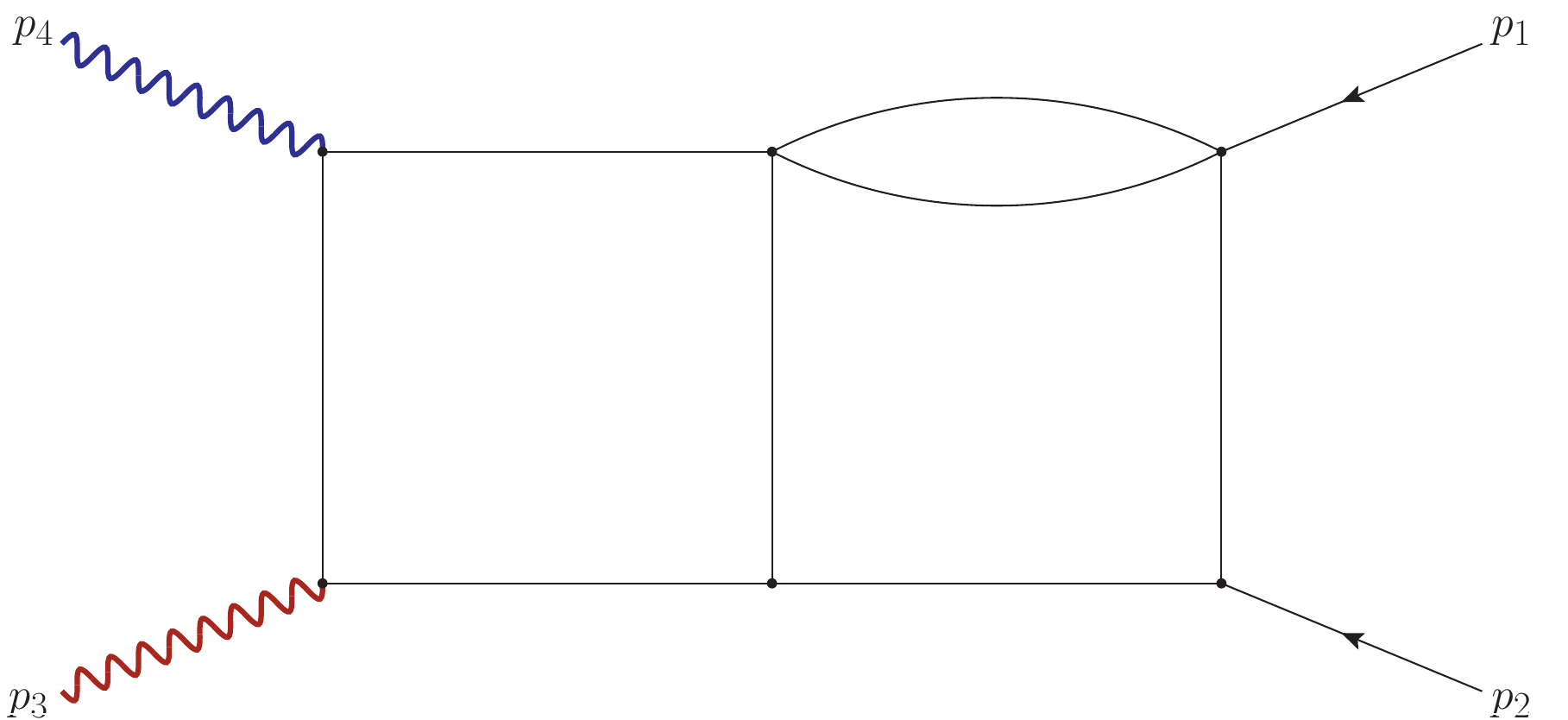}\\
\begin{equation*}
\{I_{36}^{\text{PT4}}, \, I_{37}^{\text{PT4}}, \, I_{38}^{\text{PT4}} \}
\end{equation*}
\end{multicols}

\textbf{Sector $\mathbf{F_{123}}$[1,1,1,0,0,0,0,0,1,0,1,1,1,0,1]}

\begin{multicols}{2}
\includegraphics[scale=0.195]{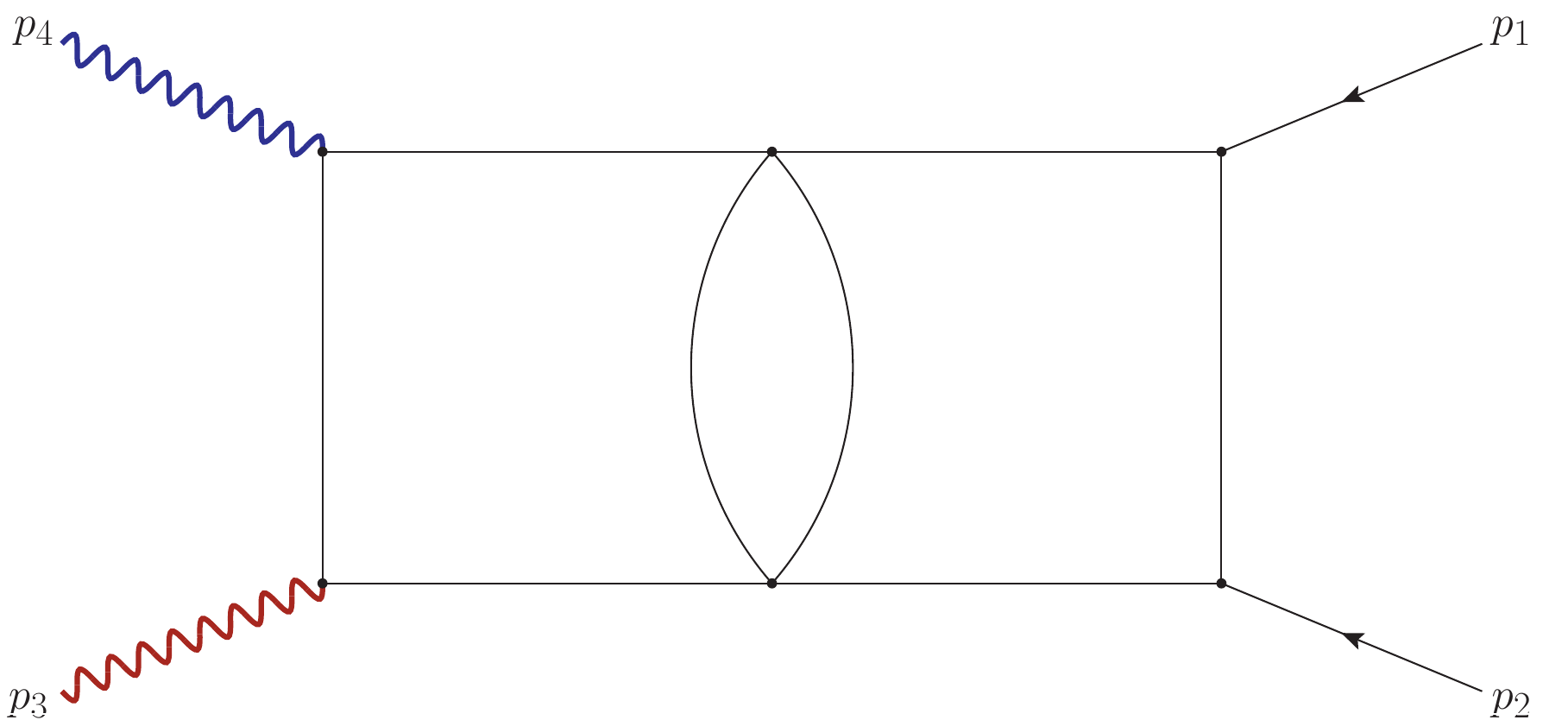}\\
\begin{equation*}
\{ I_{25}^{\text{PL1}}, \, I_{26}^{\text{PL1}}, \, I_{27}^{\text{PL1}} \}
\end{equation*}
\end{multicols}

\textbf{Sector $\mathbf{F_{123}}$[1,1,0,0,0,0,1,0,1,0,1,1,1,0,1]}

\begin{multicols}{2}
\includegraphics[scale=0.195]{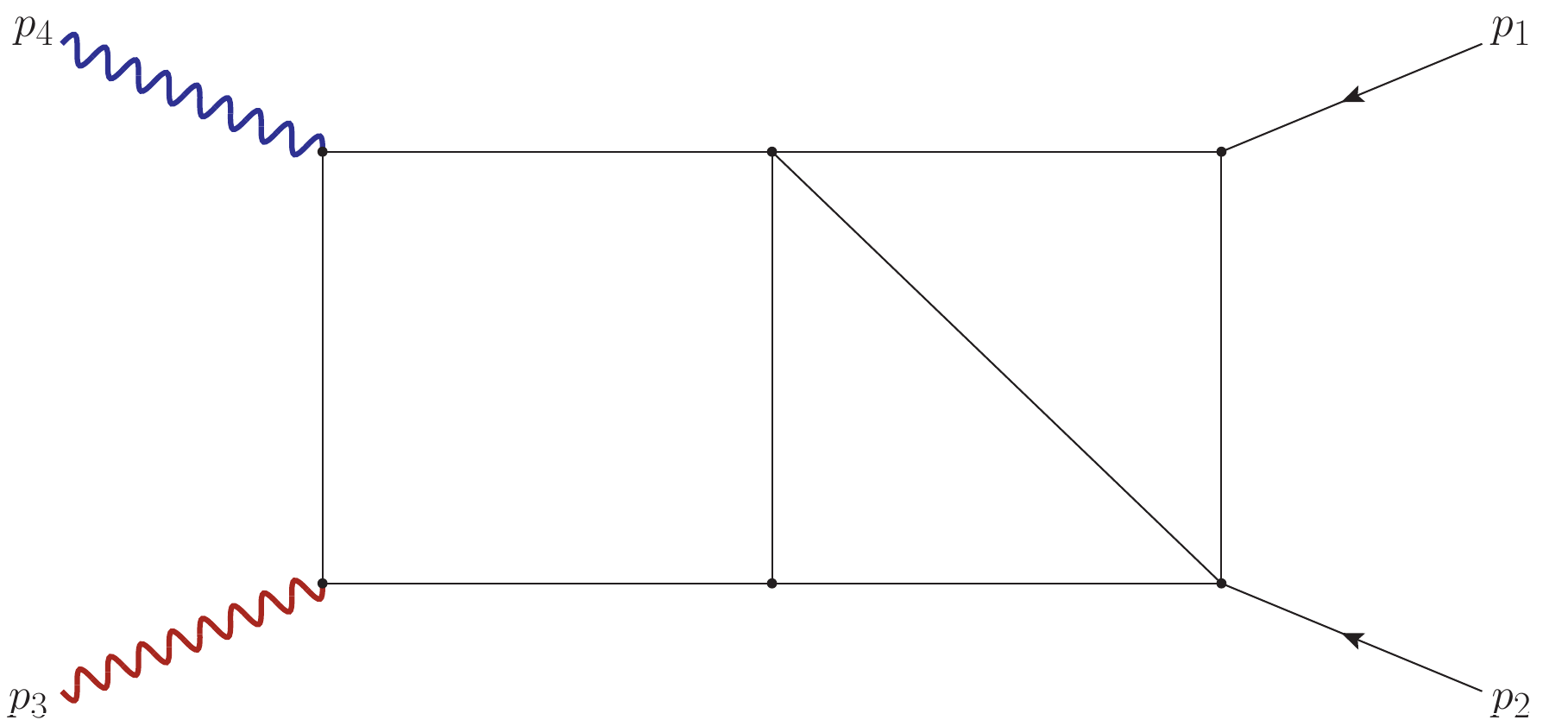}\\
\begin{equation*}
\{ I_{28}^{\text{PL1}}, \, I_{29}^{\text{PL1}}, \, I_{30}^{\text{PL1}} \}
\end{equation*}
\end{multicols}

\textbf{Sector $\mathbf{F_{123}}$[1,1,1,0,0,0,1,0,1,0,0,1,1,0,1]}

\begin{multicols}{2}
\includegraphics[scale=0.195]{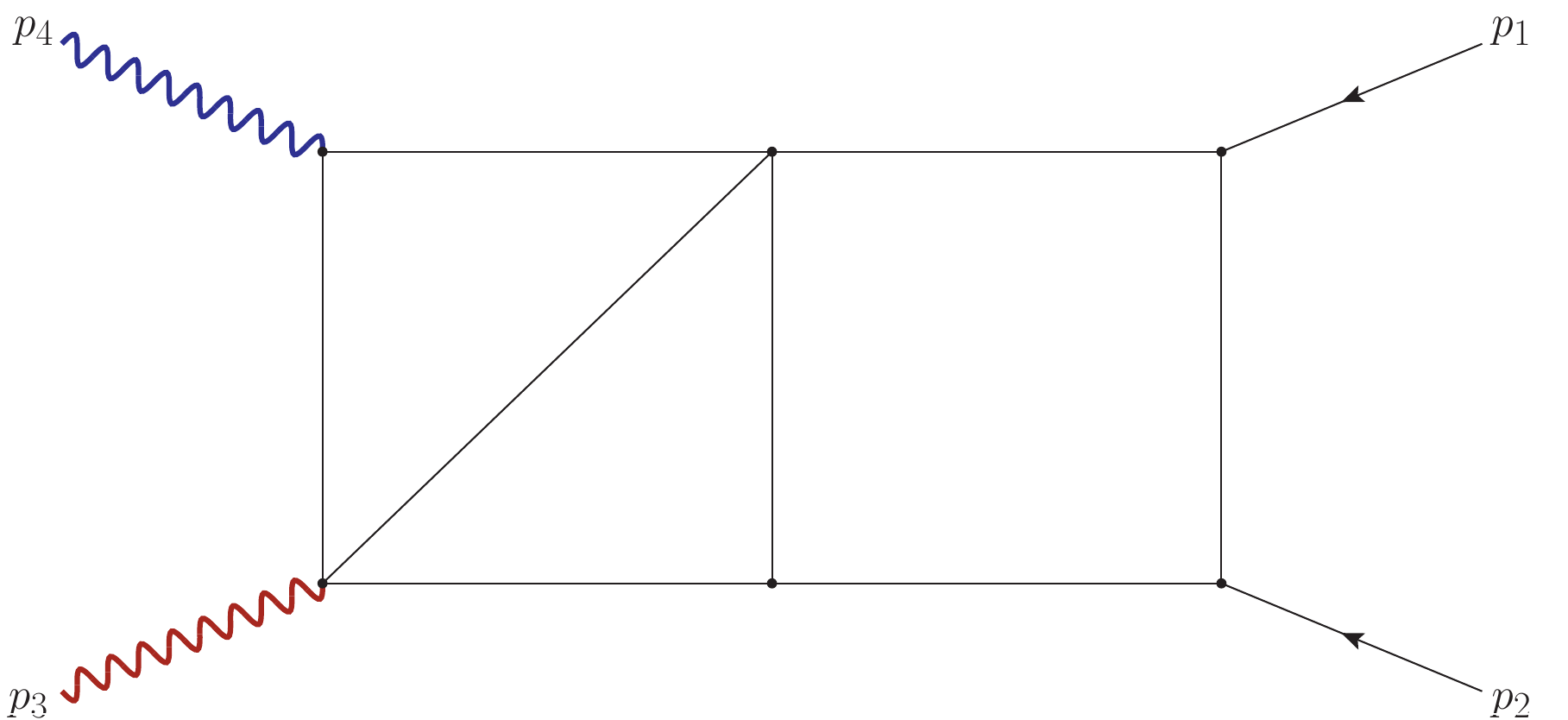}\\
\begin{equation*}
\{ I_{17}^{\text{PL1}}, \, I_{18}^{\text{PL1}}, \, I_{19}^{\text{PL1}}, \, I_{20}^{\text{PL1}}, \, I_{21}^{\text{PL1}} \}
\end{equation*}
\end{multicols}

\textbf{Sector $\mathbf{F_{123}}$[0,0,0,1,1,1,0,0,0,1,1,0,1,1,1]}

\begin{multicols}{2}
\includegraphics[scale=0.175]{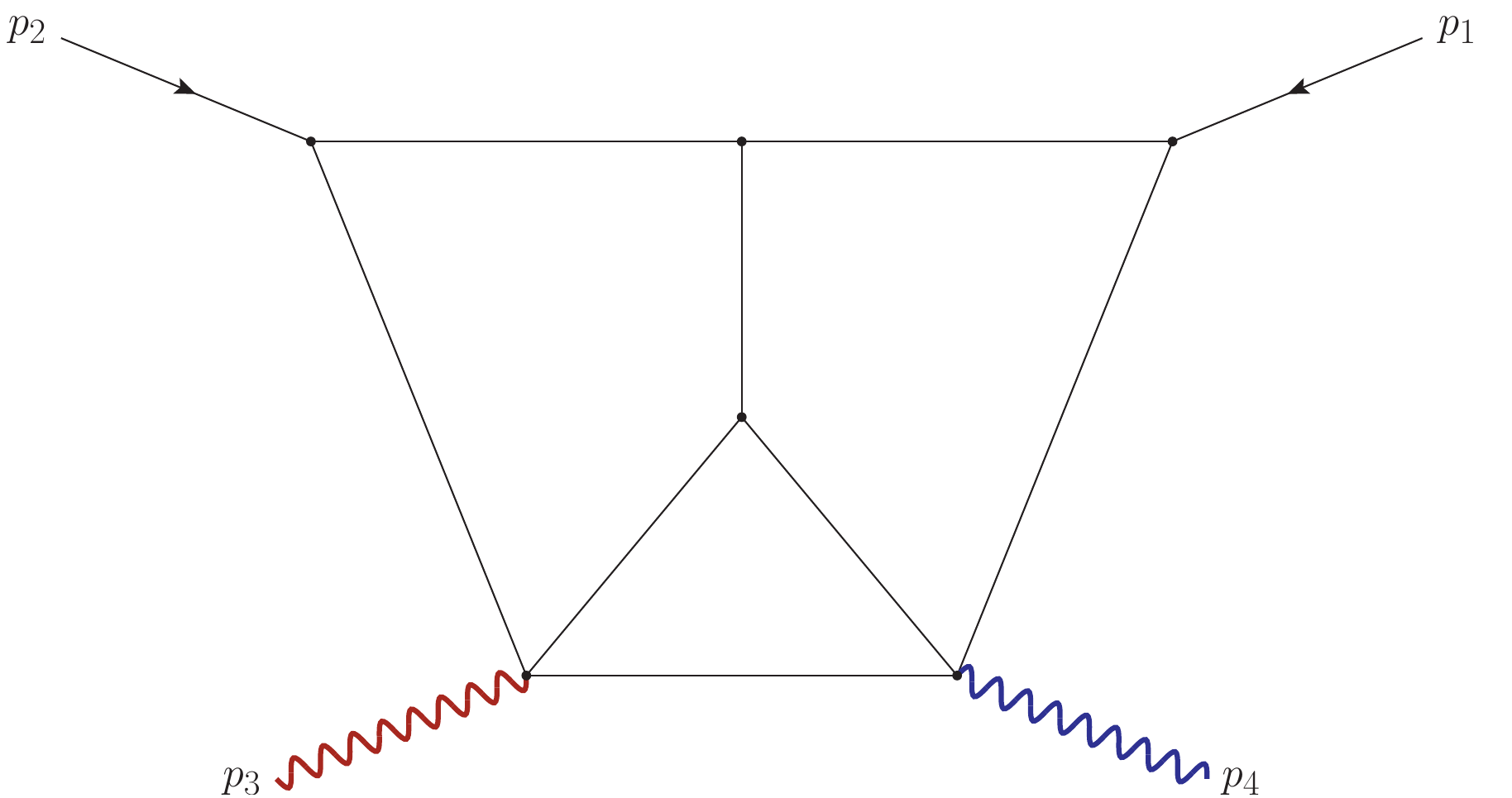}\\
\newline
\begin{equation*}
\{ I_{39}^{\text{PT4}}, \, I_{40}^{\text{PT4}} \}
\end{equation*}
\end{multicols}

\textbf{Sector $\mathbf{F_{123}}$[0,0,1,1,1,0,0,0,0,1,1,0,1,1,1]}

\begin{multicols}{2}
\includegraphics[scale=0.175]{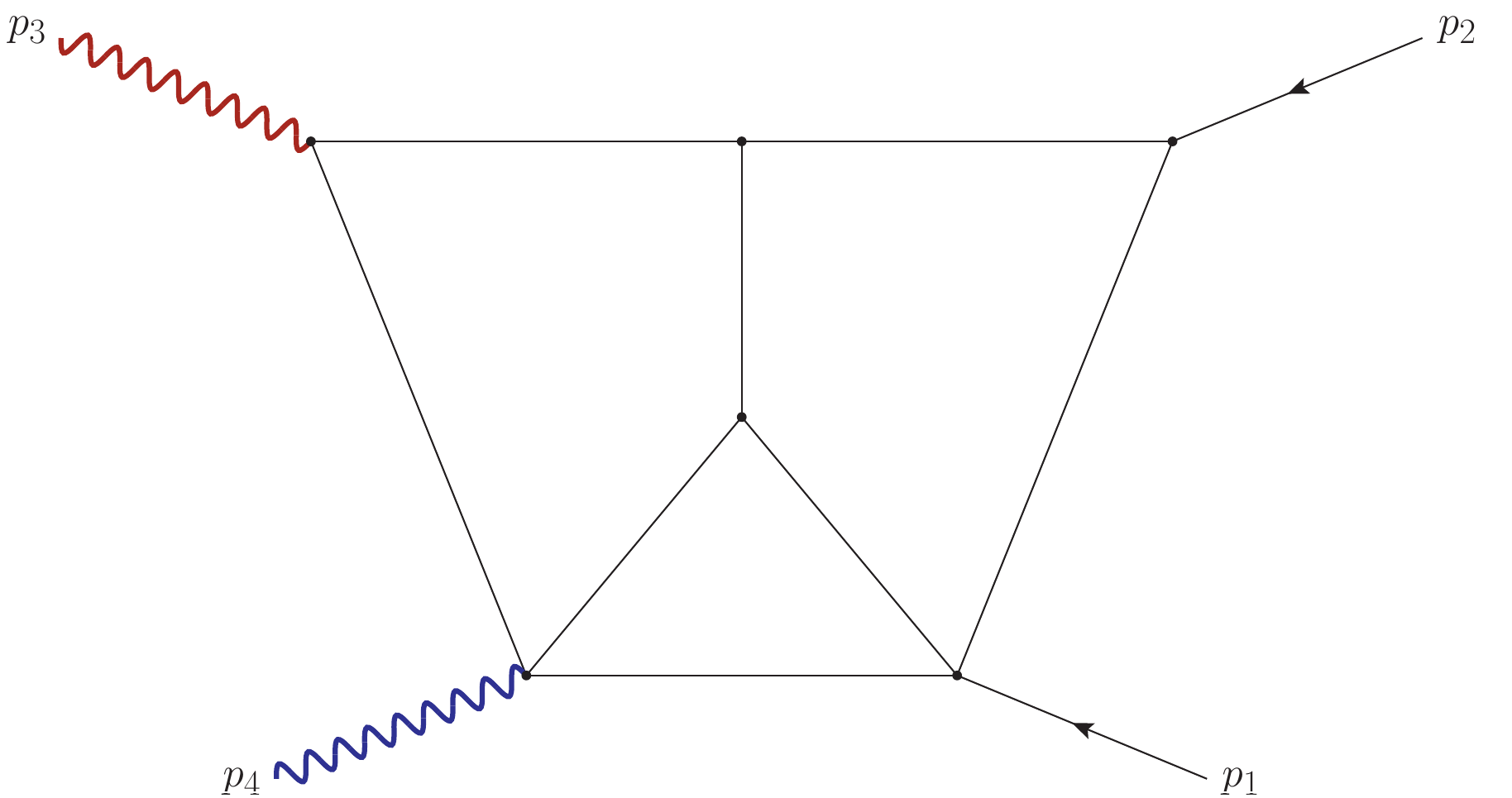}\\
\newline
\begin{equation*}
\{I_{21}^{\text{PT4}}, \, I_{22}^{\text{PT4}},  \, I_{23}^{\text{PT4}}, \, I_{24}^{\text{PT4}}, \, I_{25}^{\text{PT4}},  \, I_{26}^{\text{PT4}}\}
\end{equation*}
\end{multicols}

\textbf{Sector $\mathbf{F_{123}}$[1,1,1,1,0,0,0,0,1,0,0,1,1,0,1]}

\begin{multicols}{2}
\includegraphics[scale=0.185]{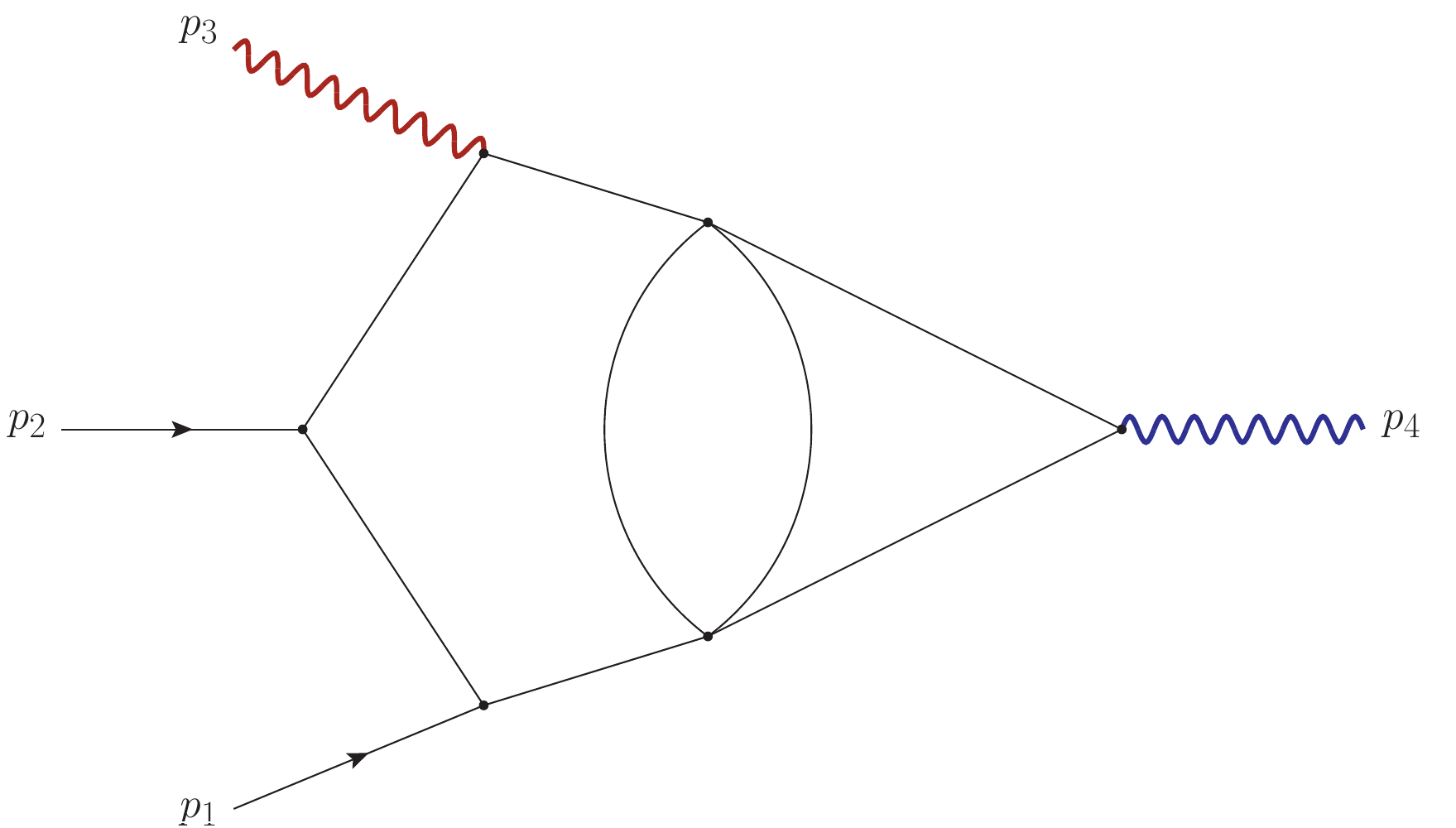}\\
\newline
\begin{equation*}
\{ I_{1}^{\text{RL1}}\}
\end{equation*}
\end{multicols}

\textbf{Sector $\mathbf{F_{132}}$[1,1,1,1,0,0,0,0,1,0,0,1,1,0,1]}

\begin{multicols}{2}
\includegraphics[scale=0.185]{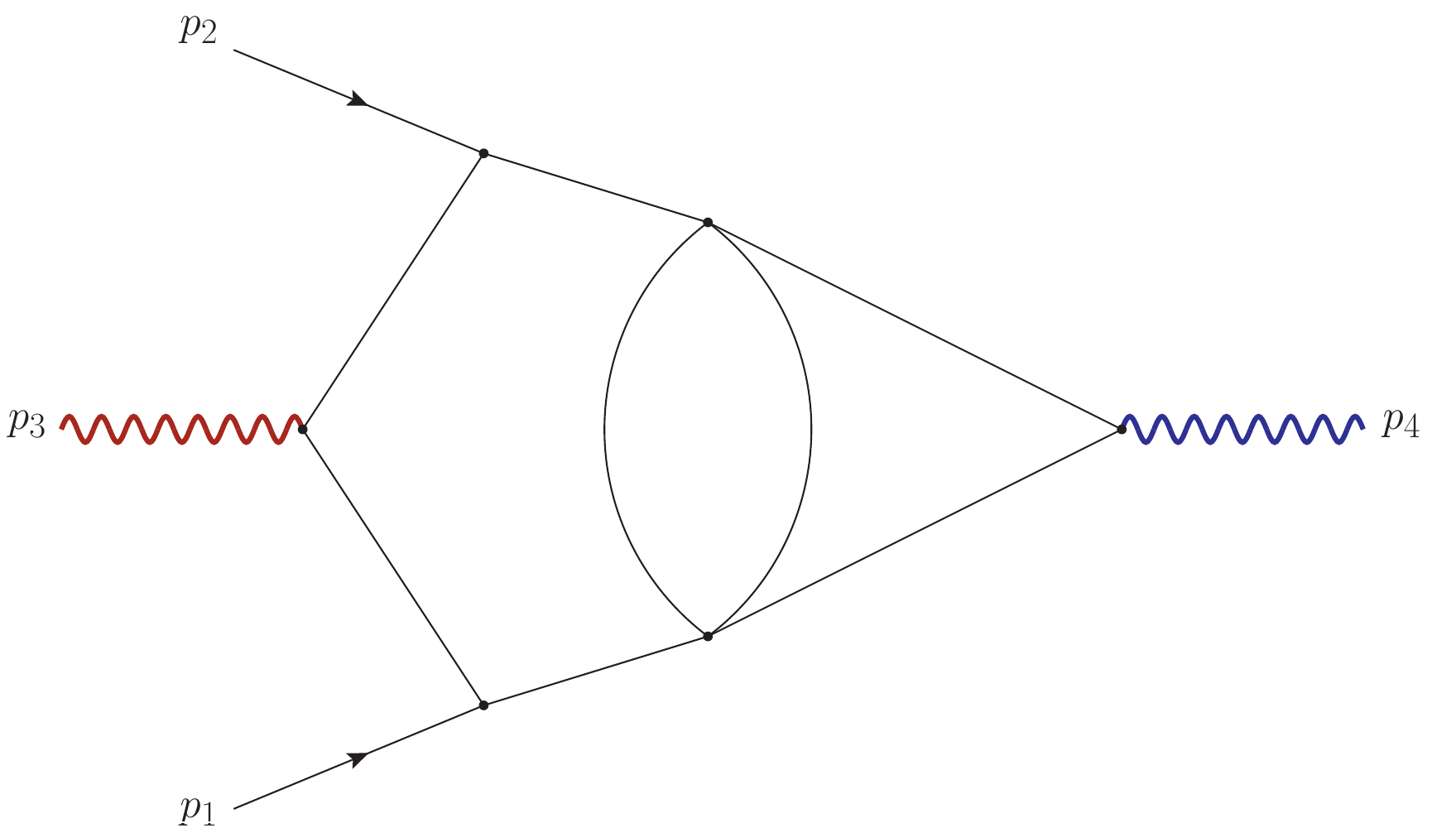}\\
\newline
\begin{equation*}
\{ I_{1}^{\text{RL2}}\}
\end{equation*}
\end{multicols}



\begin{center}
\textbf{\textit{Nine-Propagator Pure Candidates}}\\
\end{center}
\vspace{0.2cm}

\textbf{Sector $\mathbf{F_{123}}$[1,1,1,0,1,0,1,0,1,0,0,1,1,0,1]}

\begin{multicols}{2}
\includegraphics[scale=0.155]{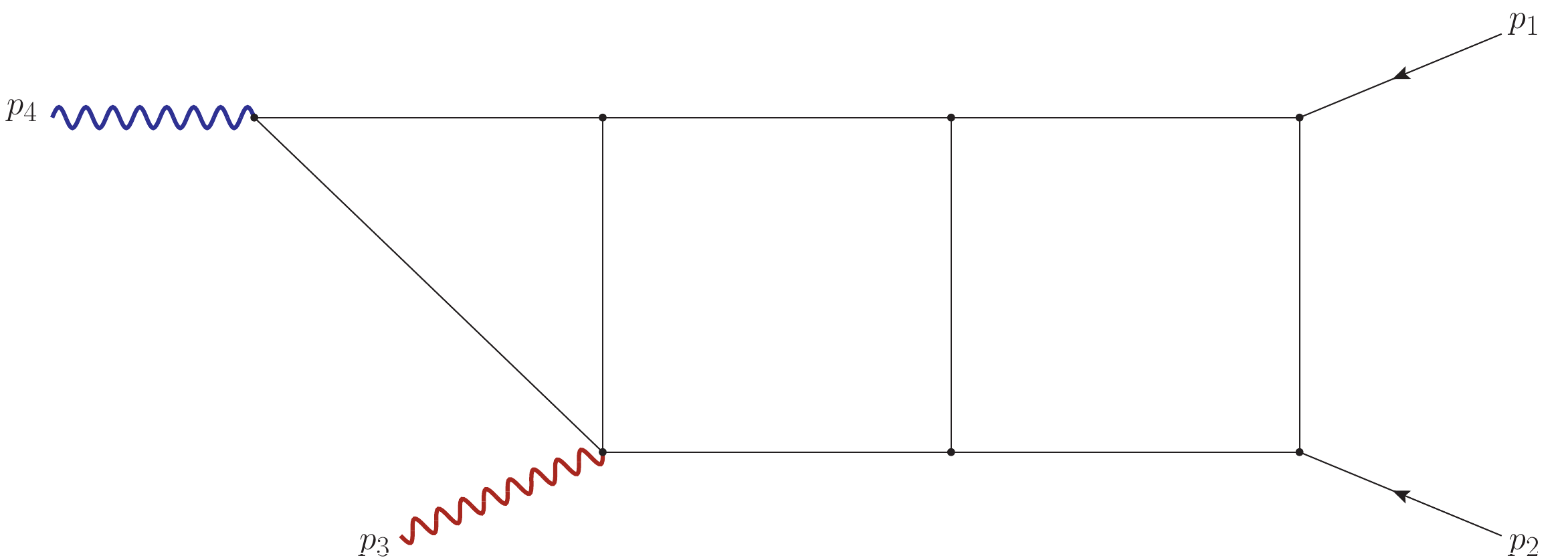}\\
\begin{equation*}
\{ I_{6}^{\text{PL1}}, \, I_{7}^{\text{PL1}}, \, I_{8}^{\text{PL1}}\}
\end{equation*}
\end{multicols}

\textbf{Sector $\mathbf{F_{123}}$[0,0,1,1,1,1,0,0,0,1,1,0,1,1,1]}

\begin{multicols}{2}
\includegraphics[scale=0.140]{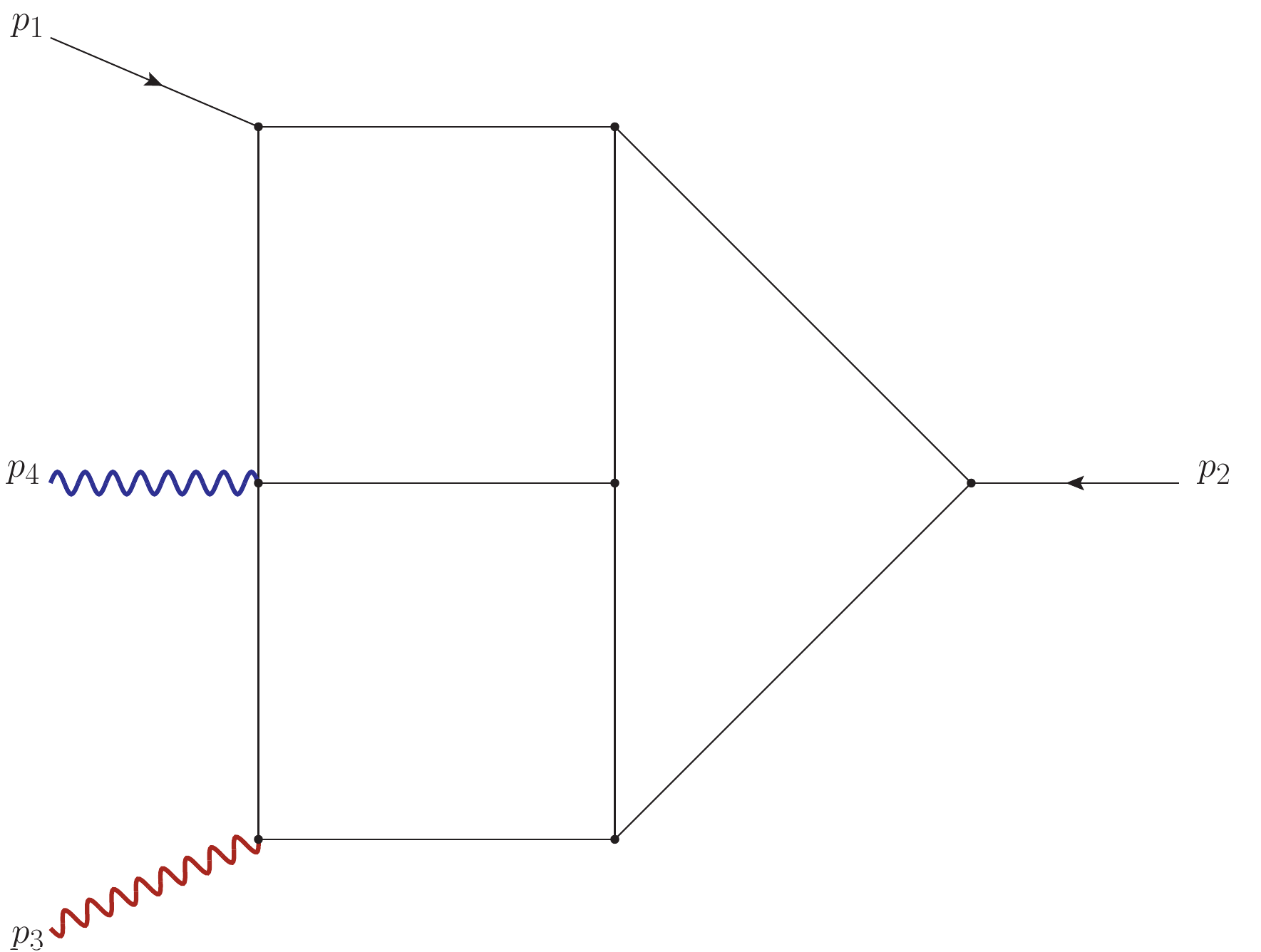}\\
\newline
\begin{equation*}
\{ I_{10}^{\text{PT4}}, \, I_{11}^{\text{PT4}}, \, I_{12}^{\text{PT4}}, \, I_{13}^{\text{PT4}} \}
\end{equation*}
\end{multicols}



\begin{center}
\textbf{\textit{Ten-Propagator Pure Candidates}}\\
\end{center}
\vspace{0.2cm}

\textbf{Sector $\mathbf{F_{123}}$[1,1,1,0,1,0,1,0,1,0,1,1,1,0,1]}

\begin{multicols}{2}
\includegraphics[scale=0.175]{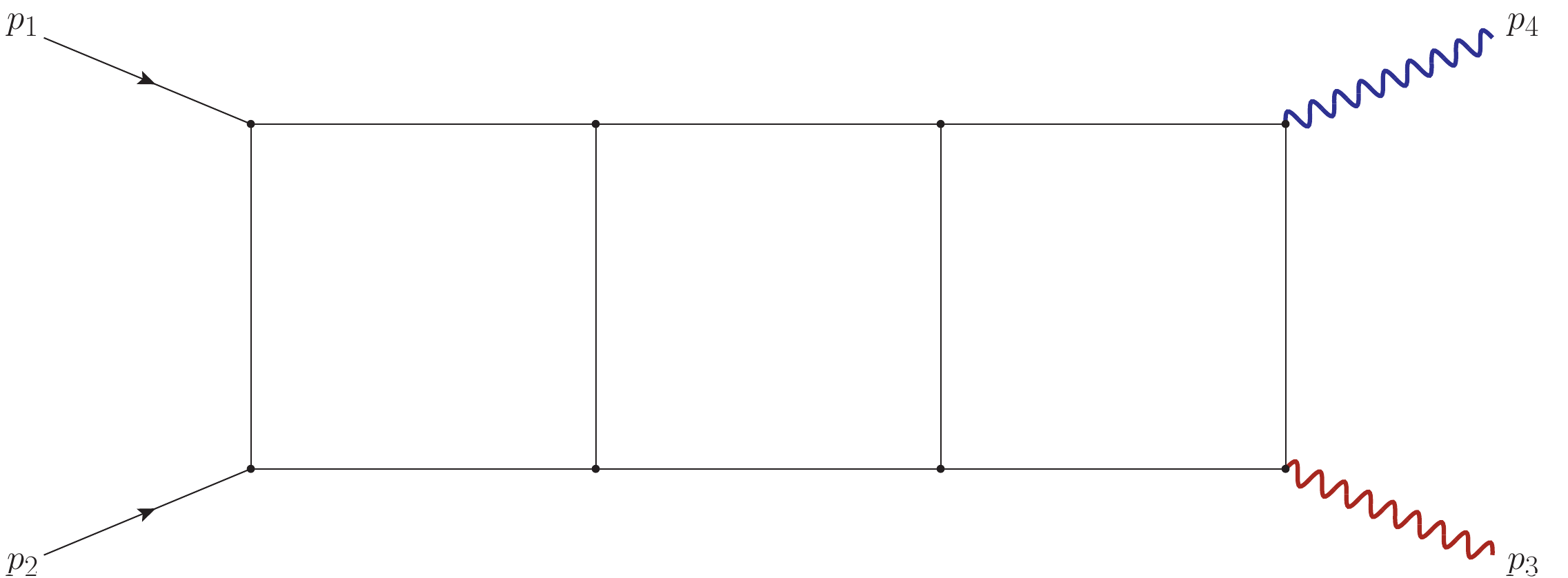}\\
\newline
\begin{equation*}
\{ I_{1}^{\text{PL1}}, \, I_{2}^{\text{PL1}}, \, I_{3}^{\text{PL1}}, \, I_{4}^{\text{PL1}}, \, I_{5}^{\text{PL1}}\}
\end{equation*}
\end{multicols}

\textbf{Sector $\mathbf{F_{123}}$[1,0,1,1,1,1,0,0,0,1,1,0,1,1,1]}

\begin{multicols}{2}
\includegraphics[scale=0.175]{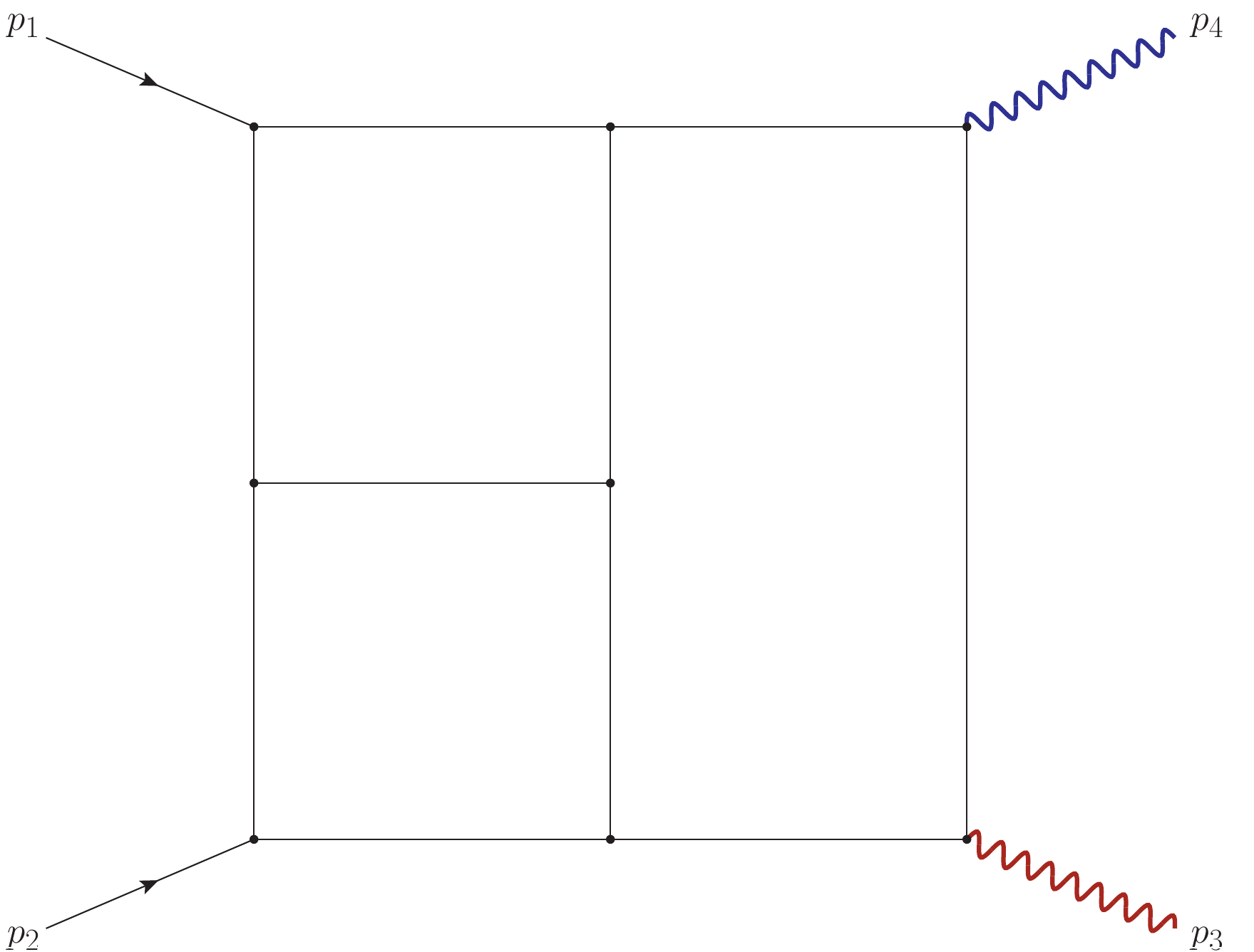}\\
\newline
\begin{equation*}
\{ I_{1}^{\text{PT4}}, \, I_{2}^{\text{PT4}}, \, I_{3}^{\text{PT4}}, \, I_{4}^{\text{PT4}}, \, I_{5}^{\text{PT4}} \}
\end{equation*}
\end{multicols}

\bibliographystyle{JHEP}
\bibliography{biblio}

\providecommand{\href}[2]{#2}\begingroup\raggedright\begin{thebibliography}{10}

\bibitem{Goncharov:1998kja}
A.~B. Goncharov, {\it {Multiple polylogarithms, cyclotomy and modular
  complexes}},  {\em Math. Res. Lett.} {\bf 5} (1998) 497--516,
  [\href{http://arxiv.org/abs/1105.2076}{{\tt arXiv:1105.2076}}].

\bibitem{Goncharov:2010jf}
A.~B. Goncharov, M.~Spradlin, C.~Vergu, and A.~Volovich, {\it {Classical
  Polylogarithms for Amplitudes and Wilson Loops}},  {\em Phys. Rev. Lett.}
  {\bf 105} (2010) 151605, [\href{http://arxiv.org/abs/1006.5703}{{\tt
  arXiv:1006.5703}}].

\bibitem{Broedel:2018qkq}
J.~Broedel, C.~Duhr, F.~Dulat, B.~Penante, and L.~Tancredi, {\it {Elliptic
  Feynman integrals and pure functions}},  {\em JHEP} {\bf 01} (2019) 023,
  [\href{http://arxiv.org/abs/1809.10698}{{\tt arXiv:1809.10698}}].

\bibitem{Bourjaily:2022bwx}
J.~L. Bourjaily et~al., {\it {Functions Beyond Multiple Polylogarithms for
  Precision Collider Physics}},  in {\em {Snowmass 2021}}, 3, 2022.
\newblock \href{http://arxiv.org/abs/2203.07088}{{\tt arXiv:2203.07088}}.

\bibitem{Weinzierl:2022eaz}
S.~Weinzierl, {\em {Feynman Integrals. A Comprehensive Treatment for Students
  and Researchers}}.
\newblock UNITEXT for Physics. Springer, 2022.

\bibitem{Badger:2023eqz}
S.~Badger, J.~Henn, J.~C. Plefka, and S.~Zoia, {\it {Scattering Amplitudes in
  Quantum Field Theory}},  {\em Lect. Notes Phys.} {\bf 1021} (2024) pp.,
  [\href{http://arxiv.org/abs/2306.05976}{{\tt arXiv:2306.05976}}].

\bibitem{Kotikov:1990kg}
A.~V. Kotikov, {\it {Differential equations method: New technique for massive
  Feynman diagrams calculation}},  {\em Phys. Lett. B} {\bf 254} (1991)
  158--164.

\bibitem{Bern:1993kr}
Z.~Bern, L.~J. Dixon, and D.~A. Kosower, {\it {Dimensionally regulated pentagon
  integrals}},  {\em Nucl. Phys. B} {\bf 412} (1994) 751--816,
  [\href{http://arxiv.org/abs/hep-ph/9306240}{{\tt hep-ph/9306240}}].

\bibitem{Remiddi:1997ny}
E.~Remiddi, {\it {Differential equations for Feynman graph amplitudes}},  {\em
  Nuovo Cim. A} {\bf 110} (1997) 1435--1452,
  [\href{http://arxiv.org/abs/hep-th/9711188}{{\tt hep-th/9711188}}].

\bibitem{Gehrmann:1999as}
T.~Gehrmann and E.~Remiddi, {\it {Differential equations for two loop four
  point functions}},  {\em Nucl. Phys. B} {\bf 580} (2000) 485--518,
  [\href{http://arxiv.org/abs/hep-ph/9912329}{{\tt hep-ph/9912329}}].

\bibitem{Tkachov:1981wb}
F.~V. Tkachov, {\it {A theorem on analytical calculability of 4-loop
  renormalization group functions}},  {\em Phys. Lett. B} {\bf 100} (1981)
  65--68.

\bibitem{Chetyrkin:1981qh}
K.~G. Chetyrkin and F.~V. Tkachov, {\it {Integration by parts: The algorithm to
  calculate $\beta$-functions in 4 loops}},  {\em Nucl. Phys. B} {\bf 192}
  (1981) 159--204.

\bibitem{Laporta:2000dsw}
S.~Laporta, {\it {High-precision calculation of multiloop Feynman integrals by
  difference equations}},  {\em Int. J. Mod. Phys. A} {\bf 15} (2000)
  5087--5159, [\href{http://arxiv.org/abs/hep-ph/0102033}{{\tt
  hep-ph/0102033}}].

\bibitem{Henn:2013pwa}
J.~M. Henn, {\it {Multiloop integrals in dimensional regularization made
  simple}},  {\em Phys. Rev. Lett.} {\bf 110} (2013) 251601,
  [\href{http://arxiv.org/abs/1304.1806}{{\tt arXiv:1304.1806}}].

\bibitem{vonManteuffel:2014ixa}
A.~von Manteuffel and R.~M. Schabinger, {\it {A novel approach to integration
  by parts reduction}},  {\em Phys. Lett. B} {\bf 744} (2015) 101--104,
  [\href{http://arxiv.org/abs/1406.4513}{{\tt arXiv:1406.4513}}].

\bibitem{Peraro:2016wsq}
T.~Peraro, {\it {Scattering amplitudes over finite fields and multivariate
  functional reconstruction}},  {\em JHEP} {\bf 12} (2016) 030,
  [\href{http://arxiv.org/abs/1608.01902}{{\tt arXiv:1608.01902}}].

\bibitem{Peraro:2019svx}
T.~Peraro, {\it {$\text{FiniteFlow}$: multivariate functional reconstruction
  using finite fields and dataflow graphs}},  {\em JHEP} {\bf 07} (2019) 031,
  [\href{http://arxiv.org/abs/1905.08019}{{\tt arXiv:1905.08019}}].

\bibitem{Klappert:2019emp}
J.~Klappert and F.~Lange, {\it {Reconstructing rational functions with
  FireFly}},  {\em Comput. Phys. Commun.} {\bf 247} (2020) 106951,
  [\href{http://arxiv.org/abs/1904.00009}{{\tt arXiv:1904.00009}}].

\bibitem{Henn:2024ngj}
J.~M. Henn, A.~Matija\v{s}i\'c, J.~Miczajka, T.~Peraro, Y.~Xu, and Y.~Zhang,
  {\it {A computation of two-loop six-point Feynman integrals in dimensional
  regularization}},  {\em JHEP} {\bf 08} (2024) 027,
  [\href{http://arxiv.org/abs/2403.19742}{{\tt arXiv:2403.19742}}].

\bibitem{Papadopoulos:2015jft}
C.~G. Papadopoulos, D.~Tommasini, and C.~Wever, {\it {The Pentabox Master
  Integrals with the Simplified Differential Equations approach}},  {\em JHEP}
  {\bf 04} (2016) 078, [\href{http://arxiv.org/abs/1511.09404}{{\tt
  arXiv:1511.09404}}].

\bibitem{Gehrmann:2018yef}
T.~Gehrmann, J.~M. Henn, and N.~A. Lo~Presti, {\it {Pentagon functions for
  massless planar scattering amplitudes}},  {\em JHEP} {\bf 10} (2018) 103,
  [\href{http://arxiv.org/abs/1807.09812}{{\tt arXiv:1807.09812}}].

\bibitem{Chicherin:2018mue}
D.~Chicherin, T.~Gehrmann, J.~M. Henn, N.~A. Lo~Presti, V.~Mitev, and
  P.~Wasser, {\it {Analytic result for the nonplanar hexa-box integrals}},
  {\em JHEP} {\bf 03} (2019) 042, [\href{http://arxiv.org/abs/1809.06240}{{\tt
  arXiv:1809.06240}}].

\bibitem{Chicherin:2018old}
D.~Chicherin, T.~Gehrmann, J.~M. Henn, P.~Wasser, Y.~Zhang, and S.~Zoia, {\it
  {All Master Integrals for Three-Jet Production at Next-to-Next-to-Leading
  Order}},  {\em Phys. Rev. Lett.} {\bf 123} (2019), no.~4 041603,
  [\href{http://arxiv.org/abs/1812.11160}{{\tt arXiv:1812.11160}}].

\bibitem{Abreu:2020jxa}
S.~Abreu, H.~Ita, F.~Moriello, B.~Page, W.~Tschernow, and M.~Zeng, {\it
  {Two-Loop Integrals for Planar Five-Point One-Mass Processes}},  {\em JHEP}
  {\bf 11} (2020) 117, [\href{http://arxiv.org/abs/2005.04195}{{\tt
  arXiv:2005.04195}}].

\bibitem{Chicherin:2020oor}
D.~Chicherin and V.~Sotnikov, {\it {Pentagon Functions for Scattering of Five
  Massless Particles}},  {\em JHEP} {\bf 20} (2020) 167,
  [\href{http://arxiv.org/abs/2009.07803}{{\tt arXiv:2009.07803}}].

\bibitem{Canko:2020ylt}
D.~D. Canko, C.~G. Papadopoulos, and N.~Syrrakos, {\it {Analytic representation
  of all planar two-loop five-point Master Integrals with one off-shell leg}},
  {\em JHEP} {\bf 01} (2021) 199, [\href{http://arxiv.org/abs/2009.13917}{{\tt
  arXiv:2009.13917}}].

\bibitem{Abreu:2021smk}
S.~Abreu, H.~Ita, B.~Page, and W.~Tschernow, {\it {Two-loop hexa-box integrals
  for non-planar five-point one-mass processes}},  {\em JHEP} {\bf 03} (2022)
  182, [\href{http://arxiv.org/abs/2107.14180}{{\tt arXiv:2107.14180}}].

\bibitem{Chicherin:2021dyp}
D.~Chicherin, V.~Sotnikov, and S.~Zoia, {\it {Pentagon functions for one-mass
  planar scattering amplitudes}},  {\em JHEP} {\bf 01} (2022) 096,
  [\href{http://arxiv.org/abs/2110.10111}{{\tt arXiv:2110.10111}}].

\bibitem{Kardos:2022tpo}
A.~Kardos, C.~G. Papadopoulos, A.~V. Smirnov, N.~Syrrakos, and C.~Wever, {\it
  {Two-loop non-planar hexa-box integrals with one massive leg}},  {\em JHEP}
  {\bf 05} (2022) 033, [\href{http://arxiv.org/abs/2201.07509}{{\tt
  arXiv:2201.07509}}].

\bibitem{Abreu:2023rco}
S.~Abreu, D.~Chicherin, H.~Ita, B.~Page, V.~Sotnikov, W.~Tschernow, and
  S.~Zoia, {\it {All Two-Loop Feynman Integrals for Five-Point One-Mass
  Scattering}},  {\em Phys. Rev. Lett.} {\bf 132} (2024), no.~14 141601,
  [\href{http://arxiv.org/abs/2306.15431}{{\tt arXiv:2306.15431}}].

\bibitem{Badger:2022hno}
S.~Badger, M.~Becchetti, E.~Chaubey, and R.~Marzucca, {\it {Two-loop master
  integrals for a planar topology contributing to pp \textrightarrow{}$
  t\overline{t}j $}},  {\em JHEP} {\bf 01} (2023) 156,
  [\href{http://arxiv.org/abs/2210.17477}{{\tt arXiv:2210.17477}}].

\bibitem{FebresCordero:2023pww}
F.~Febres~Cordero, G.~Figueiredo, M.~Kraus, B.~Page, and L.~Reina, {\it
  {Two-loop master integrals for leading-color $ pp\to t\overline{t}H $
  amplitudes with a light-quark loop}},  {\em JHEP} {\bf 07} (2024) 084,
  [\href{http://arxiv.org/abs/2312.08131}{{\tt arXiv:2312.08131}}].

\bibitem{Badger:2024fgb}
S.~Badger, M.~Becchetti, N.~Giraudo, and S.~Zoia, {\it {Two-loop integrals for
  $ t\overline{t} $+jet production at hadron colliders in the leading colour
  approximation}},  {\em JHEP} {\bf 07} (2024) 073,
  [\href{http://arxiv.org/abs/2404.12325}{{\tt arXiv:2404.12325}}].

\bibitem{Abreu:2024flk}
S.~Abreu, D.~Chicherin, V.~Sotnikov, and S.~Zoia, {\it {Two-Loop Five-Point
  Two-Mass Planar Integrals and Double Lagrangian Insertions in a Wilson
  Loop}},  \href{http://arxiv.org/abs/2408.05201}{{\tt arXiv:2408.05201}}.

\bibitem{Smirnov:2003vi}
V.~A. Smirnov, {\it {Analytical result for dimensionally regularized massless
  on shell planar triple box}},  {\em Phys. Lett. B} {\bf 567} (2003) 193--199,
  [\href{http://arxiv.org/abs/hep-ph/0305142}{{\tt hep-ph/0305142}}].

\bibitem{Henn:2013tua}
J.~M. Henn, A.~V. Smirnov, and V.~A. Smirnov, {\it {Analytic results for planar
  three-loop four-point integrals from a Knizhnik-Zamolodchikov equation}},
  {\em JHEP} {\bf 07} (2013) 128, [\href{http://arxiv.org/abs/1306.2799}{{\tt
  arXiv:1306.2799}}].

\bibitem{Henn:2013nsa}
J.~M. Henn, A.~V. Smirnov, and V.~A. Smirnov, {\it {Evaluating single-scale
  and/or non-planar diagrams by differential equations}},  {\em JHEP} {\bf 03}
  (2014) 088, [\href{http://arxiv.org/abs/1312.2588}{{\tt arXiv:1312.2588}}].

\bibitem{DiVita:2014pza}
S.~Di~Vita, P.~Mastrolia, U.~Schubert, and V.~Yundin, {\it {Three-loop master
  integrals for ladder-box diagrams with one massive leg}},  {\em JHEP} {\bf
  09} (2014) 148, [\href{http://arxiv.org/abs/1408.3107}{{\tt
  arXiv:1408.3107}}].

\bibitem{Henn:2020lye}
J.~Henn, B.~Mistlberger, V.~A. Smirnov, and P.~Wasser, {\it {Constructing d-log
  integrands and computing master integrals for three-loop four-particle
  scattering}},  {\em JHEP} {\bf 04} (2020) 167,
  [\href{http://arxiv.org/abs/2002.09492}{{\tt arXiv:2002.09492}}].

\bibitem{Canko:2020gqp}
D.~D. Canko and N.~Syrrakos, {\it {Resummation methods for Master Integrals}},
  {\em JHEP} {\bf 02} (2021) 080, [\href{http://arxiv.org/abs/2010.06947}{{\tt
  arXiv:2010.06947}}].

\bibitem{Canko:2021xmn}
D.~D. Canko and N.~Syrrakos, {\it {Planar three-loop master integrals for 2
  \textrightarrow{} 2 processes with one external massive particle}},  {\em
  JHEP} {\bf 04} (2022) 134, [\href{http://arxiv.org/abs/2112.14275}{{\tt
  arXiv:2112.14275}}].

\bibitem{Henn:2023vbd}
J.~M. Henn, J.~Lim, and W.~J. Torres~Bobadilla, {\it {First look at the
  evaluation of three-loop non-planar Feynman diagrams for Higgs plus jet
  production}},  {\em JHEP} {\bf 05} (2023) 026,
  [\href{http://arxiv.org/abs/2302.12776}{{\tt arXiv:2302.12776}}].

\bibitem{Syrrakos:2023mor}
N.~Syrrakos and D.~D. Canko, {\it {Three-loop master integrals for H+jet
  production at N3LO: Towards the non-planar topologies}},  {\em PoS} {\bf
  RADCOR2023} (2024) 044, [\href{http://arxiv.org/abs/2307.08432}{{\tt
  arXiv:2307.08432}}].

\bibitem{Long:2024bmi}
M.-M. Long, {\it {Three-loop ladder diagrams with two off-shell legs}},
  \href{http://arxiv.org/abs/2410.15431}{{\tt arXiv:2410.15431}}.

\bibitem{Gehrmann:2024tds}
T.~Gehrmann, J.~Henn, P.~Jakub\v{c}\'\i{}k, J.~Lim, C.~C. Mella, N.~Syrrakos,
  L.~Tancredi, and W.~J. Torres~Bobadilla, {\it {Graded transcendental
  functions: an application to four-point amplitudes with one off-shell leg}},
  \href{http://arxiv.org/abs/2410.19088}{{\tt arXiv:2410.19088}}.

\bibitem{Henn:2024pki}
J.~M. Henn, J.~Lim, and W.~J. Torres~Bobadilla, {\it {Analytic evaluation of
  the three-loop three-point form factor of $\operatorname{tr}\phi^3$ in
  $\mathcal{N}=4$ sYM}},  \href{http://arxiv.org/abs/2410.22465}{{\tt
  arXiv:2410.22465}}.

\bibitem{Liu:2024ont}
Y.~Liu, A.~Matija\v{s}i\'c, J.~Miczajka, Y.~Xu, Y.~Xu, and Y.~Zhang, {\it {An
  Analytic Computation of Three-Loop Five-Point Feynman Integrals}},
  \href{http://arxiv.org/abs/2411.18697}{{\tt arXiv:2411.18697}}.

\bibitem{Gehrmann:2021qex}
T.~Gehrmann and B.~Malaescu, {\it {Precision QCD Physics at the LHC}},  {\em
  Ann. Rev. Nucl. Part. Sci.} {\bf 72} (2022) 233--258,
  [\href{http://arxiv.org/abs/2111.02319}{{\tt arXiv:2111.02319}}].

\bibitem{Henn:2014lfa}
J.~M. Henn, K.~Melnikov, and V.~A. Smirnov, {\it {Two-loop planar master
  integrals for the production of off-shell vector bosons in hadron
  collisions}},  {\em JHEP} {\bf 05} (2014) 090,
  [\href{http://arxiv.org/abs/1402.7078}{{\tt arXiv:1402.7078}}].

\bibitem{Caola:2014lpa}
F.~Caola, J.~M. Henn, K.~Melnikov, and V.~A. Smirnov, {\it {Non-planar master
  integrals for the production of two off-shell vector bosons in collisions of
  massless partons}},  {\em JHEP} {\bf 09} (2014) 043,
  [\href{http://arxiv.org/abs/1404.5590}{{\tt arXiv:1404.5590}}].

\bibitem{Caola:2014iua}
F.~Caola, J.~M. Henn, K.~Melnikov, A.~V. Smirnov, and V.~A. Smirnov, {\it
  {Two-loop helicity amplitudes for the production of two off-shell electroweak
  bosons in quark-antiquark collisions}},  {\em JHEP} {\bf 11} (2014) 041,
  [\href{http://arxiv.org/abs/1408.6409}{{\tt arXiv:1408.6409}}].

\bibitem{Caola:2015ila}
F.~Caola, J.~M. Henn, K.~Melnikov, A.~V. Smirnov, and V.~A. Smirnov, {\it
  {Two-loop helicity amplitudes for the production of two off-shell electroweak
  bosons in gluon fusion}},  {\em JHEP} {\bf 06} (2015) 129,
  [\href{http://arxiv.org/abs/1503.08759}{{\tt arXiv:1503.08759}}].

\bibitem{Gehrmann:2015ora}
T.~Gehrmann, A.~von Manteuffel, and L.~Tancredi, {\it {The two-loop helicity
  amplitudes for $ q\overline{q}^{\prime}\to {V}_1{V}_2\to 4 $ leptons}},  {\em
  JHEP} {\bf 09} (2015) 128, [\href{http://arxiv.org/abs/1503.04812}{{\tt
  arXiv:1503.04812}}].

\bibitem{vonManteuffel:2015msa}
A.~von Manteuffel and L.~Tancredi, {\it {The two-loop helicity amplitudes for
  $gg \to V_1 V_2 \to 4~\mathrm{leptons}$}},  {\em JHEP} {\bf 06} (2015) 197,
  [\href{http://arxiv.org/abs/1503.08835}{{\tt arXiv:1503.08835}}].

\bibitem{Grazzini:2015wpa}
M.~Grazzini, S.~Kallweit, D.~Rathlev, and M.~Wiesemann, {\it
  {Transverse-momentum resummation for vector-boson pair production at
  NNLL+NNLO}},  {\em JHEP} {\bf 08} (2015) 154,
  [\href{http://arxiv.org/abs/1507.02565}{{\tt arXiv:1507.02565}}].

\bibitem{Grazzini:2015hta}
M.~Grazzini, S.~Kallweit, and D.~Rathlev, {\it {ZZ production at the LHC:
  fiducial cross sections and distributions in NNLO QCD}},  {\em Phys. Lett. B}
  {\bf 750} (2015) 407--410, [\href{http://arxiv.org/abs/1507.06257}{{\tt
  arXiv:1507.06257}}].

\bibitem{Caola:2015psa}
F.~Caola, K.~Melnikov, R.~R\"ontsch, and L.~Tancredi, {\it {QCD corrections to
  ZZ production in gluon fusion at the LHC}},  {\em Phys. Rev. D} {\bf 92}
  (2015), no.~9 094028, [\href{http://arxiv.org/abs/1509.06734}{{\tt
  arXiv:1509.06734}}].

\bibitem{Grazzini:2016swo}
M.~Grazzini, S.~Kallweit, D.~Rathlev, and M.~Wiesemann, {\it {$W^{\pm}Z$
  production at hadron colliders in NNLO QCD}},  {\em Phys. Lett. B} {\bf 761}
  (2016) 179--183, [\href{http://arxiv.org/abs/1604.08576}{{\tt
  arXiv:1604.08576}}].

\bibitem{Campbell:2016ivq}
J.~M. Campbell, R.~K. Ellis, M.~Czakon, and S.~Kirchner, {\it {Two loop
  correction to interference in $gg \to ZZ$}},  {\em JHEP} {\bf 08} (2016) 011,
  [\href{http://arxiv.org/abs/1605.01380}{{\tt arXiv:1605.01380}}].

\bibitem{Grazzini:2016ctr}
M.~Grazzini, S.~Kallweit, S.~Pozzorini, D.~Rathlev, and M.~Wiesemann, {\it
  {$W^+W^-$ production at the LHC: fiducial cross sections and distributions in
  NNLO QCD}},  {\em JHEP} {\bf 08} (2016) 140,
  [\href{http://arxiv.org/abs/1605.02716}{{\tt arXiv:1605.02716}}].

\bibitem{Grazzini:2017ckn}
M.~Grazzini, S.~Kallweit, D.~Rathlev, and M.~Wiesemann, {\it {$W^\pm Z$
  production at the LHC: fiducial cross sections and distributions in NNLO
  QCD}},  {\em JHEP} {\bf 05} (2017) 139,
  [\href{http://arxiv.org/abs/1703.09065}{{\tt arXiv:1703.09065}}].

\bibitem{Heinrich:2017bvg}
G.~Heinrich, S.~Jahn, S.~P. Jones, M.~Kerner, and J.~Pires, {\it {NNLO
  predictions for Z-boson pair production at the LHC}},  {\em JHEP} {\bf 03}
  (2018) 142, [\href{http://arxiv.org/abs/1710.06294}{{\tt arXiv:1710.06294}}].

\bibitem{Re:2018vac}
E.~Re, M.~Wiesemann, and G.~Zanderighi, {\it {NNLOPS accurate predictions for
  $W^+W^-$ production}},  {\em JHEP} {\bf 12} (2018) 121,
  [\href{http://arxiv.org/abs/1805.09857}{{\tt arXiv:1805.09857}}].

\bibitem{Kallweit:2018nyv}
S.~Kallweit and M.~Wiesemann, {\it {$ZZ$ production at the LHC: NNLO
  predictions for $2\ell2\nu$ and $4\ell$ signatures}},  {\em Phys. Lett. B}
  {\bf 786} (2018) 382--389, [\href{http://arxiv.org/abs/1806.05941}{{\tt
  arXiv:1806.05941}}].

\bibitem{Grazzini:2018owa}
M.~Grazzini, S.~Kallweit, M.~Wiesemann, and J.~Y. Yook, {\it {$ZZ$ production
  at the LHC: NLO QCD corrections to the loop-induced gluon fusion channel}},
  {\em JHEP} {\bf 03} (2019) 070, [\href{http://arxiv.org/abs/1811.09593}{{\tt
  arXiv:1811.09593}}].

\bibitem{Grazzini:2019jkl}
M.~Grazzini, S.~Kallweit, J.~M. Lindert, S.~Pozzorini, and M.~Wiesemann, {\it
  {NNLO QCD + NLO EW with Matrix+OpenLoops: precise predictions for
  vector-boson pair production}},  {\em JHEP} {\bf 02} (2020) 087,
  [\href{http://arxiv.org/abs/1912.00068}{{\tt arXiv:1912.00068}}].

\bibitem{Kallweit:2020gva}
S.~Kallweit, E.~Re, L.~Rottoli, and M.~Wiesemann, {\it {Accurate single- and
  double-differential resummation of colour-singlet processes with
  MATRIX+RADISH: $W^+W^-$ production at the LHC}},  {\em JHEP} {\bf 12} (2020)
  147, [\href{http://arxiv.org/abs/2004.07720}{{\tt arXiv:2004.07720}}].

\bibitem{Poncelet:2021jmj}
R.~Poncelet and A.~Popescu, {\it {NNLO QCD study of polarised W$^+$W$^-$
  production at the LHC}},  {\em JHEP} {\bf 07} (2021) 023,
  [\href{http://arxiv.org/abs/2102.13583}{{\tt arXiv:2102.13583}}].

\bibitem{Lombardi:2021rvg}
D.~Lombardi, M.~Wiesemann, and G.~Zanderighi, {\it {W$^+$W$^-$ production at
  NNLO+PS with MINNLO$_{PS}$}},  {\em JHEP} {\bf 11} (2021) 230,
  [\href{http://arxiv.org/abs/2103.12077}{{\tt arXiv:2103.12077}}].

\bibitem{Degrassi:2024fye}
G.~Degrassi, R.~Gr\"ober, and M.~Vitti, {\it {Virtual QCD corrections to gg
  \textrightarrow{} ZZ: top-quark loops from a transverse-momentum expansion}},
   {\em JHEP} {\bf 07} (2024) 244, [\href{http://arxiv.org/abs/2404.15113}{{\tt
  arXiv:2404.15113}}].

\bibitem{Jiang:2024eaj}
X.~Jiang, J.~Liu, X.~Xu, and L.~L. Yang, {\it {Symbol letters of Feynman
  integrals from Gram determinants}},
  \href{http://arxiv.org/abs/2401.07632}{{\tt arXiv:2401.07632}}.

\bibitem{canko_2024_14284044}
D.~Canko and M.~Pozzoli, ``\textit{Ancillary files for the article A first
  computation of three-loop master integrals for the production of two
  off-shell vector bosons with different masses}.''
  \url{https://doi.org/10.5281/zenodo.14284044}, 9 December 2024.

\bibitem{Gehrmann:2014bfa}
T.~Gehrmann, A.~von Manteuffel, L.~Tancredi, and E.~Weihs, {\it {The two-loop
  master integrals for $q\overline{q} \to VV$}},  {\em JHEP} {\bf 06} (2014)
  032, [\href{http://arxiv.org/abs/1404.4853}{{\tt arXiv:1404.4853}}].

\bibitem{Argeri:2014qva}
M.~Argeri, S.~Di~Vita, P.~Mastrolia, E.~Mirabella, J.~Schlenk, U.~Schubert, and
  L.~Tancredi, {\it {Magnus and Dyson Series for Master Integrals}},  {\em
  JHEP} {\bf 03} (2014) 082, [\href{http://arxiv.org/abs/1401.2979}{{\tt
  arXiv:1401.2979}}].

\bibitem{Wasser:2018qvj}
P.~Wasser, {\em {Analytic properties of Feynman integrals for scattering
  amplitudes}}.
\newblock PhD thesis, Mainz U., 2018.

\bibitem{Papadopoulos:2014lla}
C.~G. Papadopoulos, {\it {Simplified differential equations approach for Master
  Integrals}},  {\em JHEP} {\bf 07} (2014) 088,
  [\href{http://arxiv.org/abs/1401.6057}{{\tt arXiv:1401.6057}}].

\bibitem{Canko:2021hvh}
D.~D. Canko, F.~Gasparotto, L.~Mattiazzi, C.~G. Papadopoulos, and N.~Syrrakos,
  {\it {$N^3LO$ calculations for $2 \to 2$ processes using simplified
  differential equations}},  {\em SciPost Phys. Proc.} {\bf 7} (2022) 028,
  [\href{http://arxiv.org/abs/2110.08110}{{\tt arXiv:2110.08110}}].

\bibitem{Baikov:1996iu}
P.~A. Baikov, {\it {Explicit solutions of the multiloop integral recurrence
  relations and its application}},  {\em Nucl. Instrum. Meth. A} {\bf 389}
  (1997) 347--349, [\href{http://arxiv.org/abs/hep-ph/9611449}{{\tt
  hep-ph/9611449}}].

\bibitem{Baikov:1996rk}
P.~A. Baikov, {\it {Explicit solutions of the three loop vacuum integral
  recurrence relations}},  {\em Phys. Lett. B} {\bf 385} (1996) 404--410,
  [\href{http://arxiv.org/abs/hep-ph/9603267}{{\tt hep-ph/9603267}}].

\bibitem{Frellesvig:2017aai}
H.~Frellesvig and C.~G. Papadopoulos, {\it {Cuts of Feynman Integrals in Baikov
  representation}},  {\em JHEP} {\bf 04} (2017) 083,
  [\href{http://arxiv.org/abs/1701.07356}{{\tt arXiv:1701.07356}}].

\bibitem{Duhr:2011zq}
C.~Duhr, H.~Gangl, and J.~R. Rhodes, {\it {From polygons and symbols to
  polylogarithmic functions}},  {\em JHEP} {\bf 10} (2012) 075,
  [\href{http://arxiv.org/abs/1110.0458}{{\tt arXiv:1110.0458}}].

\bibitem{Vollinga:2004sn}
J.~Vollinga and S.~Weinzierl, {\it {Numerical evaluation of multiple
  polylogarithms}},  {\em Comput. Phys. Commun.} {\bf 167} (2005) 177,
  [\href{http://arxiv.org/abs/hep-ph/0410259}{{\tt hep-ph/0410259}}].

\bibitem{Badger:2023xtl}
S.~Badger, J.~Kry\'s, R.~Moodie, and S.~Zoia, {\it {Lepton-pair scattering with
  an off-shell and an on-shell photon at two loops in massless QED}},  {\em
  JHEP} {\bf 11} (2023) 041, [\href{http://arxiv.org/abs/2307.03098}{{\tt
  arXiv:2307.03098}}].

\bibitem{Jantzen:2012mw}
B.~Jantzen, A.~V. Smirnov, and V.~A. Smirnov, {\it {Expansion by regions:
  revealing potential and Glauber regions automatically}},  {\em Eur. Phys. J.
  C} {\bf 72} (2012) 2139, [\href{http://arxiv.org/abs/1206.0546}{{\tt
  arXiv:1206.0546}}].

\bibitem{Liu:2022chg}
X.~Liu and Y.-Q. Ma, {\it {AMFlow: A Mathematica package for Feynman integrals
  computation via auxiliary mass flow}},  {\em Comput. Phys. Commun.} {\bf 283}
  (2023) 108565, [\href{http://arxiv.org/abs/2201.11669}{{\tt
  arXiv:2201.11669}}].

\bibitem{Liu:2017jxz}
X.~Liu, Y.-Q. Ma, and C.-Y. Wang, {\it {A Systematic and Efficient Method to
  Compute Multi-loop Master Integrals}},  {\em Phys. Lett. B} {\bf 779} (2018)
  353--357, [\href{http://arxiv.org/abs/1711.09572}{{\tt arXiv:1711.09572}}].

\bibitem{Hidding:2020ytt}
M.~Hidding, {\it {DiffExp, a Mathematica package for computing Feynman
  integrals in terms of one-dimensional series expansions}},  {\em Comput.
  Phys. Commun.} {\bf 269} (2021) 108125,
  [\href{http://arxiv.org/abs/2006.05510}{{\tt arXiv:2006.05510}}].

\bibitem{pslq}
H.~Ferguson and D.~Bailey, {\it {A Polynomial Time, Numerically Stable Integer
  Relation Algorithm}},  {\em RNR Technical Report RNR-91-032} (1992).

\bibitem{Lyndon}
D.~Radford, {\it {A natural ring basis for the shuffle algebra and an
  application to group schemes}},  {\em J. Algebra} {\bf 58} (1979) 432--454.

\bibitem{Duhr:2019tlz}
C.~Duhr and F.~Dulat, {\it {PolyLogTools \textemdash{} polylogs for the
  masses}},  {\em JHEP} {\bf 08} (2019) 135,
  [\href{http://arxiv.org/abs/1904.07279}{{\tt arXiv:1904.07279}}].

\bibitem{Moriello:2019yhu}
F.~Moriello, {\it {Generalised power series expansions for the elliptic planar
  families of Higgs + jet production at two loops}},  {\em JHEP} {\bf 01}
  (2020) 150, [\href{http://arxiv.org/abs/1907.13234}{{\tt arXiv:1907.13234}}].

\bibitem{ginac}
\texttt{GiNaC}, {\it http://www.ginac.de},  2019.

\bibitem{Naterop:2019xaf}
L.~Naterop, A.~Signer, and Y.~Ulrich, {\it {handyG \textemdash{}Rapid numerical
  evaluation of generalised polylogarithms in Fortran}},  {\em Comput. Phys.
  Commun.} {\bf 253} (2020) 107165,
  [\href{http://arxiv.org/abs/1909.01656}{{\tt arXiv:1909.01656}}].

\bibitem{Chicherin:2020umh}
D.~Chicherin, J.~M. Henn, and G.~Papathanasiou, {\it {Cluster algebras for
  Feynman integrals}},  {\em Phys. Rev. Lett.} {\bf 126} (2021), no.~9 091603,
  [\href{http://arxiv.org/abs/2012.12285}{{\tt arXiv:2012.12285}}].

\bibitem{Aliaj:2024zgp}
R.~Aliaj and G.~Papathanasiou, {\it {An Exceptional Cluster Algebra for Higgs
  plus Jet Production}},  \href{http://arxiv.org/abs/2408.14544}{{\tt
  arXiv:2408.14544}}.

\end{thebibliography}\endgroup

\end{document}